\documentclass[12pt]{article}
\usepackage{amssymb,amsmath}
\usepackage{color,graphicx}
\usepackage{cite}
\usepackage{yfonts}
\newcommand\fverb{\setbox\pippobox=\hbox\bgroup\verb}
\newcommand\fverbdo{\egroup\medskip\noindent%
                              \fbox{\unhbox\pippobox}\ }
\newcommand\fverbit{\egroup\item[\fbox{\unhbox\pippobox}]}
\newbox\pippobox

\newcommand{\be} {\begin{equation}}
\newcommand{\ee} {\end{equation}}
\newcommand{\beq} {\begin{equation}}
\newcommand{\eeq} {\end{equation}}
\newcommand{\bea} {\begin{eqnarray}}
\newcommand{\eea} {\end{eqnarray}}
\newcommand{\bear}{\begin{eqnarray}}
\newcommand{\eear}{\end{eqnarray}}

\newcommand{\rc}{\nonumber\\}
\newcommand{\Tr}{\mbox{Tr}}    
\def\tr{{\rm Tr}}
\def\ie{{\em i.e.}}

\def\om{\omega}

\begin{document}
 
\begin{flushright}
HIP-2015-9/TH
\end{flushright}

\begin{center}


\centerline{\Large {\bf Universal properties of cold holographic matter}}

\vspace{8mm}

\renewcommand\thefootnote{\mbox{$\fnsymbol{footnote}$}}
Niko Jokela${}^{1,2}$\footnote{niko.jokela@helsinki.fi}
and Alfonso V. Ramallo${}^{3,4}$\footnote{alfonso@fpaxp1.usc.es}

\vspace{4mm}
${}^1${\small \sl Department of Physics} and ${}^2${\small \sl Helsinki Institute of Physics} \\
{\small \sl P.O.Box 64} \\
{\small \sl FIN-00014 University of Helsinki, Finland} 

\vskip 0.2cm
${}^3${\small \sl Departamento de  F\'\i sica de Part\'\i  culas} \\
{\small \sl Universidade de Santiago de Compostela} \\
{\small \sl and} \\
${}^4${\small \sl Instituto Galego de F\'\i sica de Altas Enerx\'\i as (IGFAE)} \\
{\small \sl E-15782 Santiago de Compostela, Spain}

\end{center}

\vspace{8mm}
\numberwithin{equation}{section}
\setcounter{footnote}{0}
\renewcommand\thefootnote{\mbox{\arabic{footnote}}}

\begin{abstract}
\noindent
We study the collective excitations of holographic quantum liquids formed in the low energy theory living at the intersection of two sets of D-branes. The corresponding field theory dual is a supersymmetric Yang-Mills theory with massless matter hypermultiplets in the fundamental representation of the gauge group which generically live on a defect of the unflavored theory. Working in the quenched (probe) approximation, we focus on determining the universal properties of these systems.  We analyze their thermodynamics, the speed of first sound, the diffusion constant, and the speed of zero sound. We study the influence of temperature, chemical potential, and magnetic field on these quantities, as well as on the corresponding collisionless/hydrodynamic crossover.  We also generalize the alternative quantization for all conformally $AdS_4$ cases and study the anyonic correlators.
\end{abstract}

\newpage
\tableofcontents

\renewcommand{\theequation}{{\rm\thesection.\arabic{equation}}}
%
%


\section{Introduction}

The understanding of new phases of matter is one of the main goals of fundamental physics. At low temperature and non-zero density, Fermi liquids are commonly described by the Landau's phenomenological theory \cite{Landau}, in which the elementary excitations take place near the Fermi surface and  are fermionic quasiparticles. Landau's theory has been very successful in dealing with an ample variety of low temperature materials \cite{LFL}. However, there are examples of systems whose behaviors are not well described by the Landau theory. The gauge/gravity duality \cite{AdS_CFT_reviews} is a new principle which could shed light on in establishing new paradigms in the case of systems with strong interactions and without a quasiparticle description. In particular, there is some hope that holography could help to classify compressible states of matter, \ie, states with non-zero charge density which varies continuously with the chemical potential $\mu$. 

In this paper we adopt a top-down approach to this problem and explore the properties of matter engineered with intersections of D-branes of different dimensionalities at non-zero density and low temperature. 
We will consider generic intersections in which $N_c$ D$p$-branes intersect $N_f$ D$q$-branes, with $q\ge p$, along $n$ common spatial directions. We will denote by $(n\, |\, p\perp q)$ such a D-brane intersection (for example $(3\, |\, 3\perp 7)$ for the familiar supersymmetric D3-D7 configuration). The gauge theory dual of this $(n\, |\, p\perp q)$ configuration corresponds to a $(p+1)$-dimensional $SU(N_c)$  gauge theory in which one adds $N_f$ fundamental hypermultiplets living on a $(n+1)$-dimensional defect \cite{Karch:2002sh}.

Our description will be valid in the 't Hooft large-$N_c$ limit, with large 't Hooft coupling. In addition, in our approach we will assume that $N_f\ll N_c$ and we will treat the D$q$-branes as probes in the gravitational background created by the D$p$-branes, which corresponds to the quenched approximation on the gauge theory side.

By turning on  a suitable worldvolume gauge field  we will add a non-zero baryonic charge density \cite{Kobayashi:2006sb}. In addition we will switch on a magnetic field along two of the spatial directions of the worldvolume. The embedding of the brane probe is characterized by a function, which measures the distance between the two sets of branes. The gauge theory dual of this distance is the mass of the hypermultiplet fields. In the present paper we will only consider massless fundamentals for which the embedding function is trivial. We leave the analysis of the massive case for a future work. 

We aim to determine universal properties for this holographic matter, which do not depend much on the particular intersections and are common to all of them. 
It turns out that the different observables studied depend only on  the dimensionality $p$ of the bulk theory, as well as on an index $\lambda$, defined in terms of $p$, $q$,  and $n$ as:
\beq
\lambda\,=\,2n\,+\,{1\over 2}\,(p-3)\,(p+q-2n-8)\,\,.
\label{lambda_n}
\eeq
This index $\lambda$ is the same for several intersections (for example, $\lambda=6$ for all  $(p\, |\, p\perp (p+4))$ configurations). Both $p$ and $\lambda$ determine a universality class in the set of intersections analyzed.

Among the observables we study are the thermodynamic properties (entropy and specific heat), speed of first sound, as well as the excitation spectra. The latter can be obtained by looking at the quasinormal fluctuation modes of the system. At zero temperature we look at the holographic zero sound, which is a collective mode first found in \cite{Karch:2008fa,Karch:2009zz} and then generalized to finite temperature in \cite{Bergman:2011rf} and to magnetic fields in \cite{Jokela:2012vn}. The holographic zero sound has been on focus in many different contexts \cite{Kulaxizi:2008kv,Kulaxizi:2008jx,Kim:2008bv,Hung:2009qk,Lee:2010uy,Edalati:2010pn,Lee:2010ez,HoyosBadajoz:2010kd,Ammon:2011hz,Davison:2011ek,Goykhman:2012vy,Gorsky:2012gi,Davison:2013bxa,Pang:2013ypa,Dey:2013vja,Edalati:2013tma,Davison:2013uha,DiNunno:2014bxa,Brattan:2012nb,Jokela:2012se,Arias:2014msa} and, in particular, for some of the D-brane intersections included in our analysis. Our results reproduce the values found previously for the speed and the attenuation of the zero sound and generalize them for arbitrary intersections. The zero sound mode is the dominant one at sufficiently low temperature, where quantum effects dominate over the thermal effects. 

At enough high temperature thermal effects become more important than quantum effects and the system enters into a hydrodynamic regime. We will find that the dominant collective mode in this regime is a diffusion mode and we will be able to calculate an analytic expression for the corresponding diffusion constant. At temperatures between the collisionless and hydrodynamic regimes there is a crossover transition which we will study numerically for the different values of $(p,\lambda)$. 

One of the main objectives of this paper is to study the influence of the magnetic field on the collective excitations of the different systems; previous studies include D3-D7' \cite{Jokela:2012vn}, D2-D8' \cite{Jokela:2012se}, D3-D5 \cite{Goykhman:2012vy,Brattan:2012nb}, and D3-D7 \cite{Brattan:2012nb}. It is known that, for sufficiently large $B$ at $T=0$ and non-zero density, a quantum phase transition takes place in the D3-D7 model \cite{Jensen:2010vd}. Similar transitions may happen in other intersections, we thus assume that $B$ is below the critical value of the possible phase transition.\footnote{$\#ND=6$ models: Sakai-Sugimoto, D3-D7', and D2-D8', are subject to modulated instabilities, so they need to be studied at large enough temperatures (relative to charge densities).}

In the presence of the magnetic field $B$ parity is broken  and  the longitudinal and transverse oscillations of the brane probe are coupled. This coupling corresponds to the mixing of the dual field theory operators that is induced when $B\not=0$. For small values of $B$ we will be able to decouple the equations of the zero sound mode. The result of this analysis is that the zero sound is now gapped, with the gap being given by $B/\mu$ for any values of $(p, \lambda)$ at $T=0$. This result generalizes the ones, {\emph{e.g.}}, in \cite{Jokela:2012vn,Brattan:2012nb} and is in agreement with Kohn's theorem \cite{Khon}. At non-zero temperature there is a critical magnetic field above which the zero sound becomes massive \cite{Jokela:2012vn}. As in \cite{Brattan:2012nb}, the location of the collisionless/hydrodynamic crossover is insensitive to the strength of the magnetic field. We will also study the diffusion mode and determine the magnetic field dependence of the diffusion constant. 

When the intersection is ($2+1$)-dimensional there exists the possibility of imposing mixed Dirichlet-Neumann UV boundary conditions \cite{Jokela:2013hta}, \ie, to perform an alternative quantization, to the quasinormal modes. This alternative quantization corresponds to rendering the charge carrying excitations to anyons (particles of fractional statistics) and is characterized by some constant 
$\textfrak{n}$, which measures the degree of mixing of the boundary conditions. We will study the collective excitations of the anyonic fluids in the presence of the $B$ field and we will find that the zero sound is generically gapped, although the parameter $\textfrak{n}$ can be fine-tuned to produce gapless spectrum precisely when the anyons feel a vanishing effective magnetic field. This parallels the findings \cite{Jokela:2013hta,Jokela:2014wsa} in the anyonic superfluid formed in the D3-D7' model in the incompressible phase. We will also study the diffusion constant and the DC conductivities of the anyonic systems. Related works in the D3-D5 model appeared in \cite{Brattan:2013wya,Brattan:2014moa}.

The rest of this paper is organized as follows. In section \ref{setup} we present our general setup and analyze its thermodynamical properties, both at zero and non-zero temperature. In section \ref{sec:fluctuations}  we begin to study the fluctuations of the probe. In section \ref{zeroB} we concentrate on the case of vanishing magnetic field. We find analytic equations for the diffusion constant and the spectrum of the zero sound. These analytic expressions are compared with numerical calculations. Section \ref {nonzeroB} is devoted to the study of the effects due to the magnetic field in the diffusion constant and the zero sound. In section  \ref{anyons} we perform the alternative quantization for the ($2+1$)-dimensional intersections. 
Section \ref{sec:discussions} contains a discussion on the scaling behavior of different dimensionful quantities. For example, we argue that $\lambda$ determines the scaling dimension of the charge density.
Section  \ref{conclusions} contains our conclusions and discusses future lines of research not included in the present work. 

The paper is completed with several appendices that complement the analysis made in the main text. In appendix  \ref{appendixA} we collect the equations of motion for the fluctuations. 
In appendix  \ref{appendixB} we solve the indicial equation for the fluctuations around the horizon.  In appendix  \ref{Trasnverse_correlator} we find the correlator of two transverse currents, in the absence of the magnetic field. Finally, in appendix \ref{appendixD} we detail the method used to solve the coupled equations of the $B\not=0$  zero sound.


\section{D$p$-D$q$ systems with charge and magnetic field at non-zero temperature}
\label{setup}
In this section we introduce our setup and study some of its properties. 
Let us consider a generic ten-dimensional background metric at zero temperature (we will soon relax this assumption) of the type:
\beq
ds_{10}^2 = g_{tt}(r)dt^2+g_{xx}(r)\big[(dx^1)^2+\ldots+(dx^p)^2\big]+g_{rr}(r) d\vec y\cdot  d\vec y \ ,
\eeq
where $\vec y\,=\,(y^1, \ldots, y^{9-p})$ are the coordinates transverse to the D$p$-brane and the functions $g_{tt}$, $g_{xx}$, and 
$g_{rr}$ depend on the transverse radial direction $r$ ($r^2=\vec y\cdot \vec y$). We note, that these metric components have explicit functional forms in the case of (black) D$p$-brane background, but we
retain from revealing them until Section \ref{sec:intersections} to stress that the following treatment is rather general. 
We now embed a D$q$-brane probe extended along the directions 
\beq
(t, x^1,\ldots, x^n,y^1,\ldots, y^{q-n})\,\,.
\eeq
We will refer to this configuration as a $(n\, |\, p\perp q)$ intersection ($n$ is the number of common spatial directions of the D$p$- and D$q$-branes). This intersection is represented by the array:
\beq
\begin{array}{cccccccccccccl}
 &x^1&\cdots&x^n&x^{n+1}& \cdots& x^p&y^1 &\cdots&y^{q-n}&y^{q-n+1}&\cdots &y^{9-p}& \nonumber \\
Dp: & \times &\cdots &\times&\hspace{-0.2truein}\times&\cdots &\times & -&\cdots &-&\hspace{-0.2truein}-&\cdots  &\hspace{-0.2truein}-    \nonumber \\
Dq: &\times&\cdots&\times&\hspace{-0.2truein}-&\cdots&-&\times&\cdots&\times&\hspace{-0.2truein}-&\cdots&\hspace{-0.2truein}-&
\end{array}\,\,.
\label{D3D5intersection}
\eeq
We will denote by $\vec z$ the coordinates $\vec y$ transverse to the D$q$-brane:
\beq
\vec z\,=\,(z^1,\ldots,z^{9+n-p-q})\,\,,
\eeq
with $z^m\,=\,y^{q-n+m}$ for $m=1,\ldots, 9+n-p-q$. Moreover, let us define the coordinate $\rho$ as:
\beq
\rho^2\,=\,(y^1)^2+\ldots+(y^{q-n})^2\,\,.
\eeq
Since:
\beq
(\,\vec dy\,)^2\,=\,d\rho^2\,+\,\rho^2\,d\Omega^2_{q-n-1}\,+\,d\vec z^{\,\,2}
\eeq
the background metric in these coordinates can thus be written as:
\bear
&&ds_{10}^2\,=\,g_{tt}(r)\,dt^2\,+\,g_{xx}(r)\,\Big[
(dx^1)^2\,+\,\ldots +(dx^n)^2\,+\,
(dx^{n+1})^2\,+\,\ldots +(dx^p)^2\,\Big]\rc
&&\qquad\qquad\qquad\qquad\qquad\qquad
+\,g_{rr}(r)\,\Big[\,d\rho^2\,+\,\rho^2\,d\Omega^2_{q-n-1}\,+\,d\vec z^{\,\,2}\,\Big]\,\,.
\eear
We will consider embeddings with $x^{n+1}=\cdots =x^p={\rm constant}$ and
$|\vec z|\,=\,L$, with $L$ also constant. Then, $r^2=\rho^2+L^2$ and the induced metric on the D$q$-brane worldvolume takes the form:
\beq
ds_{q+1}^2\,=\,g_{tt}(r)\,dt^2\,+\,g_{xx}(r)\,\big[
(dx^1)^2\,+\,\ldots + (dx^n)^2\,\big]
\,+\,g_{rr}(r)\,\big[\,d\rho^2\,+\,\rho^2\,d\Omega^2_{q-n-1}\,\big]\,\,.
\eeq
In what follows we will consider massless embeddings with $L=0$ (we will return to $L\ne 0$ case elsewhere). In this case 
the $r$ and $\rho$ variables are equal. Moreover, we will switch on a non-zero temperature, which amounts to including a blackening factor $f_p$, in such a way that the induced metric becomes:
\beq
ds_{q+1}^2\,=\,g_{tt}(\rho)\,f_p(\rho)dt^2+g_{xx}(\rho)\,\big[
(dx^1)^2\,+\,\ldots +(dx^n)^2\,\big]
+g_{rr}(\rho)\,\big[\,{d\rho^2
\over f_p(\rho)}\,+\,\rho^2\,d\Omega^2_{q-n-1}\,\big]\,\,.
\eeq

Let us now compute the DBI action of the D$q$-brane with a non-zero worldvolume gauge field $F$  with components along $\rho$, $ t$. Moreover, when the number $n$ of Cartesian coordinates on the D$q$-brane worldvolume is greater than or equal to 2, we will allow a non-zero constant magnetic field $B$ along the directions $x^1$ and $x^2$. Thus, we will take $F$ to be given by:
\beq
F\,=\,A_t'\, d\rho\wedge dt  \,+\,B\, dx^1\wedge dx^2\,\,,
\label{F_unperturbed}
\eeq
where $A_t'=\partial_{\rho} A_t$ and we have chosen a gauge for $A$ such that $A_{\rho}=0$.  The DBI action for this configuration becomes:
\beq
S_{Dq}=-T_{Dq}\,\int d^{q+1}\xi \,e^{-\phi}\,
\sqrt{-\det(g+F)}\,=\,
-{\cal N}\,V_{{\mathbb R}^{(n,1)}}\,\int d\rho
\sqrt{H}\,\sqrt{g_{rr}\,|g_{tt}|\,-\,A_t'^{\,2}}\,\,,
\label{Dq_DBI_action}
\eeq
where $V_{{\mathbb R}^{(n,1)}}$ is the volume of the ($n+1$)-dimensional Minkowski space, 
$\phi$ is the dilaton of the background, ${\cal N}$ the constant:
\beq
{\cal N}\,=\,T_{Dq}\,{\rm Vol} \big({\mathbb S}^{q-n-1}\big)\,\,,
\eeq
and $H$ is the function:
\beq
H(\rho)\equiv \rho^{2(q-n-1)}\,g_{xx}^{n-2}\,g_{rr}^{q-n-1}\,e^{-2\phi}\,
\big( g_{xx}^2\,+\,B^2\big)\,\,.
\eeq
The gauge field component $A_t$ is a cyclic variable in the action (\ref{Dq_DBI_action})  and has an associated constant of motion. Then, we can write:
\beq
d\,=\,{\sqrt{H}\,A_t'\over 
\sqrt{g_{rr}\,|g_{tt}|\,-\,A_t'^{\,2}}}\,\,,
\eeq
where $d$ is a constant (the charge density). Inverting this relation we get:
\beq
A_t'\,=\,{d\,\sqrt{g_{rr}\,|g_{tt}|}\over {\sqrt{H+d^2}}}\,\,.
\eeq
By using this result we obtain the on-shell Lagrangian density:
\beq
{\cal L}_{DBI}^{(on-shell)}\,=\,-{\cal N}\,V_{{\mathbb R}^{(n,1)}}\,
{\sqrt{g_{rr}\,|g_{tt}|}\over \sqrt{H+d^2}}\,\,H\,\,.
\label{on-shell-calL}
\eeq

\subsection{Intersections in the D$p$-brane background }\label{sec:intersections}

Let us now write some of the equations for the background corresponding to a stack of  D$p$-branes.  The metric and dilaton  in this case are given by:
\beq
-g_{tt}\,=\,g_{xx}\,=\,\Big({r\over R}\Big)^{{7-p\over 2}}\,\,,
\qquad
g_{rr}\,=\,\Big({R\over r}\Big)^{{7-p\over 2}}\,\,,
\qquad
e^{-2\phi}\,=\,\Big({R\over r}\Big)^{{(7-p)(p-3)\over 2}}\,\,,
\label{Dp-metric_dilaton}
\eeq
where $R$ is a constant radius. The blackening factor $f_p$ is given by:
\beq
f_p\,=\,1\,-\,\Big({r_h\over r}\Big)^{7-p}\,\,,
\label{blackeningDp}
\eeq
where $r_h$ is the horizon radius, related to the temperature $T$ as follows:
\beq
T\,=\,{7-p\over 4\pi}\,r_h^{{5-p\over 2}}\,\,.
\label{T_rh}
\eeq

In the following we will scale out the constant $R$ or, equivalently, we will directly take $R=1$. In this case, the function $H$ corresponding to a $(n\, |\, p\perp q)$ intersection takes the form:
\beq
H\,=\,\rho^{\lambda}+\rho^{\lambda+p-7}\,B^2\,\,,
\label{H_Dp}
\eeq
where the constant exponent $\lambda$ is the combination of $p$, $q$, and $n$ written in (\ref{lambda_n}). We will show below that the numbers $p$ and $\lambda$ determine the  thermodynamics and collective excitations of the intersection.

\subsection{Thermodynamics at zero temperature}

We begin by studying the thermodynamics of the intersections at zero temperature in the absence of a magnetic field by following \cite{Karch:2009eb}. According to the standard AdS/CFT dictionary,  the chemical potential $\mu$ at zero temperature for the brane intersection is given by the boundary value of $A_t(\infty)$ and for D$q$-brane probes entering the Poincar\'e horizon can be written as an integral:
\beq
\mu\equiv A_t(\infty) =\,\int_{0}^{\infty}\,d\rho\,A_t'\,=\,d\,\int_{0}^{\infty}{\sqrt{g_{rr}\,|g_{tt}|}\over {\sqrt{H+d^2}}}\,d\rho\,\,.
\eeq
When the metric is given by (\ref{Dp-metric_dilaton}) and $H=\rho^{\lambda}$ (see (\ref{H_Dp})), the chemical potential becomes:
\beq\label{mu_gamma}
\mu\,=\,d\,\int_{0}^{\infty}\,{d\rho\over \sqrt{\rho^{\lambda}\,+\,d^2}} = \,\gamma\,d^{{2\over \lambda}} \ ,
\eeq
where $\gamma$ is a constant
\beq
\gamma\,=\,{1\over \sqrt{\pi}}\,\Gamma\Big({1\over 2}-{1\over \lambda}\Big)\,
\Gamma\Big(1+{1\over \lambda}\Big)\,\,.
\label{gamma_def}
\eeq
The on-shell action of the probe (without the Minkowski volume factor $V_{{\mathbb R}^{(n,1)}}$) is given by the integral of the on-shell Lagrangian density (\ref{on-shell-calL}):
\beq
S_{on-shell}\,=\,-{\cal N}\,\,\int_{0}^{\infty}\,{\rho^{\lambda}\over 
 \sqrt{\rho^{\lambda}\,+\,d^2}}\,d\rho\,\,,
\eeq
which is divergent and must be regulated. We will do it by subtracting  the same integral with $d=0$. We get:
\beq
S_{on-shell}^{reg}\,=\,-{\cal N}\,\,\int_{0}^{\infty}\,
\rho^{{\lambda\over 2}}\,\Bigg[\,
{\rho^{{\lambda\over 2}}\over  \sqrt{\rho^{\lambda}\,+\,d^2}}\,-\,1
\Bigg]\,d\rho\,\,.
\label{S_on-shell}
\eeq
The grand potential density $\Omega=\Omega(\mu)$ is given by minus the regulated on-shell action:
\beq
\Omega\,=\,-S_{on-shell}^{reg}\,\,.
\eeq
After performing explicitly the integral in (\ref{S_on-shell}) we arrive at:
\beq
\Omega\,=\,-{2\over \lambda+2}\,{\cal N}\,\gamma\,d^{1+{2\over \lambda}} = -{2\over \lambda+2}\,{\cal N}\,\gamma^{-{\lambda\over 2}}\,
\mu^{1+{\lambda\over 2}} \ .
\eeq
From $\Omega$ we can obtain the density $\rho$ as:
\beq
\rho\,=\,-{\partial\Omega\over \partial \mu} = {\cal N}\,d \ .
\eeq
Moreover, the energy density $\epsilon$ is given by $\epsilon\,=\,\Omega+\mu\,\rho$:
\beq
\epsilon\,=\,{\lambda\over \lambda+2}\,{\cal N}\,\gamma\,d^{1+{2\over \lambda}}\,\,.
\eeq
The pressure is just $p=-\Omega$ and therefore:
\beq\label{eq:polytrope}
p\,=\,{2\over \lambda+2}\,{\cal N}\,\gamma\,d^{1+{2\over \lambda}} = {2\over \lambda}\,\epsilon \ .
\eeq
We notice that $\lambda$ is thus related to the polytropic index of the equation of state.
Finally, the speed of (first) sound is defined as:
\beq\label{speed_first_sound}
u_s^2\equiv {\partial p\over \partial \epsilon} = {2\over \lambda} \ ,
\eeq
which follows immediately from the relation between $p$ and $\epsilon$. We wish to warn the reader though, that obtaining the speed of sound at $T=0$ is a little bit of a stretch, as being beyond the applicability
of the hydrodynamics.

We have thus found that the zero temperature thermodynamic behavior is determined uniquely by the index $\lambda$ defined in (\ref{lambda_n}). In subsection \ref{sec:bound} we will discuss which models violate the bound on the speed of sound as expected for strongly coupled field theories with a gravity dual.

\subsection{Thermodynamics at non-zero temperature}
We now study the thermodynamics of the probe at $T\not=0$ and $B=0$. The chemical potential is given by:
\beq 
\mu\,=\,\int_{r_h}^{\infty}\,d\rho\,A_t'\,=\,d\,\int_{r_h}^{\infty}{1\over {\sqrt{\rho^{\lambda}+d^2}}}\,d\rho =\gamma\,d^{{2\over \lambda}}\,-\,r_h\,
F\Big({1\over 2}, {1\over \lambda};1+{1\over \lambda};-{r_h^{\lambda}\over d^2}\Big)\,\,.
\label{chemical_pot_nonzeroT}
\eeq
The grand potential  $\Omega$ at $T\not=0$ is given by:
\bea
\Omega & = & -S_{on-shell}^{reg}={\cal N}\,\,\int_{r_h}^{\infty}\rho^{{\lambda\over 2}}\,\Bigg[\,{\rho^{{\lambda\over 2}}\over  \sqrt{\rho^{\lambda}\,+\,d^2}}\,-\,1\Bigg]\,d\rho \rc  
 & = &  -{2\over \lambda+2}\,{\cal N}\,\gamma\,d^{1+{2\over \lambda}}-{{\cal N}\,r_h^{\lambda+1}\over (\lambda+1)\,d}\,
F\Big({1\over 2}, 1+{1\over \lambda};2+{1\over \lambda};-{r_h^{\lambda}\over d^2}\Big)\,+\,{2{\cal N}\over \lambda+2}\,r_h^{{\lambda\over 2}+1}\label{Omega_nonzeroT} \ .
\eea
The last  term in (\ref{Omega_nonzeroT}) is independent of the density and therefore of chemical potential. We define the $\mu$-dependent part $\Delta\Omega(\mu)$  of the grand potential as:
\beq
\Delta\Omega\equiv \Omega-
{2{\cal N}\over \lambda+2}\,r_h^{{\lambda\over 2}+1}\,\,.
\eeq

In order to study the behavior of the system with  the temperature, let us consider the case in which $r_h$ is small. Expanding the chemical potential (\ref{chemical_pot_nonzeroT}) at next-to-leading order in $r_h$, we get:
\beq
\mu\,=\,\gamma\,d^{{2\over \lambda}}\,-\,r_h\,+\,
{1\over 2(\lambda+1)}\,{r_h^{\lambda+1}\over d^2}\,+\,{\mathcal O}(r_h^{2\lambda+1})\,\,.
\label{chemical_pot_lowT}
\eeq
The expansion of $\Delta\Omega$ is:
\beq
\Delta\Omega\,=\,-{2\over \lambda+2}\,{\cal N}\,\gamma\,d^{1+{2\over \lambda}}\,-\,
{{\cal N}\over (\lambda+1)\,d}\,r_h^{\lambda+1}\,+\,\ldots\,.
\label{DeltaOmega_lowT}
\eeq
Plugging the expression of $d$ (\ref{chemical_pot_lowT}) into (\ref{DeltaOmega_lowT}) we find:
\beq
\Delta\Omega\,=\,-{2\over \lambda+2}\,{\cal N}\,\gamma^{-{\lambda\over 2}}\,
\Big(\mu+r_h-{1\over 2(\lambda+1)}\,{r_h^{\lambda+1}\over d^2}\Big)^{1+{\lambda\over 2}}\,-\,
{{\cal N}\over (\lambda+1)\,d}\,r_h^{\lambda+1}
\,+\,\ldots\,\,.
\label{DeltaOmega_lowT_rh}
\eeq
The entropy depending on the density is given by the following derivative:
\beq
s=-{\partial (\Delta\Omega)\over \partial T}\Bigg|_{\mu}=-{\partial (\Delta\Omega)\over \partial r_h}\Bigg|_{\mu}\,
{\partial\, r_h\over \partial T}=
-{2\over 5-p}\,\Big({4\pi\over 7-p}\Big)^{{2\over 5-p}}\,T^{{p-3\over 5-p}}\,
{\partial (\Delta\Omega)\over \partial \, r_h}\Bigg|_{\mu}
\,\,.
\eeq
From (\ref{DeltaOmega_lowT_rh}), we obtain
\beq
{\partial (\Delta\Omega)\over \partial r_h}\Bigg|_{\mu}\,=\,-{\cal N}\,d\,-\,
{{\cal N}\over 2d}\,r_h^{\lambda}\,+\,\ldots\,\,.
\eeq
Therefore, we get:
\beq
s\,=\,
{2\over 5-p}\,\Big({4\pi\over 7-p}\Big)^{{2\over 5-p}}\,
{\cal N}\,d\,
T^{{p-3\over 5-p}}\,\Big(1+{1\over 2\,d^2}\,\Big({4\pi \,T\over 7-p}\Big)^{{2\lambda\over 5-p}}\,+\,\ldots\Big)\,\,.
\label{entropy_lowT}
\eeq
Let us now compute the specific heat $c_V$ from the formula:
\beq
c_V\,=\,T\,{\partial s\over \partial T}\Big|_{d}\,\,.
\eeq
For $p\not=3$ we only need to keep the leading term in (\ref{entropy_lowT}) to find the 
behavior at low $T$, which only depends on $p$. We find:
\beq
c_V(p\not=3)\,=\,{2(p-3)\over (5-p)^2}\,\Big({4\pi\over 7-p}\Big)^{{2\over 5-p}}\,
{\cal N}\,d\,
T^{{p-3\over 5-p}}\,+\,\ldots\,\,.
\eeq
Notice that $c_V$ is linear in $T$ (as for the Landau-Fermi liquid) only for $p=4$. For $p=3$ the entropy is instead:
\beq
s(p=3)\,=\,\pi\,{\cal N}\,d\,+\,{\pi\over 2}\,{{\cal N}\over d}\,(\pi T)^{\lambda}\,+\,\ldots\,\,,
\eeq
and the specific heat depends on $\lambda$ in the following form:
\beq
c_V(p=3)\,=\,{\lambda\over 2}\,\pi^{\lambda+1}\,{\cal N}\, {T^{\lambda}\over d}\,+\,\ldots\,\,.
\eeq
Notice that $\lambda=2n$ for $p=3$, as follows from (\ref{lambda_n}). Notice also that the entropy at $T=0$ is non-vanishing in the $p=3$ case, as pointed out in \cite{Karch:2008fa,Karch:2009zz}. 
This degeneracy suggests that there is an instability towards a non-degenerate ground state, see, {\emph{e.g.}}, \cite{Hartnoll:2009ns}. However, to date such instabilities have not been found. As discussed
in \cite{Bigazzi:2013jqa}, the backreaction of the flavor D-branes may play a significant role in understanding this puzzle.

\subsection{Models encompassed}
In this section we will list the probe brane intersection models in which our results apply.
We begin by recalling that in our notation $(n\, |\, p\perp  q)$ denotes the intersection of two stacks of D$p$- and D$q$-branes  (with $q\ge p$) along $n$ common directions. Let us also recall that the embeddings considered in this paper are the ones corresponding to massless quarks, in which the embedding function is trivial. 
We will first list the intersections which preserve some amount of supersymmetry.

\subsubsection{Supersymmetric intersections: \#ND=4}
The  intersections $(n\, |\, p\perp  q)$  which preserve some amount of supersymmetry are those for which $n$ is related to $p$ and $q$ as follows:\footnote{One can easily find this relation by imposing the no-force condition between the two stacks of D-branes (see, for example, \cite{Arean:2006pk}).}
\beq
n\,=\,{p+q-4\over 2}\,\,.
\label{SUSY_condition}
\eeq
It is straightforward to verify that the condition (\ref{SUSY_condition}) selects the following three series of intersections:
\beq
(p\, |\, p\perp  p+4)\,\,,
\qquad
(p-1\, |\, p\perp  p+2)\,\,,
\qquad
(p-2\, |\, p\perp  p) \ .
\label{SUSY_series}
\eeq
Let us evaluate the index $\lambda$ for the three series of SUSY intersections  (\ref{SUSY_series}). In these cases (\ref{lambda_n}) simplifies drastically and we simply get:
\beq
\lambda\,=\,2(q-n-1)\,=\,q-p+2\,\,.
\eeq
In other words $\lambda=6,4,2$ for the intersections D$p$-D$(p+4)$, D$p$-D$(p+2)$, and 
D$p$-D$p$, respectively, as illustrated in  table \ref{table:nd4}. 
\begin{table}[ht]\begin{center}
\begin{tabular}{|l|l|l|l|l|}
\hline
Model & $\lambda$ & $p$ & $q$ & $n$ \\
\hline\hline
D$p$-D$(p+4)$ & 6 & $p$ & $p+4$ & $p$ \\
\hline
D$p$-D$(p+2)$ & 4 & $p$ & $p+2$ & $p-1$ \\
\hline
D$p$-D$p$ & 2 & $p$ & $p$ & $p-2$ \\
\hline
\end{tabular}
\end{center}
\caption{The supersymmetric models for which our results apply.}\label{table:nd4}
\end{table}



\subsubsection{Non-supersymmetric intersections: \#ND=6}

The models which break all the supersymmetries can be subject to instabilities. In flat space, the electromagnetic and gravitational forces of the two sets of D$p$- and D$q$-branes do not cancel out. This then typically manifests itself as a tachyonic mode below the Breitenlohner-Freedman bound in the open string spectrum in the near-horizon limit. However, certain circumstances may render the brane configuration perturbatively stable; topology in the Sakai-Sugimoto model \cite{Sakai:2004cn} or turning on an internal flux in the worldvolume of the probe D-branes \cite{Myers:2008me,Bergman:2010gm}. The non-supersymmetric models where some of our results apply are listed in Table \ref{table:nd6}. 

\begin{table}[ht]\begin{center}
\begin{tabular}{|l|l|l|l|l|}
\hline
Model & $\lambda$ & $p$ & $q$ & $n$ \\
\hline\hline
Sakai-Sugimoto D4-D8/$\overline{\rm{D8}}$ \cite{Sakai:2004cn} & 5 & 4 & 8 & 3 \\
\hline
D3-D7' \cite{Bergman:2010gm} & 4 & 3 & 7 & 2 \\
\hline
D2-D8' \cite{Jokela:2011eb} & 5 & 2 & 8 & 2 \\
\hline
\end{tabular}
\end{center}
\caption{The non-supersymmetric models for which some of our results apply.}\label{table:nd6}
\end{table}

We do not plan to detail which of our results are directly applicable to the above non-supersymmetric models. This would invoke a separate involved study, so we just warn the reader by recalling a few facts.  The Sakai-Sugimoto
model is to be treated only in the deconfined parallel phase. 
The presence of the internal flux may affect the speed of zero sound and the attenuation; this expectation was recently confirmed for the D3-D5 model in \cite{Itsios:2015kja}. 
One should also keep in mind that all these models are subject to striped instability at non-zero density. Furthermore, in all the models, a further inclusion of the magnetic field will have a major effect due to Chern-Simons action contributions and hardly anything will apply.

\subsection{Bound on the speed of sound}\label{sec:bound}

It was argued in \cite{Hohler:2009tv,Cherman:2009tw,Cherman:2009kf} that the speed of sound in a strongly-coupled theory with gravity dual is always smaller than the conformal value. In a $(n+1)$-dimensional theory this value is $u_s^2=1/n$. We will assume in what follows that $n\ge 1$. Let us explore under which circumstances our system violates this bound. Since we have obtained $u_s^2=2/\lambda$ (\ref{speed_first_sound}), it is clear that the bound is violated if
\beq
{\lambda\over 2}<n\,\,.
\label{bound_violation}
\eeq
From the general value of $\lambda$  written in (\ref{lambda_n}), we conclude that (\ref{bound_violation}) holds if:
\beq
(p-3)(p+q-2n-8)<0\,\,.
\label{violation_condition}
\eeq
In the SUSY case $p+q-2n-8=-4$ (see eq. (\ref{SUSY_condition})). Therefore, the bound is always violated for the intersections listed in Table~\ref{table:nd4} with $p>3$. For the non-SUSY intersections listed in Table~\ref{table:nd6}, only the Sakai-Sugimoto model violates the bound. The violation of the bound was also discussed in \cite{Karch:2009eb} for such intersections. 

Actually, one can probe in full generality that the bound is always violated for $p>3$ except for two particular intersections. In this case the condition (\ref{violation_condition}) requires that $p+q-2n-8<0$. But, since the total number of spatial dimensions is 9, the integers $p$, $q$, and $n$ must satisfy $p+q-n-9\le 0$. Let us now write
\beq
p+q-2n-8\,=\,(p+q-n-9)+(1-n)\,\,.
\eeq
The two numbers in parenthesis are less or equal to zero. Unless when they are simultaneously zero, the bound is violated. This only happens when $n=1$ and $p+q=10$. The only intersections that satisfy these conditions for $q\ge p>3$ are $(1|4\perp 6)$ and $(1|5\perp 5)$ for which cases $u_s^2=1$.


\section{Fluctuations}
\label{sec:fluctuations}

We now want to analyze the excitation spectrum of our holographic system. These excitations correspond to density waves in the dual field theory which appear as poles of the retarded Green's functions. In the holographic context finding these poles is equivalent to obtaining the quasinormal modes of the gravitational system. Accordingly,  we  now  assume that $T$ and $B$ are non-zero and  allow fluctuations of the gauge field along the Minkowski directions of the intersection, in the form:
\beq
A\,=\,A^{(0)}\,+\,a (\rho, x^{\mu})\,\,,
\eeq
where 
 $A^{(0)}=A_\nu^{(0)}\,dx^{\nu}=A_t\,dt\,+\,B\,x^1\,dx^2$ and $a(\rho,x^{\mu})=a_{\nu}(\rho,x^{\mu})dx^{\nu}$. The total gauge field strength is:
\beq
F\,=\,F^{(0)}\,+\,f\,\,,
\eeq
with  $F^{(0)}=dA^{(0)}$ being  the two-form written in (\ref{F_unperturbed}). We will choose the gauge in which $a_{\rho}=0$. Moreover, we will consider fluctuation fields  $a_{\nu}$ which depend on $\rho$, $t$, and $x^1$. In this case it is possible to restrict to the case in which $a_{\nu}\not=0$ only when $\nu=t, x^1\equiv x$, and $ x^2\equiv y$. It follows that the non-vanishing components of $f$ are:
\bear
&&
f_{t\rho}\,=\,-a_t'\,,
\qquad
f_{x\rho}\,=\,-a_x'\,,
\qquad
f_{y\rho}\,=\,-a_{y}'\, \ ,\rc
&&f_{xy}\,=\,\partial_x\,a_y\,,\qquad
f_{tx}\,=\,\partial_t\,a_x\,-\,\partial_x\,a_t\,,\quad f_{ty}\,=\,\partial_t\,a_y\ ,
\eear
where the prime denotes derivative with respect to $\rho$. 
In order to write down the Lagrangian for the fluctuations, let us define the matrix $X$ as:
\beq
X\,\equiv\,\Big(\,g^{(0)}\,+\,F^{(0)}\Big)^{-1}\,f\,\,.
\eeq
Then, the DBI determinant can be expanded in powers of $X$ as:
\beq
\sqrt{-\det (g+F)}\,\,=\,
\sqrt{-\det (g^{(0)}+F^{(0)})}\,\Big[1\,+\,{1\over 2}\Tr X\,-\,{1\over 4}\,
\Tr X^2\,+\,{1\over 8}\,\Big(\Tr X\Big)^2\,+{\cal{O}}(X^3)\Big] \ .
\label{DBI_expansion}
\eeq
Let us split the inverse of the matrix $g^{(0)}\,+\,F^{(0)}$ as:
\beq
\Big(\,g^{(0)}\,+\,F^{(0)}\Big)^{-1}\,=\,{\cal G}^{-1}\,+\,{\cal J}\,\,,
\eeq
where ${\cal G}^{-1}$ is the symmetric part and ${\cal J}$ is the antisymmetric part (${\cal G}$ is the so-called open string metric). It follows that:
\beq
X^{a}_{\,\,\,\,\, b}\,=\, {\cal G}^{ac}\,f_{cb}\,+\,{\cal J}^{ac}\,f_{cb}\,\,,
\eeq
where the Latin indices take values in $a,b,c\in \{t,x,y,\rho\}$. The traces needed in the expansion (\ref{DBI_expansion}) up to second order in $X$ are:
\bear
&&\Tr X\,=\,{\cal J}^{ab}\,f_{ba}\,\,,\rc\rc
&&\Tr X^2\,=\,-{\cal G}^{ac}\,{\cal G}^{bd}\,f_{cd}\,f_{ab}\,+\,
{\cal J}^{ac}\,{\cal J}^{bd}\,f_{cd}\,f_{ab}\,\,.
\eear
In our case the relevant  elements of the open string metric are:
\bear
&&{\cal G}^{tt}\,=\,-{1\over f_p}\,{g_{rr}\over g_{rr}\,|g_{tt}|\,-\,A_t'^{\,2}}\,\,,
\qquad\qquad
{\cal G}^{\rho\rho}\,=\,f_p\,{|g_{tt}|\over g_{rr}\,|g_{tt}|\,-\,A_t'^{\,2}}\,\,,\rc\rc
&&{\cal G}^{x^1\,x^1}\,=\,{\cal G}^{x^2\,x^2}\,=\,{g_{xx}\over g_{xx}^2+B^2}\,\,,
\eear
while those of the antisymmetric matrix ${\cal J}$ are:
\beq
{\cal J}^{t\rho}=-{\cal J}^{\rho t}\,=\,-{A_t'\over g_{rr}\,|g_{tt}|\,-\,A_t'^{\,2}}\,\,,
\qquad\qquad
{\cal J}^{x^1\,x^2}=-{\cal J}^{x^2\,x^1}\,=\,-{B\over g_{xx}^2+B^2}\,\,.
\eeq
Let us write the components along $t,\rho$ of these matrices in terms of the charge density $d$. We get for the open string metric:
\beq
{\cal G}^{tt}\,=\,-{H+d^2\over f_p\,|g_{tt}|\,H}\,\,,
\qquad\qquad
{\cal G}^{\rho\rho}\,=\,f_p\,{H+d^2\over g_{rr}\,H}\,,
\eeq
while ${\cal J}^{t\rho}$ takes the form
\beq
{\cal J}^{t\rho}=-{\cal J}^{\rho t}\,=\,-{d\over \sqrt{g_{rr}\,|g_{tt}|}}\,\,
{\sqrt{H+d^2}\over H}\,\,.
\eeq
From these values we can immediately calculate 
\beq
\tr X \,=\,2\,{d\over \sqrt{g_{rr}\,|g_{tt}|}}\,\,
{\sqrt{H+d^2}\over H}\,f_{t\rho}\,+\,2{B\over g_{xx}^2+B^2}\,f_{x y}\,\,,
\eeq
and one can demonstrate  that the linear term with $\Tr X$ in the action does not contribute to the equations of motion, as it should. Moreover, it is straightforward to verify that, up to a multiplicative constant, the Lagrangian density for the fluctuations  at second order in $f$ is given by:
\beq
{\cal L}\,\sim\,{\sqrt{g_{rr}\,|g_{tt}|}\over \sqrt{H+d^2}}\,\,H\,\,
\Big(\,{\cal G}^{ac}\,{\cal G}^{bd}\,-\,
{\cal J}^{ac}\,{\cal J}^{bd}\,+\,{1\over 2}\,{\cal J}^{cd}\,{\cal J}^{ab}
\Big)f_{cd}\,f_{ab}\,\,.
\eeq
The corresponding equation of motion  for $a^d$ is:
\beq
\partial_{c}\,\Bigg[{\sqrt{g_{rr}\,|g_{tt}|}\over \sqrt{H+d^2}}\,\,H
\Big(\,{\cal G}^{ca}\,{\cal G}^{db}\,-\,
{\cal J}^{ca}\,{\cal J}^{db}\,+\,{1\over 2}\,{\cal J}^{cd}\,{\cal J}^{ab}
\Big)\,f_{ab}\Bigg]\,=\,0\,\,.
\label{eom_general}
\eeq
From the equation of motion for $a_{\rho}$ (with $a_{\rho}=0$) we get the  transversality condition:
\beq
\partial_t\,a_t'\,-\,u^2(\rho)\,\partial_x\,a_x'\,=\,0\,\,,
\label{transversality_xt}
\eeq
where $u(\rho)$ is the function
\beq
u^2(\rho)\,=\,-{{\cal G}^{xx}\over {\cal G}^{tt}}=\,{g_{xx}\,|g_{tt}|\,f_p\over g_{xx}^2\,+B^2}\,\,{H\over H+d^2} \ .
\eeq
Let us Fourier transform  the gauge field to momentum space as:
\beq
a_\nu(\rho, t, x)\,=\,\int {d\omega\,dk\over (2\pi)^2}\,
a_\nu(\rho, \omega, k)\,e^{-i\omega\,t\,+\,i k x}\,\,.
\eeq
In momentum space the transversality condition (\ref{transversality_xt}) takes the form:
\beq
\omega\,a_t'\,+\,u^{2}(\rho)\,k\,a_x'\,=\,0\,\,.
\label{transversality_momentum}
\eeq
We now define the electric field $E$ as the gauge-invariant combination:
\beq
E\,=\,k\,a_t\,+\,\omega\,a_x\,\,.
\label{E_at_ax}
\eeq
Using the transversality condition, we can obtain $a_t'$ and $a_x'$ in terms of $E'$ as follows:
\beq
a_t'\,=\,-{k\,u^2\over \omega^2\,-\,k^2\,u^2}\,E'\,\,,
\qquad\qquad
a_x'\,=\,{\omega\over \omega^2\,-\,k^2\,u^2}\,E'\,\,.
\label{at_ax_E}
\eeq
In terms of $u$ the Lagrangian density of the fluctuations takes the form:
\beq
{\cal L}\,\sim\,{\sqrt{g_{rr}\,(g_{xx}^2+B^2)\,H}
\over \sqrt{g_{xx}\,f_p}}\,\,u\,\,
\Big(\,{\cal G}^{ac}\,{\cal G}^{bd}\,-\,
{\cal J}^{ac}\,{\cal J}^{bd}\,+\,{1\over 2}\,{\cal J}^{cd}\,{\cal J}^{ab}
\Big)f_{cd}\,f_{ab}\,\,,
\label{Lagrangian_general_u}
\eeq
and the elements of the matrices ${\cal G}$ and ${\cal J}$ along $t\rho$ are:
\bear
&&{\cal G}^{tt}\,=\,-{g_{xx}\over g_{xx}^2+B^2}\,{1\over u^2}\,\,,
\qquad\qquad
{\cal G}^{\rho\rho}\,=\,{g_{xx}\,|g_{tt}|\,f_p^2\over g_{rr}(g_{xx}^2+B^2)}\,
\,{1\over u^2}\,\,,\rc\rc
&&{\cal J}^{t\rho}=-{\cal J}^{\rho t}\,=\,-{d\over u}\,\,
{\sqrt{g_{xx}\,f_p}\over \sqrt{g_{rr}\,(g_{xx}^2\,+\,B^2)\,H}}\,\,.
\eear
It is also interesting to write the explicit expression of the function 
$u$ for the D$p$-brane background:
\beq
u^2 = \frac{g_{xx}|g_{tt}|f_p}{g_{xx}^2+B^2}\frac{H}{H+d^2} = {\rho^{\lambda} \,f_p(\rho)\over \rho^{\lambda} +\rho^{\lambda+p-7}\,B^2 +d^2}\,\,.
\eeq
The equations of motion derived from (\ref{Lagrangian_general_u}) for the D$p$-brane background have been explicitly written in appendix \ref{appendixA} (eqs. (\ref{eom_E_Bnonzero}) and (\ref{eom_ay_general})). After fixing the $a_{\rho}=0$ gauge and  using (\ref{at_ax_E}) these equations reduce to a system of two coupled second-order differential equations for $E$ and $a_y$ which we study in the next three sections, both analytically and numerically. 

The numerical methods we employ to solving ordinary differential equations are by now standard. The only non-trivial complication comes from the fact that the fluctuations are generically coupled and one needs to find normalizable solutions for all the fields at once. Implementing such a method, though, is straightforward (see, {\emph{e.g.}}, \cite{Bergman:2011rf}) and was first introduced in the holographic context in \cite{Amado:2009ts,Kaminski:2009dh}.


\section{Vanishing magnetic field}
\label{zeroB}
We will start our analysis of the fluctuation equations by considering the case in which the magnetic field vanishes. In this case the equations of motion for the longitudinal and transverse excitations decouple. Indeed, when $B=0$ the momentum space equation (\ref{eom_E_Bnonzero}) for the electric field $E$ becomes:
\beq
E''+\partial_{\rho}\log
{(\rho^{\lambda}+d^2)^{{3\over 2}}\,f_p\over
(\omega^2-f_p\,k^2)\,\rho^{\lambda}+\omega^2\,d^2}\,E'\,+\,
{1\over \rho^{7-p}\,f_p^2}\,
{(\omega^2-f_p\,k^2)\,\rho^{\lambda}+\omega^2\,d^2\over 
\rho^{\lambda}+d^2}\,E=0\,\,.
\label{eom_E_Bzero}
\eeq
Similarly,  when $B=0$, the equation  for $a_y$ written in (\ref{eom_ay_general}) is:
\beq
a_y''\,+\,\partial_{\rho}\,\log\Big[\sqrt{\rho^{\lambda}+d^2}\,f_p\Big]\,a_y'\,+\,
{1\over \rho^{7-p}\,f_p^2}\,
{(\omega^2-f_p\,k^2)\,\rho^{\lambda}+\omega^2\,d^2\over 
\rho^{\lambda}+d^2}\,a_y\,=\,0\,\,.
\label{ay_eq_Bzero}
\eeq
In the rest of this section we will analyze the solutions of (\ref{eom_E_Bzero}) in different regimes. We will begin by analyzing in the next subsection the diffusive solutions of  (\ref{eom_E_Bzero})
in which $\omega$ is purely imaginary and is related to the momentum $k$ as $\omega=-i D\,k^2$, with $D$ being the so-called diffusion constant.  We leave for appendix \ref{Trasnverse_correlator} the analysis of (\ref{ay_eq_Bzero}) and of the calculation of the  corresponding transverse correlators.

\subsection{Diffusion constant}
\label{diffusion_zeroB}

Let us determine the diffusion constant which follows from the equation of motion of $E$ at zero magnetic field (\ref{eom_E_Bzero}). With this purpose we first expand this equation near the horizon $\rho=r_h$. To begin with we expand the blackening factor $f_p(\rho)$  near $\rho=r_h$. We get:
\beq
f_p(\rho)= {7-p\over r_h}\,(\rho-r_h)\,+\,\ldots\,\,.
\eeq
It  follows that the coefficients of $E'$ and $E$ near $\rho=r_h$ can be represented as:
\bear
\partial_{\rho}\log
{(\rho^{\lambda}+d^2)^{{3\over 2}}\,f_p\over
(\omega^2-f_p\,k^2)\,\rho^{\lambda}+\omega^2\,d^2} & = & {1\over \rho-r_h}\,+\,c_1
\,+\,\ldots   \rc
{1\over \rho^{7-p}\,f_p^2}\,
{(\omega^2-f_p\,k^2)\,\rho^{\lambda}+\omega^2\,d^2\over 
\rho^{\lambda}+d^2} & = & {A\over (\rho-r_h)^2}\,+\,{c_2\over \rho-r_h}\,+\,\ldots\,\,,
\label{nh_expansion_Eeq}
\eear
where the constants $A$, $c_1$, and $c_2$ are given by:
\bear
&& A\,=\,{\omega^2\over (7-p)^2\,r_h^{5-p}}  \rc
&& c_1\,=\,(7-p)\,{r_h^{\lambda-1}\over d^2+r_h^{\lambda}}\,
{k^2\over  \omega^2}\,+\,{p-8\over 2r_h}\,+\,{\lambda\over 2}\,
{r_h^{\lambda-1}\over d^2+r_h^{\lambda}}  \rc
&&c_2\,=\,-{r_h^{p+\lambda-6}\over (7-p)\,( d^2+r_h^{\lambda})}\,k^2\,+\,
{1\over (7-p)^2\,r_h^{6-p}}\,\omega^2\,\,.
\label{A_c1_c2}
\eear
Therefore, the near-horizon equation for $E$ takes the form studied in  appendix \ref{appendixB} (eq. 
(\ref{E_eq_near-horizon})) and can be solved in a Frobenius series as in (\ref{Frobenius_E_nh}), \ie, as $E(\rho)\sim(\rho-r_h)^{\alpha}[1+\beta(\rho-r_h)+\ldots\,]$. It follows from (\ref{sol_indicial_eq})  that the solution of the indicial equation with infalling boundary condition is:
\beq
\alpha\,=\,-{i\omega\over (7-p)\,r_h^{{5-p\over 2}}}\,\,.
\label{exponent_alpha}
\eeq
Let us next perform  a low frequency expansion by considering $k\sim {\mathcal O}(\epsilon)$, $\omega\sim {\cal O}(\epsilon^2)$. Then
\beq
\alpha\sim {\cal O}(\epsilon^2)\,\,,
\qquad\qquad
c_1\sim {\cal O}(\epsilon^{-2})\,\,,
\qquad\qquad
c_2\sim {\cal O}(\epsilon^{2})\,\,.
\eeq
Since $\alpha\,c_1\sim {\cal O}(1)$, we get that:
\beq
\beta\,\approx\,-\alpha\,c_1\,\,.
\eeq
More explicitly:
\beq
\beta\,=\,i\,{k^2\over \omega}\,\,{r_h^{\lambda+{p-7\over 2}}\over 
d^2\,+\,r_h^{\lambda}}\,\,.
\eeq
Notice that $\beta \sim {\cal O}(1)$. Moreover, since $\alpha\sim\omega\sim {\cal O}(\epsilon^2)$, we can neglect the prefactor $(\rho-r_h)^{\alpha}$ in the near-horizon expansion of $E$ and write:
\beq
E\approx\,E_{nh}\,\big[1\,+\,\beta (\rho-r_h)\,\big]\,\,,
\label{nh_low_freq_E}
\eeq
where $E_{nh}$ is the value of $E$ at $\rho=r_h$.

Let us now perform the limits in the opposite order and expand the equation (\ref{eom_E_Bzero}) for $E$ in frequency first. One easily sees that the term without derivatives of $E$ is of higher order in this expansion and can be neglected. Moreover, in the term with $E'$ we approximate 
$(\omega^2-f_p\,k^2)\,\rho^{\lambda}+\omega^2\,d^2\approx -k^2\,f_p\,\rho^{\lambda}$ and then:
\beq
\log
{(\rho^{\lambda}+d^2)^{{3\over 2}}\,f_p\over
(\omega^2-f_p\,k^2)\,\rho^{\lambda}+\omega^2\,d^2}\,\approx\,-
\log {\rho^{\lambda}\over 
(\rho^{\lambda}+d^2)^{{3\over 2}}}\,+\,{\rm constant}\,\,.
\eeq
Therefore, the equation of $E$ in this regime becomes:
\beq
E''\,-\,\partial_{\rho}\,\log {\rho^{\lambda}\over 
(\rho^{\lambda}+d^2)^{{3\over 2}}}\,E'\,=\,0\,\,.
\eeq
For $\lambda>2$ this equation can be integrated as:
\beq
E(\rho)\,=\,E^{(0)}\,+\,c_E\,\int_{\rho}^{\infty}d\bar\rho\,
{\bar\rho^{\lambda}\over \big(\bar\rho^{\lambda}+d^2\big)^{{3\over 2}}}\,\,,
\eeq
where $E^{(0)}=E(\rho=\infty)$
(for $\lambda=2$ the previous integral is not convergent, see subsection \ref{lambda2_case} for a detailed treatment for this). Next, we expand $E(\rho)$ near the horizon:
\beq
E(\rho)\,\approx\,E^{(0)}\,+\,c_E\,I_{\lambda}\,-\,
{r_h^{\lambda}\,c_E\over \big(r_h^{\lambda}+d^2\big)^{{3\over 2}}}\,
(\rho-r_h)\,\,,
\label{low_freq_nh_E}
\eeq
with
\beq
I_{\lambda}\,\equiv\,\int_{r_h}^{\infty}
d\bar\rho\,
{\bar\rho^{\lambda}\over \big(\bar\rho^{\lambda}+d^2\big)^{{3\over 2}}} = \,{2\over \lambda-2}\,{r_h^{1-{\lambda\over 2}}}\,
F\Big(\,{3\over 2}, {1\over 2}-{1\over\lambda}; {3\over 2}\,-\,{1\over\lambda};
-{d^2\over r_h^{\lambda}}\,\Big)\,\,.
\eeq
By comparing (\ref{nh_low_freq_E}) and (\ref{low_freq_nh_E}) we get that 
$E^{(0)}\,=\,E_{nh}-c_E\,I_{\lambda}$ and that the integration constant $c_E$ is given by:
\beq
c_E\,=\,-\,\beta\,{(r_h^{\lambda}+d^2\big)^{{3\over 2}}\,E_{nh}\over 
r_h^{\lambda}}\,=\,-\,i\,{k^2\over \omega}\,r_h^{{p-7\over 2}}\,
(r_h^{\lambda}+d^2\big)^{{1\over 2}}\,E_{nh}\,\,.
\eeq
Thus, we can write the UV asymptotic value $E^{(0)}$ as:
\beq
E^{(0)}\,=\,E_{nh}\,\Big[1\,+\,i\,{k^2\over \omega}\,r_h^{{p-7\over 2}}\,
(r_h^{\lambda}+d^2\big)^{{1\over 2}}\,I_{\lambda}\,\Big]\,\,.
\eeq
By  requiring that $E^{(0)}=0$,\footnote{This is the widely used Dirichlet condition for the gauge field. We will consider other possibilities leading to anyonic correlations in section \ref{anyons}.} we find the following dispersion relation
\beq
\omega\,=\,-i\,D\,k^2\,\,,
\eeq
where the diffusion constant $D$ is given by:
\beq
D\,=\,r_h^{{p-7\over 2}}\,
(r_h^{\lambda}+d^2\big)^{{1\over 2}}\,I_{\lambda}
=\,{2\over \lambda-2}\,r_h^{{p-5-\lambda\over 2}}\,
(r_h^{\lambda}+d^2\big)^{{1\over 2}}\,
F\Big(\,{3\over 2}, {1\over 2}-{1\over\lambda}; {3\over 2}\,-\,{1\over\lambda};
-{d^2\over r_h^{\lambda}}\,\Big)\,\,.
\eeq
In order to write $D$  in a more  convenient way, let us define the rescaled density 
$\hat d$ as follows:
\beq
\hat d\,=\,{d\over r_h^{{\lambda\over 2}}}\,=\,\Big({7-p\over 4\pi}\Big)^{{\lambda\over 5-p}}\,
{d\over T^{{\lambda\over 5-p}}}
\,\,.
\label{hat_d_def}
\eeq
Moreover, we also rescale the frequency and momentum as:
\beq
\hat \omega\,=\,{\omega\over  r_h^{{5-p\over 2}}}\,=\,{7-p\over 4\pi}\,{\omega\over T}
\,\,,
\qquad\qquad
\hat k\,=\,{k\over  r_h^{{5-p\over 2}}}\,=\,{7-p\over 4\pi}\,{k\over T}
\,\,,
\label{hat_omega_k}
\eeq
and define the rescaled diffusion constant $\hat D$ as the one which satisfies:
\beq
\hat \omega\,=\,-i\,\hat D\,\hat k^{2}\,\,.
\eeq
It follows that $\hat D$ and $D$ are related as
\beq
\hat D\,=\, r_h^{{5-p\over 2}}\,D\,=\,{4\pi\over 7-p}\,T\,D
\,\,.
\label{hatD_def}
\eeq
Moreover, $\hat D$ only depends of $\hat d$ and $\lambda$ through the expression:
\beq
\hat D\,=\,{2\over \lambda-2}\,(1+\hat d^{\,2}\big)^{{1\over 2}}\,
F\Big(\,{3\over 2}, {1\over 2}-{1\over\lambda}; {3\over 2}\,-\,{1\over\lambda};
-\hat d^2\,\Big)\,\,.
\label{hat_D_value}
\eeq

\begin{figure}[ht]
\center
 \includegraphics[width=0.7\textwidth]{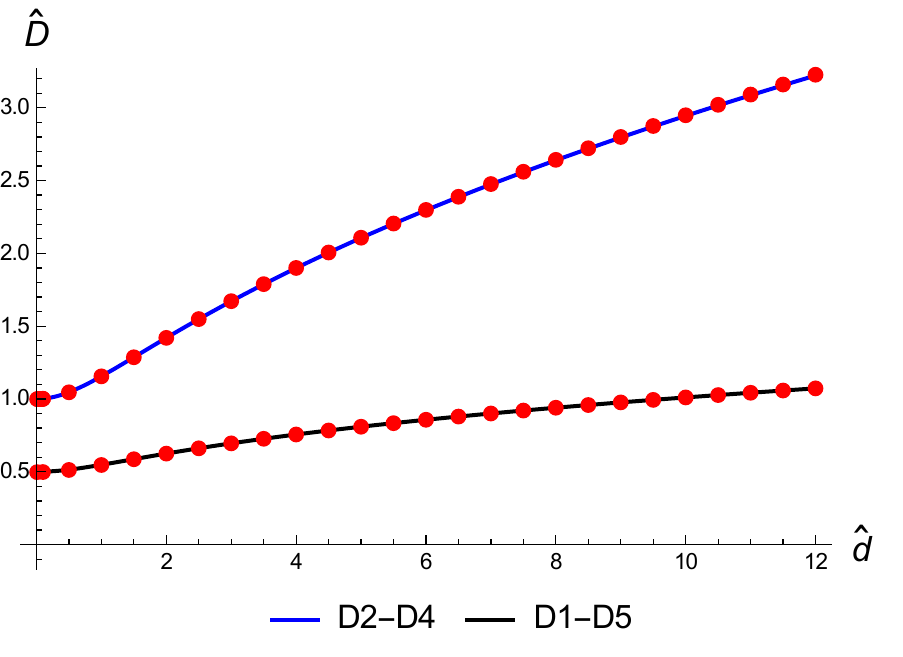}
 \caption{The diffusion constant $\hat D$ in the D2-D4 model ($\lambda=4$, top) and D1-D5 model ($\lambda=6$, bottom) as a function of $\hat d$. The red points represent the numerical data and the continuous  curves are plotted using the formula (\ref{hat_D_value}).}
 \label{diffusion_zero_B}
\end{figure}

We now analyze different limits of (\ref{hat_D_value}).  First, we consider the case in which 
$\hat d\to 0$, which is equivalent to the large temperature limit (see (\ref{hat_d_def})). 
It follows  immediately from (\ref{hat_D_value})  that:
\beq
\lim_{T\to\infty}\,\hat D\,=\,{2\over \lambda-2}\,.
\label{hatD_Tinfnity}
\eeq
Let us write this behavior in terms of the temperature $T$. Recall the relation between $T$ and the horizon radius (\ref{T_rh}) (when the radius $R$ is taken to be one). Therefore, it follows from (\ref{hatD_def}) that the  relation between $\hat D$ and $D$ can be written in terms of the temperature as:
\beq
D\,=\,{7-p\over 4\pi\, T}\,\hat D\,\,.
\eeq
Thus, we have at large $T$ that the diffusion constant behaves as:
\beq
D\,\approx \,{7-p\over \lambda-2}\,{1\over 2\pi\,T}\,\,,
\qquad\qquad
T\to\infty\,\,.
\label{D_large_T}
\eeq

Let us now analyze the opposite regime in which $\hat d$ is large or, equivalently, when the temperature $T$ is small for fixed charge density $d$. To obtain the behavior of the hypergeometric function in (\ref{hat_D_value}) in this regime, we use the following relation:
\bear
&&F(\alpha,\beta;\gamma; z)\,=\,
{\Gamma(\gamma)\Gamma(\beta-\alpha)\over 
\Gamma(\beta)\Gamma(\gamma-\alpha)}\,\,(-z)^{-\alpha}\,\,
F(\alpha,\alpha+1-\gamma;\alpha+1-\beta; {1\over z})\,+\,\rc\rc
&&
\qquad\qquad\qquad\qquad
+\,
{\Gamma(\gamma)\Gamma(\alpha-\beta)\over \Gamma(\alpha)\Gamma(\gamma-\beta)}
\,\,(-z)^{-\beta}\,\,
F(\beta,\beta+1-\gamma;\beta+1-\alpha; {1\over z})\,\,.
\qquad
\eear
It follows that, for large $\hat d$ we approximately have:
\beq
F\Big(\,{3\over 2}, {1\over 2}-{1\over\lambda}; {3\over 2}\,-\,{1\over\lambda};
-\hat d^2\,\Big)\,\approx\,
{2\over \sqrt{\pi}}\,\Gamma\Big({3\over 2}-{1\over \lambda}\Big)\,
\Gamma\Big(1+{1\over \lambda}\Big)\,
\hat d^{\,{2\over \lambda}-1}\,\,,
\eeq
and, therefore $\hat D$ takes the following approximate value for large $\hat d$:
\beq
\hat D\,\approx\,{4\over (\lambda-2)\,\sqrt{\pi}}\,
\Gamma\Big({3\over 2}-{1\over \lambda}\Big)\,
\Gamma\Big(1+{1\over \lambda}\Big)\,
\hat d^{\,{2\over \lambda}}
={4\over (\lambda-2)\,\sqrt{\pi}}\,
\Gamma\Big({3\over 2}-{1\over \lambda}\Big)\,
\Gamma\Big(1+{1\over \lambda}\Big)\,
 d^{\,{2\over \lambda}}\,\,r_h^{-1}\,\,.\label{low_T_D_hat_d}
\eeq
Since $D= r_h^{{p-5\over 2}}\,\hat D$ and 
$r_h\sim T^{2\over 5-p}$, it follows that $D$ behaves with the temperature as:
\beq
D\,\sim T^{-{7-p\over 5-p}}\,\,,
\qquad\qquad
T\sim 0\,\,.
\label{low_T_D}
\eeq
The numerical solution of (\ref{eom_E_Bzero}) shows that the diffusion mode is the dominant one for high enough temperature and allows the diffusion constant to be extracted.  In figure \ref{diffusion_zero_B} we compare the prediction (\ref{hat_D_value}) to the results obtained numerically as a function on $\hat d$ for two different intersections. The numerical results for $\hat D$ agree very well with the prediction (\ref{hat_D_value}) which, in particular,  shows that the relation between $\hat D$ and $\hat d$ only depends on the index $\lambda$.

When the temperature is low enough, the dominant excitation mode is the zero sound. In the next subsection we will study analytically this mode following the method of \cite{Karch:2008fa}. The transition between the collisionless and hydrodynamic regimes will be explored in subsection~\ref{crossover_zeroB}.

\subsection{Zero sound}
\label{zero_sound_zeroB}

The equation of motion of the electric field fluctuation $E$ at zero temperature and magnetic field can be obtained from (\ref{eom_E_Bzero}) by taking $f_p=1$, namely:
\beq
E''+\partial_{\rho}\log
{(\rho^{\lambda}+d^2)^{{3\over 2}}\over
(\omega^2-k^2)\,\rho^{\lambda}+\omega^2\,d^2}\,E'\,+\,
 \rho^{p-7}\,
{(\omega^2-k^2)\,\rho^{\lambda}+\omega^2\,d^2\over 
\rho^{\lambda}+d^2}\,E=0\,\,.
\label{eom_E_BTzero}
\eeq
Let us study the solutions of (\ref{eom_E_BTzero}) near the horizon $\rho= 0$ at low frequency and momentum (\ie, when $\omega\sim k\sim {\cal O}(\epsilon)$).  
Near the horizon $\rho=0$, (\ref{eom_E_BTzero})  can be approximated as:
\beq
E''+\omega^2\, \rho^{p-7}\,E\,\approx\,0\,\,.
\label{nh_eq_E_zeroT}
\eeq
When $p<5$, the two independent solutions of this equation are given by Hankel functions:
\beq
E(\rho)\,=\,\rho^{{1\over 2}}\,H_{{1\over 5-p}}^{(\alpha)}\,
\Big({2\omega\over 5-p}\,\rho^{{p-5\over 2}}\Big)\,\,,
\qquad\qquad (\alpha=1,2\,,\,p<5)\,\,.
\label{Hankel_nh}
\eeq
For $p=5$ the two  independent solutions of the differential equation (\ref{nh_eq_E_zeroT}) are:
\beq
E(\rho)=\sqrt{\rho}\,\rho^{\pm {1\over 2}\sqrt{1-4\omega^2}}\,\,,
\qquad\qquad (p=5) \ .
\eeq
In what follows we will only consider the case $p<5$. According to the standard prescription, the retarded Greens' functions correspond to modes with incoming boundary conditions at the horizon.  Actually, in our case, when $\rho\to 0$ the argument of the Hankel functions  in (\ref{Hankel_nh}) grows ($\rho^{{p-5\over 2}}\to\infty$ for $p<5$). As the asymptotic behavior of the Hankel functions is:
\beq
H_{\nu}^{(1)}(z)\sim \sqrt{{2\over \pi z}}\,\,e^{i(z-{1\over 2}\,\nu\,\pi\,-{\pi\over 4})}\,\,,
\qquad
H_{\nu}^{(2)}(z)\sim \sqrt{{2\over \pi z}}\,\,e^{-i(z-{1\over 2}\,\nu\,\pi\,-{\pi\over 4})}\,\,,
\quad (z\to\infty)\,\,,
\eeq
it follows that the incoming wave corresponds to the function $H_{\nu}^{(1)}$. Thus, we select the solutions with $\alpha=1$ in (\ref{Hankel_nh}).  Moreover, 
when the index $
\nu$ of $H_{\nu}^{(1)}$ is not integer (in our case this corresponds to $\nu<1$ and 
$p<4$), the Hankel function has the following expansion near the origin:
\beq
H_{\nu}^{(1)}(\alpha x)= -{2^{\nu}\,\Gamma(\nu)\over \pi\,\alpha^{\nu}}\,i\,
\Big[{1\over x^{\nu}}\,+\,{\pi\over \Gamma(\nu)\,\Gamma(\nu+1)}\,
\Big({\alpha\over 2}\Big)^{2\nu}\,
\Big(i\,-\,\cot (\pi\nu)\Big)\,x^{\nu}\,+\,\ldots\Big]\,\,,
\eeq
for any constant $\alpha$, when $\alpha x$ is small and $\nu<1$. In our case:
\beq
\alpha={2\omega\over 5-p}\,\,,
\qquad\qquad
x\,=\,\rho^{{p-5\over 2}}\,\,,
\qquad\qquad
\nu\,=\,{1\over 5-p}\,\,.
\label{alpha_x_nu}
\eeq
Therefore, after absorbing the multiplicative constant factor, we can write:
\beq
E(\rho)\,=\,A\,\rho+A\,c_p\,\omega^{{2\over 5-p}}\,+\,\ldots\,\,,
\qquad\qquad
(p<4)\,\,,
\label{E_nh_low}
\eeq
where $A$ is a constant and the coefficient $c_p$ is:
\beq
c_p\,=\,\pi\,{(5-p)^{{3-p\over 5-p}}
\over 
\,\Big[\Gamma\Big({1\over 5-p}\Big)\Big]^2}\,\Big[
i\,-\,
\cot\Big({\pi\over 5-p}\Big)\Big]\,\,,
\qquad\qquad 
(p<4)\,\,.
\label{cp}
\eeq
Notice that $c_p$ has both real and imaginary parts.

Let us next consider the case $p=4$. The incoming solution contains in this case the function $H_{1}^{(1)}$. The corresponding expansion for small values of the argument of the function takes the form:
\beq
H_{1}^{(1)}(\alpha x)= -{2i\over \pi\alpha}\,\Big[
\,
{1\over x}\,+\,{\alpha^2\over 4}\,\Big(i\pi\,+\,1-2\gamma_E-2\log\big(\alpha x/ 2\big)\,\Big)x\,+\,\ldots\Big]
\,\,,
\eeq
where $\gamma_E=.577\cdots$ is the Euler-Mascheroni constant. 
In our case $x=\rho^{-1/2}$ and $\alpha\,=2\omega$ and we can write the expansion of $E$ as
\beq
E(\rho)\,=\,A\,\rho\,+\,A\,c_4\,\omega^2\,+A\,\omega^2\log \Big(\rho/\omega^2\Big)\,+\,\ldots\,\,,
\label{E_E_nh_low_p4}
\eeq
where $c_4$ is given by:
\beq
c_4\,=\,i\pi+1-2\gamma_E\,\,.
\label{c4}
\eeq

Let us now perform the near-horizon and low frequency limits in the opposite order. For low frequencies and momentum, the equation for $E$ becomes:
\beq
E''+\partial_{\rho}\log
{(\rho^{\lambda}+d^2)^{{3\over 2}}\over
(\omega^2-k^2)\,\rho^{\lambda}+\omega^2\,d^2}\,E'\,\approx\,0\,\,.
\label{eom_E_low_freq}
\eeq
This equation can be readily integrated:
\beq
E'\,=\,c_E\,{(\omega^2-k^2)\,\rho^{\lambda}+\omega^2\,d^2\over
(\rho^{\lambda}+d^2)^{{3\over 2}}}\,\,,
\label{Eprime_boundary}
\eeq
where $c_E$ is a constant. A second  integration gives:
\beq
E(\rho)\,=\,E^{(0)}\,-\,c_E\,
\int_{\rho}^{\infty}\,
{(\omega^2-k^2)\,\bar\rho^{\lambda}+\omega^2\,d^2\over
(\bar\rho^{\lambda}+d^2)^{{3\over 2}}}\,d\bar\rho\,\,,
\label{E_low_frequency}
\eeq 
where $E^{(0)}=E(\rho\to\infty)$ is the value of the electric field at the UV boundary.
Let us now define the following functions:
\beq
{\cal J}_1(\rho)\,\equiv\,\int_{\rho}^{\infty}\,
{\bar\rho^{\lambda}\over
(\bar\rho^{\lambda}+d^2)^{{3\over 2}}}\,d\bar\rho\,\,,
\qquad\qquad
{\cal J}_2(\rho)\,\equiv\,\int_{\rho}^{\infty}\,
{d\bar\rho\over
(\bar\rho^{\lambda}+d^2)^{{3\over 2}}}\,\,.
\label{calJ1_calJ2}
\eeq
Then it follows
\beq
E(\rho)\,=\,E^{(0)}\,-\,c_E\,\Big[
(\omega^2-k^2)\,{\cal J}_1(\rho)\,+\,\omega^2\,d^2\,{\cal J}_2(\rho)\,\Big]\,\,.
\label{E_Js}
\eeq
For $\lambda>2$ these integrals are convergent and can be computed analytically:
\bear
&&{\cal J}_1(\rho)\,=\,{2\over \lambda-2}\,\rho^{1-{\lambda\over 2}}\,
F\Big(\,{3\over 2}, {1\over 2}-{1\over\lambda}; {3\over 2}\,-\,{1\over\lambda};
- {d^2\over \rho^{\lambda}}\,\Big)\,\,,\rc\rc
&&{\cal J}_2(\rho)\,=\,{2\over 3\lambda-2}\,\rho^{1-{3\lambda\over 2}}\,
F\Big(\,{3\over 2}, {3\over 2}-{1\over\lambda}; {5\over 2}\,-\,{1\over\lambda};
- {d^2\over \rho^{\lambda}}\,\Big)\,\,.
\eear
Moreover, the hypergeometric functions in $(\omega^2-k^2)\,{\cal J}_1\,+\,\omega^2\,d^2\,{\cal J}_2$ can be combined as
\bear
&&(\omega^2-k^2)\,{\cal J}_1(\rho)\,+\,\omega^2\,d^2\,{\cal J}_2(\rho)=
-{2\over \lambda}\,\Bigg[\,
{\rho\over (\rho^{\lambda}+d^2)^{{1\over 2}}}\,k^2\rc
&&\qquad\qquad\qquad\qquad
+{2k^2-\lambda \omega^2\over \lambda-2}\,\rho^{1-{\lambda\over 2}}\,
F\Big(\,{1\over 2}, {1\over 2}-{1\over\lambda}; {3\over 2}\,-\,{1\over\lambda};
- {d^2\over \rho^{\lambda}}\,\Big)\,\Bigg]\,\,.
\eear
Therefore, for $\lambda>2$, we have:
\bear
&&E(\rho)\,=\,E^{(0)}\,+\,{2\,c_E\over \lambda}\,
\Bigg[\,
{\rho\over (\rho^{\lambda}+d^2)^{{1\over 2}}}\,k^2\rc
&&\qquad\qquad\qquad\qquad
+{2k^2-\lambda \omega^2\over \lambda-2}\,\rho^{1-{\lambda\over 2}}\,
F\Big(\,{1\over 2}, {1\over 2}-{1\over\lambda}; {3\over 2}\,-\,{1\over\lambda};
- {d^2\over \rho^{\lambda}}\,\Big)\,\Bigg]\,\,.
\label{E_low_freq_sol}
\eear
Let us now expand $E(\rho)$  in (\ref{E_low_freq_sol}) near $\rho\approx 0$. With this purpose it is better to deal directly with the definition  (\ref{calJ1_calJ2}) of the integrals ${\cal J}_1$ and ${\cal J}_2$. One can prove easily that:
\beq
{\cal J}_1(\rho)={2\over \lambda}\,\gamma\,d^{{2\over \lambda}-1}\,+\,{\mathcal O}(\rho^2)\,\,,
\qquad\qquad
{\cal J}_2(\rho)={\lambda-2\over \lambda}\,\gamma\,d^{{2\over \lambda}-3}\,-\,
{\rho\over d^3}\,+\,{\mathcal O}(\rho^2)\,\,,
\eeq
where $\gamma$ is the constant defined in (\ref{gamma_def}). Plugging these equations into (\ref{E_Js}) we arrive at:
\beq
E\,=\,c_E\,{\omega^2\over d}\,\rho\,+\,E^{(0)}\,+\,{2c_E\over \lambda}\,
d^{{2\over \lambda}-1}\,\gamma\,[\,k^2\,-\,{\lambda\over 2}\,\omega^2\,]\,\,.
\eeq
Moreover, taking into account (\ref{mu_gamma}), we can rewrite $E$ in terms of the chemical potential $\mu$ as:
\beq
E\,=\,c_E\,{\omega^2\over d}\,\rho\,+\,E^{(0)}\,+\,{2c_E\over \lambda}\,
{\mu\over d}\,[\,k^2\,-\,{\lambda\over 2}\,\omega^2\,]\,\,.
\label{E_low_nh}
\eeq
Let us now match the two expressions we have found for $E$ when $p<4$ (eqs. (\ref{E_nh_low}) and (\ref{E_low_nh})). From the terms linear in $\rho$ we get the following relation between  the constants $A$ and $c_E$:
\beq
A\,=\,c_E\,{\omega^2\over d}\,\,.
\label{A_c_E}
\eeq
We can use this last relation to eliminate $A$. By comparing the constant terms in 
 (\ref{E_nh_low})  and (\ref{E_low_nh}) we get
\beq
E^{(0)}\,=\,{c_E\over d}\,\,\Big[
{2\mu\over \lambda}\,\big({\lambda\over 2}\,\omega^2\,-\,k^2\Big)\,+\,c_p\,
\omega^{2{6-p\over 5-p}}\,\Big]\,\,,
\qquad\qquad (p<4, \lambda>2)\,\,.
\label{E0_cE}
\eeq
We now impose the Dirichlet boundary condition $E^{(0)}=0$, which leads to the following relation:
\beq
k^2\,-\,{\lambda\over 2}\,\omega^2\,=\,{\lambda\over 2\mu}\,c_p\,
\omega^{2{6-p\over 5-p}}\,\,.
\label{disp_zero_general}
\eeq
At lowest order, the right-hand-side of this equation can be neglected and we arrive at the following dispersion relation:
\beq
\omega\,=\,\pm \sqrt{{2\over \lambda}}\,k\,\,,
\qquad\qquad
(\lambda>2)\,\,.
\label{zero_sound_first_order}
\eeq
Thus, the speed of zero sound coincides with the speed of first sound for the probe that we have determined in (\ref{speed_first_sound}) and only depends on the index $\lambda$ of the intersection. Notice that the result (\ref{zero_sound_first_order}) is also valid for $p=4$ since the term on the right-hand-side of (\ref{disp_zero_general})  is higher order. To avoid clutter, let us consider the solution with plus sign in (\ref{zero_sound_first_order}) and  expand $\omega$ as:
\beq
\omega\,=\, \sqrt{{2\over \lambda}}\,k\,+\,\delta \omega\,\,.
\eeq
Plugging this ansatz in (\ref{disp_zero_general}) and expanding at zero order in $\delta\omega$, we arrive at an equation for $\delta\omega$:
\beq
\delta\omega\,=\,-{\sqrt{\lambda}\over 2\sqrt{2}}\,{c_p\over \mu}\,
\Bigg[\Big({2\over \lambda}\Big)^{{6-p\over 5-p}}\,k^{{7-p\over 5-p}}\,+\,
2\,{6-p\over 5-p}\,\Big({2\over \lambda}\Big)^{{7-p\over 2(5-p)}}\,
k^{{2\over 5-p}}\,\delta\omega\,\Bigg]\,\, ,
\eeq
which we can solve for $\delta\omega$ in powers of $k$. The second term on the right-hand-side is clearly of higher  order, thus
\beq
\delta\omega\,=\,-{c_p\over 2\mu}\,
\Big({2\over \lambda}\Big)^{{7-p\over 2(5-p)}}
\,k^{{7-p\over 5-p}}\,\,.
\eeq
Taking into account the expression of $c_p$ in (\ref{cp}), we find the following expression for the imaginary part of $\omega$:
\beq
{\rm Im} \,\omega=0 +{\rm Im} \,\delta\omega\,=\,-{\pi\over 2\mu}\,
{(5-p)^{{3-p\over 5-p}}\over \Big[\Gamma\Big({1\over 5-p}\Big)\Big]^2}\,
\Big({2\over \lambda}\Big)^{{7-p\over 2(5-p)}}
\,k^{{7-p\over 5-p}}\,\,.
\label{Im_omega}
\eeq
Similarly, we get a correction to the real part of $\omega$, namely:
\beq
{\rm Re} \,\delta\omega\,=\,{\pi\over 2\mu}\,
{(5-p)^{{3-p\over 5-p}}\over \Big[\Gamma\Big({1\over 5-p}\Big)\Big]^2}\,
\cot\Big({\pi\over 5-p}\Big)\,
\Big({2\over \lambda}\Big)^{{7-p\over 2(5-p)}}
\,k^{{7-p\over 5-p}}\,\,.
\label{Re_delta_omega}
\eeq
\begin{figure}[ht]
\center
 \includegraphics[width=0.40\textwidth]{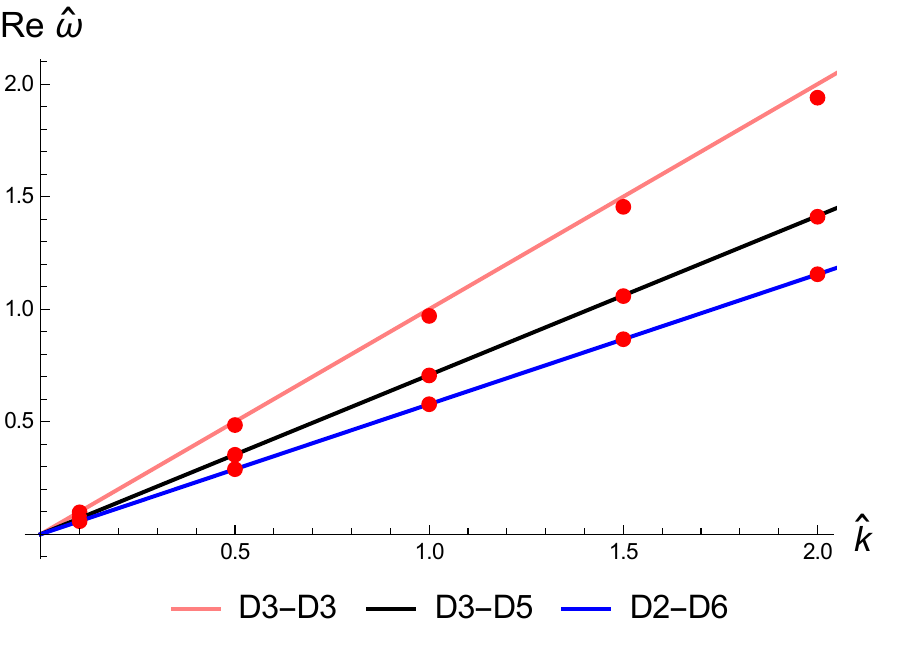}
 \qquad\qquad
 \includegraphics[width=0.40\textwidth]{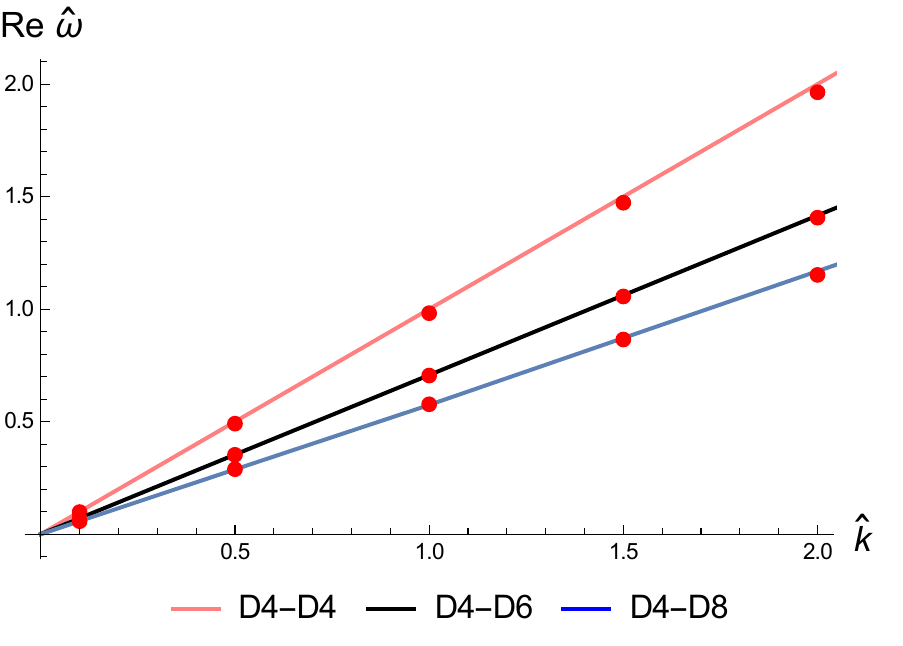}
 \caption{In this plot we represent the real part of $\hat \omega$ versus $\hat k$ for two classes of systems. On the left we depict the dispersion relation for the intersections D3-D3 ($p=3$, $\lambda=2$), D3-D5 ($p=3$, $\lambda=4$), and D2-D6 ($p=2$, $\lambda=6$). On the right we plot the same quantity for D4-D4 ($p=4$, $\lambda=2$), D4-D6 ($p=4$, $\lambda=4$), and D4-D8 ($p=4$, $\lambda=6$). The models are listed top-down. The points correspond to numerical values at $\hat d= 10^{3\lambda/2}$ and the continuous curves are the analytical results.}
 \label{Zero_sound_velocity}
\end{figure}
In Fig.~\ref{Zero_sound_velocity} we plot our prediction for the real part of $\hat\omega$ and we compare it with the results found numerically for different values of $\hat k$. We notice that the numerical dispersion relation for the zero sound mode is well reproduced by our analytic equations.

\subsubsection{The $p=4$ case}
\label{p4_case}
When $p=4$ the general formula (\ref{E_low_nh}) does not match (\ref{E_E_nh_low_p4}), due to the presence of the logarithmic term in the latter (which is absent in (\ref{E_low_nh})).  In order to include these terms we have to compute the next order correction to (\ref{E_low_nh})  near the horizon, following the procedure of \cite{Kulaxizi:2008jx}.  First of all, we recall that the near-horizon equation (\ref{nh_eq_E_zeroT}) for $E$ for $p=4$ is given by:
\beq
E''\,=\,-{\omega^2\over \rho^3}\,E\,\,.
\label{nh-eq-p4}
\eeq
Neglecting the right-hand side in this equation and integrating it gives rise to a linear solution as in (\ref{E_low_nh}), which is second order in the expansion parameter $\epsilon$.  To go beyond this order we just take into account the right-hand side of (\ref{nh-eq-p4}) and substitute the value of $E$, taken from (\ref{E_low_nh}), in this term. We arrive at the equation:
\beq
E''\,=\,-\omega^2\,\big[\Lambda_0\,\rho^{-3}\,+\,\Lambda_1\,\rho^{-2}\big]\,\,,
\eeq
where $\Lambda_0$ and $\Lambda_1$ are:
\beq
\Lambda_0\,=\,E^{(0)}\,+\,{2c_E\over \lambda}\,
{\mu\over d}\,[\,k^2\,-\,{\lambda\over 2}\,\omega^2\,]\,\,,
\qquad\qquad
\Lambda_1\,=\,c_E\,{\omega^2\over d}\,\,.
\eeq
($\Lambda_0\,+\,\Lambda_1\,\rho$ is just the second-order expression (\ref{E_low_nh})). Integrating this expression twice, we get
\beq
E\,=\,\Lambda_0+\Lambda_1\rho+\Lambda_1\,\omega^2\,\log\rho\,-\,{\Lambda_0\over 2}\,{\omega^2\over \rho}\,+\,\ldots\,\,,
\label{E_with_log_p4}
\eeq
where we took into account that the solution of the homogeneous equation is just 
(\ref{E_low_nh}). Recall also that the solution (\ref{nh-eq-p4}) with in-falling boundary conditions is given by $\rho^{{1\over 2}}\,H_{1}^{(1)} (2\omega/\rho^{{1\over 2}})$ and that the expansion (\ref{E_E_nh_low_p4}) is valid when the argument of the Hankel function is small, \ie, when $\rho\gg \omega^2$. Therefore, we can neglect the last term in (\ref{E_with_log_p4}). The resulting expression can be written as:
\beq
E\,=\,c_E\,{\omega^2\over d}\,\rho\,+\,E^{(0)}\,+\,{2c_E\over \lambda}\,
{\mu\over d}\,[\,k^2\,-\,{\lambda\over 2}\,\omega^2\,]
\,+\,c_E\,{\omega^4\over d}\,\log\rho\,\,+\,\ldots
\,\,.
\label{E_low_nh_p4}
\eeq
Let us now match (\ref{E_low_nh_p4}) and (\ref{E_E_nh_low_p4}). By comparing the terms linear in $\rho$ and those logarithmic in $\rho$, we find that the constant $A$ is related to $c_E$ as in (\ref{A_c_E}). Moreover, from the identification of the constant terms, we get that $E^{(0)}$ is given by:
\beq
E^{(0)}\,=\,{c_E\over d}\,\,\Big[
{2\mu\over \lambda}\,\big({\lambda\over 2}\,\omega^2\,-\,k^2\Big)\,+\,(c_4-\log \omega^2)\,
\omega^{4}\,\Big]\,\,,
\qquad\qquad (p=4, \lambda>2)\,\,,
\label{E0_cE_p4}
\eeq
where $c_4$ has been written in (\ref{c4}). 
By imposing that $E^{(0)}=0$ we get the following dispersion relation:
\beq
k^2\,-\,{\lambda\over 2}\,\omega^2\,=\,{\lambda\over 2\mu}\,(c_4\,-\,\log\omega^2)\,
\omega^{4}\,\,,
\qquad\qquad (p=4, \lambda>2)\,\,.
\label{disp_zero_p4}
\eeq
Let us solve this polynomial equation at different orders in the expansion parameter $\epsilon$. At leading order we find that $\omega$ is real and given by (\ref{zero_sound_first_order}), \ie, this general expression is also valid for $p=4$. 
The next-to-leading contribution $\delta\omega$ is given by:
\beq
\delta\omega\,=\,-{1\over 2\mu}\,\Big({2\over \lambda}\Big)^{{3\over 2}}\,
\Big(c_4-\log \big({2k^2\over \lambda}\big)\Big)\,k^3\,\,.
\qquad\qquad (p=4, \lambda>2)\,\,.
\eeq
Separating the real and imaginary parts using (\ref{c4}), we get:
\bear
{\rm Im}\,\omega & =& 0+{\rm Im}\,\delta\omega = -{\pi\over 2\mu}\,\Big({2\over \lambda}\Big)^{{3\over 2}}\,k^3\rc
{\rm Re}\,\delta\omega & = & {2\gamma_E\,-\,1+\log\big(2k^2/ \lambda\big)\over
2\mu}\,\Big({2\over \lambda}\Big)^{{3\over 2}}\,k^3\,\,,
\qquad\qquad (p=4, \lambda>2)\,\,.
\eear
Notice that, curiously, the leading term to ${\rm Im}\,\omega$ written above is exactly the same as the one obtained by taking $p=4$  in (\ref{Im_omega}).   For $\lambda=5$ these formulas were obtained  in \cite{Kulaxizi:2008jx}.

\subsubsection{The $\lambda=2$ case}\label{lambda2_case}

When $\lambda=2$ the integral ${\cal J}_1(\rho)$ defined in (\ref{calJ1_calJ2}) is not convergent and, therefore, the integral of (\ref{Eprime_boundary}) written in (\ref{E_Js})  is not valid. In this case we can define a new integral $\bar{\cal J}_1(\rho)$  as:
\beq
\bar{\cal J}_1(\rho)\,\equiv\,\int_{\rho}^{\infty}\,\Big[
{\bar\rho^2\over (\bar\rho^2+d^2)^{{3\over 2}}}-{1\over \bar\rho}\Big]=
{\rho\over \sqrt{\rho^2+d^2}}-1\,+\,\log{2\rho\over \sqrt{\rho^2+d^2}+\rho}\,\,.
\eeq
Then, the solution of (\ref{Eprime_boundary}) for $\lambda=2$ can be written as:
\beq
E(\rho)\,=\,E^{(0)}\,-\,c_E\,\Big[
(\omega^2-k^2)\,(
\bar {\cal J}_1(\rho)\,-\,\log\rho)
+\,\omega^2\,d^2\,{\cal J}_2(\rho)\,\Big]\,\,.
\label{E_Js_lambda2}
\eeq

At the UV (\ie,  when $\rho$ is large) the integrals $\bar {\cal J}_1(\rho)$ and 
${\cal J}_2(\rho)$ vanish by construction and $E(\rho)$ therefore behaves as:
\beq
E(\rho)\,=\,E^{(0)}\,+\,c_E\,(\omega^2-k^2)\,\log\rho\,+\,\ldots\,\,,
\qquad\qquad
(\rho\to\infty)\,\,.
\eeq
When this logarithmic behavior is present, the source in the AdS/CFT correspondence is determined by the coefficient of the logarithm, which should vanish. This can be achieved either by requiring $c_E=0$ or:
\beq
\omega\,=\,\pm k\,\,.
\label{dispersion_lambda2}
\eeq
Let us see that if for $c_E=0$ the solution of the fluctuation equation is trivial. Indeed, in this case 
$E(\rho)$ is constant and the matching with the near-horizon results (\ref{E_nh_low}) or (\ref{E_E_nh_low_p4}) is only possible if the constant $A$ in those solutions is zero. This, in turn, implies that the whole solution vanishes, as claimed. Therefore, the dispersion relation in this case should be given by (\ref{dispersion_lambda2}), which corresponds to having no dissipation  and the speed of zero sound is equal to one. Notice that, when (\ref{dispersion_lambda2}) holds, $E(\rho)$ behaves near $\rho=0$ as:
\beq
E(\rho)\,=\,E^{(0)}\,+c_E\,{\omega^2\over d}\,\rho+{\mathcal O}(\rho^2)\,\,.
\label{E_low_nh_lambda2}
\eeq
One can then match  (\ref{E_low_nh_lambda2}) with  eq. (\ref{E_nh_low}) (for $p<4$). From the comparison of the linear terms we get that $A$ is given by the same expression as in (\ref{A_c_E}) and that $E^{(0)}$ is related to $c_E$ as:
\beq
E^{(0)}\,=\,{c_E\over d}\,\omega^{2{6-p\over 5-p}}\,\,.
\label{E0_cE_lambda4}
\eeq
Notice that (\ref{E0_cE_lambda4}) is just the same as (\ref{E0_cE}) when $\lambda=2$ and $\omega=\pm k$. However, in the present $\lambda=2$ case we do not have to impose that $E^{(0)}=0$ and, thus, the dispersion relation is not given by (\ref{disp_zero_general}) but instead we have (\ref{dispersion_lambda2}) (see \cite{Hung:2009qk} for a similar analysis in the D3-D3 intersection).

When $p=4$ the matching of the logarithmic term in the near-horizon expansion requires to go beyond the leading term in $\omega$,  as in subsection \ref{p4_case}. It is easy to see that (\ref{E_low_nh_lambda2}) can be corrected in such a way that it matches (\ref{E_E_nh_low_p4}) and  that $E^{(0)}$ is related to $c_E$ as in  (\ref{E0_cE_p4}) with $\lambda=2$ and $\omega=\pm k$.

The dispersion relation (\ref{dispersion_lambda2}) is the natural one for a massless excitation in $1+1$ dimensions. Accordingly, let us determine all possible intersections with $\lambda=2$ and $n=1$. Taking  $\lambda=2$ and $n=1$ in (\ref{lambda_n}) we find the following relation between $p$ and $q$:
\beq
(p-3)\,q\,=\,(p-3)\,(10-p)\,\,,
\qquad\qquad
(\lambda=2\,,\,n=1) \ .
\eeq
When $p=3$ this equation is satisfied by any $q$, whereas for $p\ne 3$ it requires that $p+q=10$. Therefore, we have the following two series of intersections  having $\lambda=2$ and $n=1$:
\bear
&& (1|3\perp q)\,\,,\qquad\qquad q=3,5,7\,\,,\rc\rc
&& (1|p\perp (10-p))\,\,,\qquad\qquad p=1,2,4\,\,,
\eear
where we have required that $q\ge p$.

\subsection{The collisionless/hydrodynamic crossover}\label{crossover_zeroB}

\begin{figure}[!ht]
\center
 \includegraphics[width=0.65\textwidth]{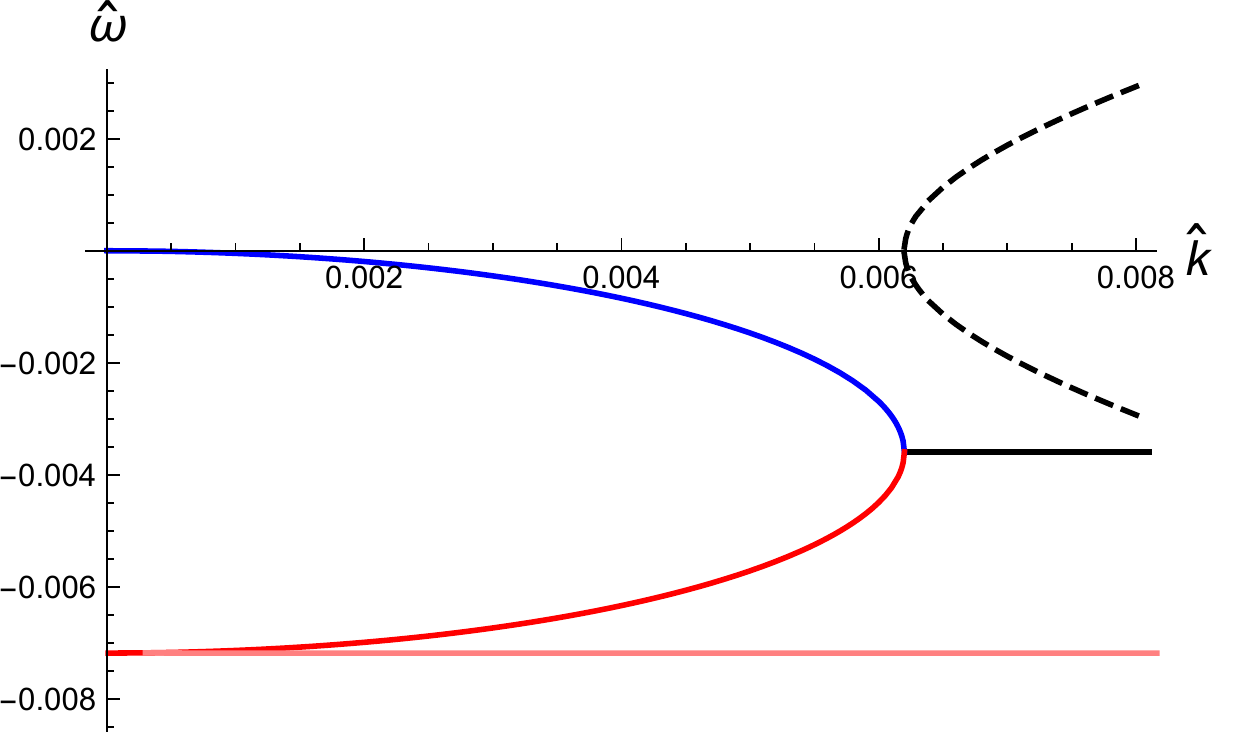}
 \caption{We are presenting typical dispersions in the D2-D6 model at $\hat d=10^6$ and $\hat B=0$. The dashed curves are real parts and continuous curves are imaginary parts. Around $\hat k\sim 0.006$ the diffusive mode merges with another purely imaginary mode, which defines what is called the crossover location ($\hat k_{cr}$,$\hat\om_{cr}$). Notice, that we have included the first excited purely imaginary mode from the transverse sector (lowest pink curve).}
 \label{fig:dispersion_zeroB}
\end{figure}

\begin{figure}[!ht]
\center
 \includegraphics[width=0.75\textwidth]{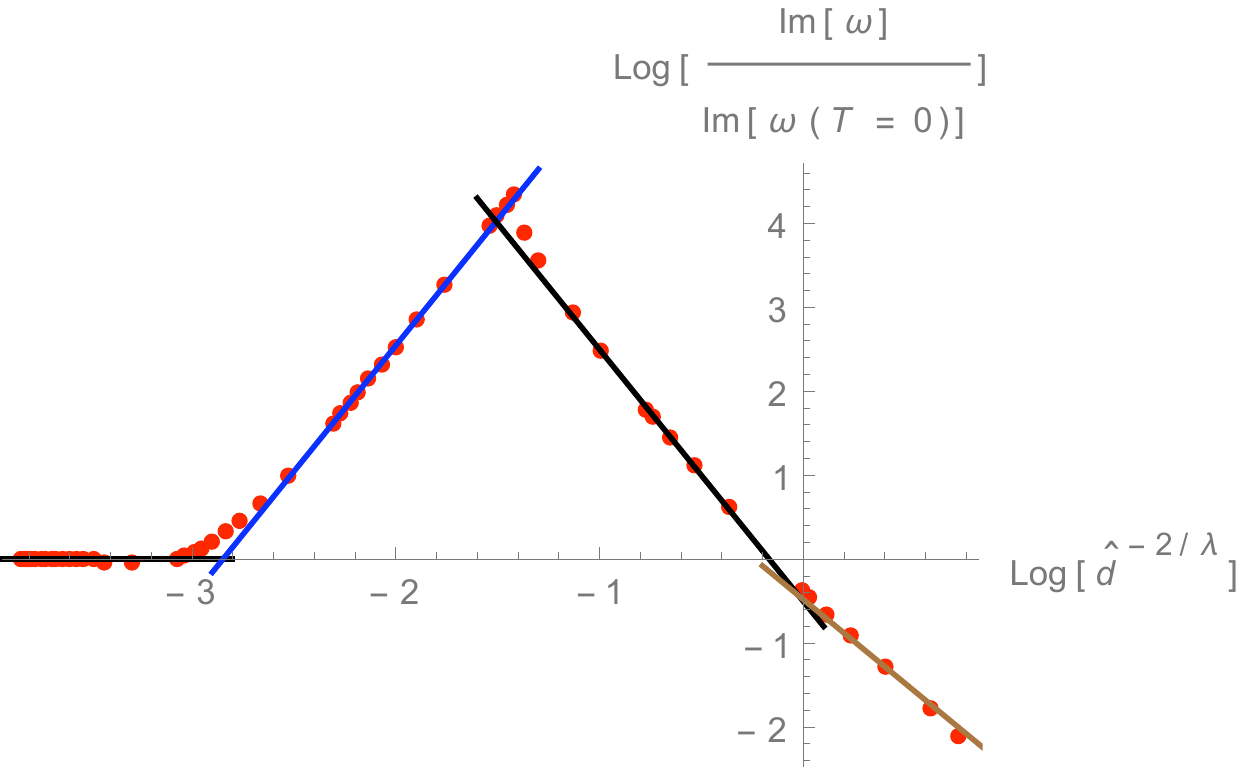}
 \caption{The imaginary part of the lowest excitation mode in the D1-D5 model at fixed $\frac{\hat k}{\hat d^{(5-p)/\lambda}}=0.01$. The red point represent the numerical results and the line segments have slopes 0,3,-3,-2 from left to right. These are consistent with the analytic results. The 0 conforms with the notion that the zero sound is robust against temperature variations and only for sufficiently hot environment will the decay rate behave as $T^{3/2}$. The latter two slopes correspond to the diffusive regime with temperature behaviors $D\sim T^{-3/2}$ (eq. (\ref{low_T_D})) and $D\sim T^{-1}$ (eq. (\ref{D_large_T})).}
 \label{decay_rate}
\end{figure}

Our analytic treatment of the zero sound in subsection \ref{zero_sound_zeroB} was carried out at $T=0$. The numerical analysis of the fluctuation equation shows that the zero sound persists at $T\not=0$ if $T$ is low enough. 
A typical dispersion relation is depicted in Fig.~\ref{fig:dispersion_zeroB}.
In this regime the real part of this mode is independent of the temperature, while its imaginary part receives corrections which are proportional to $T^{{7-p\over 5-p}}$. This behavior is illustrated in Fig.~\ref{decay_rate} for the D1-D5 intersection.

\begin{figure}[!ht]
\center
 \includegraphics[width=0.7\textwidth]{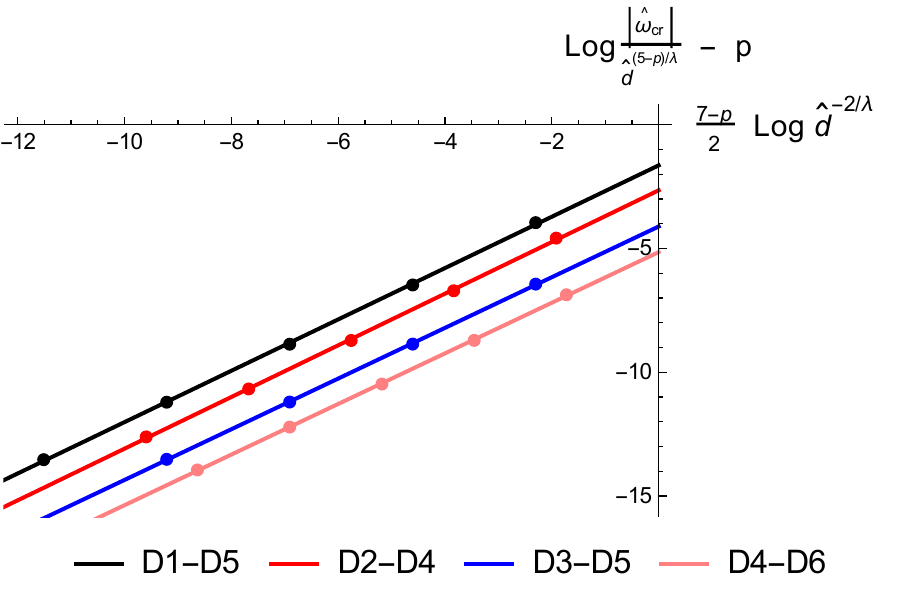}
 \qquad
  \includegraphics[width=0.7\textwidth]{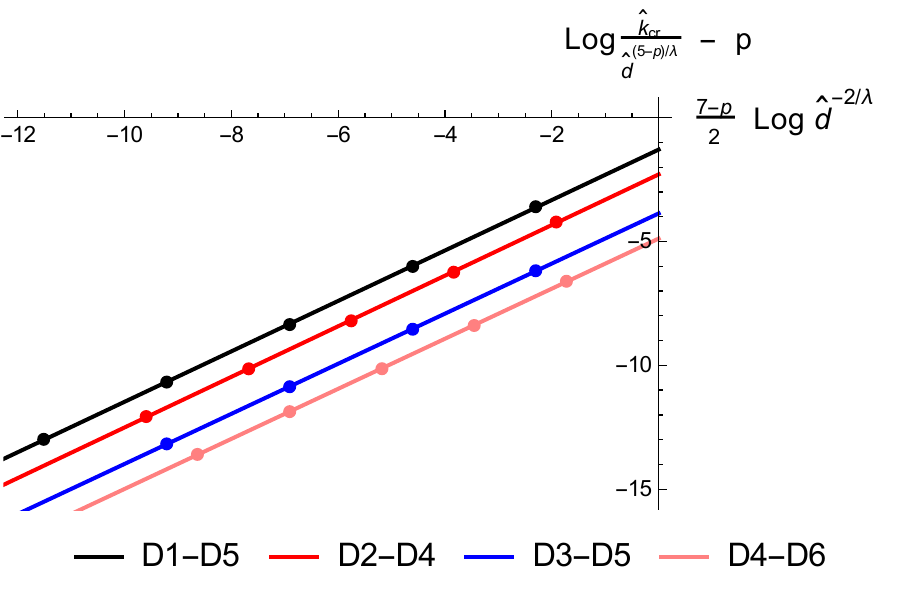}
 \caption{We plot the frequency (top panel) and the momentum (bottom panel) for the location of the transition from the hydrodynamic to collisionless regime. The scalings are chosen such that the slopes for all the models is $1$ (best fit values). The models are D1-D5, D2-D6, D3-D5, and D4-D6 top-down. This translates to scalings $\om_{cr}\sim r_h^{(7-p)/2}/\mu\sim T^{(7-p)/(5-p)}/\mu$ and $k_{cr}\sim r_h^{(7-p)/2}/\mu\sim T^{(7-p)/(5-p)}/\mu$. Notice that we have subtracted a constant $p$ to separate the lines.}
 \label{transition}
\end{figure}

If we continue increasing the temperature, at some point there is going to be a crossover to a hydrodynamic diffusive regime, in which the $D$ behaves as in eqs. (\ref{low_T_D}) and (\ref{D_large_T}). The frequency $\omega_{cr}$ and momentum $k_{cr}$ at which this collisionless/hydrodynamic crossover takes place depends on the temperature and chemical potential. We have determined this dependence numerically (see Fig.~\ref{transition}). From this analysis we conclude that $\omega_{cr}$ and $k_{cr}$ scale as:
\beq
\omega_{cr}\sim {T^{{7-p\over 5-p}}\over \mu} \ \ \ \ , \ \ \ \ k_{cr}\sim {T^{{7-p\over 5-p}}\over \mu} \ .
\eeq
Notice that, for the D3-D5 and D3-D7 system, $\omega_{cr}\sim T^2/\mu$, in agreement with the analysis of \cite{Davison:2011ek} and \cite{Brattan:2012nb}, respectively, and for the Sakai-Sugimoto model $\omega_{cr}\sim T^3/\mu$, in agreement with the analysis of \cite{DiNunno:2014bxa}.

\section{Non-vanishing $B$ field}
\label{nonzeroB}
In this section we study the influence of the magnetic field $B$ in the collective excitations of our brane intersection. The equations of motion of the brane fluctuations have been written in appendix \ref{appendixA} (eqs. (\ref{eom_E_Bnonzero}) and (\ref{eom_ay_general})).  Notice that, when the magnetic field is non-vanishing, the longitudinal and transverse fluctuations are coupled. We first study the magnetized system at $T\not=0$ in the diffusive regime.

\subsection{Diffusion constant}
\label{diffusion_nonzeroB}

To obtain the diffusion constant in the presence of the magnetic field we follow the same steps as in subsection \ref{diffusion_zeroB}. First we expand the equations of motion around the horizon $\rho=r_h$. As in ref. \cite{Brattan:2012nb} the equations decouple and we have to expand the terms multiplying $E'$ and $E$ in (\ref{eom_E_Bnonzero}). These expansions are:
\bear
&&\partial_{\rho}\log\,\Bigg[
{\rho^{7-p}\over \rho^{7-p}+B^2}\,
{(\rho^{\lambda}+\rho^{\lambda+p-7}\,B^2\,+d^2)^{{3\over 2}}\,f_p\over
(\omega^2-f_p\,k^2)\,\rho^{\lambda}+\omega^2\,B^2\,\rho^{\lambda+p-7}+
\omega^2\,d^2}\Bigg]\,=\,{1\over \rho-r_h}\,+\,c_1
\,+\,\ldots\rc
&&{1\over \rho^{7-p}\,f_p^2}\,
{(\omega^2-f_p\,k^2)\,\rho^{\lambda}+\omega^2\,B^2\,\rho^{\lambda+p-7}+
\omega^2\,d^2
\over 
\rho^{\lambda}+\rho^{\lambda+p-7}\,B^2+d^2}
\,=\,{A\over (\rho-r_h)^2}\,+\,{c_2\over \rho-r_h}\,+\,\ldots\,\,,\qquad
\label{nh_expansion_Eeq_nonzeroB}
\eear
where the constants $A$, $c_1$, and $c_2$ are given by:
\bear
&& A\,=\,{\omega^2\over (7-p)^2\,r_h^{5-p}}\rc
&& c_1\,=\,(7-p)\,{r_h^{\lambda-1}\over d^2+r_h^{\lambda}+B^2\,r_h^{p+\lambda-7}}\,
{k^2\over  \omega^2}\,+{d^2\over 2 r_h}\,
{p-8+(6-p) B^2 r_h^{p-7}\over
(1+B^2\,r_h^{p-7})(
 d^2+r_h^{\lambda}+B^2\,r_h^{p+\lambda-7})}\rc
&&\qquad\qquad\qquad\qquad
+{r_h^{\lambda-1}\over 2}{p-8+\lambda+(\lambda-1)\,B^2\,r_h^{p-7}\over
 d^2+r_h^{\lambda}+B^2\,r_h^{p+\lambda-7}}\rc
&&c_2\,=\,-{r_h^{p+\lambda-6}\over (7-p)\,( d^2+r_h^{\lambda}+B^2\,r_h^{p+\lambda-7})}\,k^2\,+\,
{1\over (7-p)^2\,r_h^{6-p}}\,\omega^2\,\,.
\label{A_c1_c2_nonzeroB}
\eear
The resulting fluctuation equations can be solved in Frobenius series as in (\ref{Frobenius_E_nh}), in terms of two parameters $\alpha$ and $\beta$. In the low frequency limit 
with $k\sim {\mathcal O}(\epsilon)$, $\omega\sim {\mathcal O}(\epsilon^2)$, the exponent $\alpha$ is given by the same expression as in (\ref{exponent_alpha}), while $\beta$ is given by:
\beq
\beta\,=\,i\,{k^2\over \omega}\,\,{r_h^{\lambda+{p-7\over 2}}\over 
d^2\,+\,r_h^{\lambda}+B^2\,r_h^{p+\lambda-7}}\,\,.
\label{Beta_nonzeroB}
\eeq

We now expand first in frequency. The equation for $E$ decouples from the one for $a_y$ in this limit and becomes:
\beq
E''\,-\,\partial_{\rho}\,\log\Big[{\rho^{\lambda}+\rho^{p+\lambda-7}\,B^2\over 
(\rho^{\lambda}+\rho^{p+\lambda-7}\,B^2\,+d^2)^{{3\over 2}}}\Big]\,E'\,=\,0\,\,.
\eeq
This equation can be integrated as:
\beq
E(\rho)\,=\,E^{(0)}\,+\,c_E\,\int_{\rho}^{\infty}d\bar\rho\,\,
{\bar\rho^{\lambda}+\bar\rho^{p+\lambda-7}\,B^2\
\over \big(\bar\rho^{\lambda}+\bar\rho^{p+\lambda-7}\,B^2+d^2\big)^{{3\over 2}}}\,\,,
\label{E_low_freq_nonzeroB}
\eeq
where $c_E$ is a constant of integration. Let us expand the function $E(\rho)$ in (\ref{E_low_freq_nonzeroB}) near the horizon as:
\beq
E(\rho)= E^{(0)}\,+\,c_E\,{\cal I}_{\lambda, p}\,-\,
c_E\,{r_h^{\lambda}+r_h^{\lambda+p-7}\,B^2\over
(r_h^{\lambda}+r_h^{\lambda+p-7}\,B^2\,+d^2)^{{3\over 2}}}\,(\rho-r_h)\,+\,\ldots\,\,,
\label{low_freq_nh_E_nonzeroB}
\eeq
where ${\cal I}_{\lambda, p}$ is the integral:
\beq
{\cal I}_{\lambda, p}\,=\,
\int_{r_h}^{\infty}d\bar\rho\,\,
{\bar\rho^{\lambda}+\bar\rho^{p+\lambda-7}\,B^2\
\over \big(\bar\rho^{\lambda}+\bar\rho^{p+\lambda-7}\,B^2+d^2\big)^{{3\over 2}}} = \,r_h^{1-{\lambda\over 2}}\,
\int_{1}^{\infty}dx\,\,
{x^{\lambda}+x^{p+\lambda-7}\,\hat B^2\
\over \big(x^{\lambda}+x^{p+\lambda-7}\,\hat B^2+\hat d^2\big)^{{3\over 2}}} \ ,
\eeq
and where we have rescaled the magnetic field $\hat B$ as:
\beq\label{hat_B_def}
\hat B\,=\,{B\over r_h^{{7-p\over 2}}} = \Big({7-p\over 4\pi}\Big)^{{7-p\over 5-p}}\,\,
{B\over T^{{7-p\over 5-p}}} \ .
\eeq
We now match (\ref{low_freq_nh_E_nonzeroB}) with the near-horizon expression (\ref{nh_low_freq_E}). By comparing the constant terms in both expressions we conclude that
$E^{(0)}=E_{nh}-c_E\,{\cal I}_{\lambda, p}$, whereas the constant $c_E$ can be determined by looking at the linear  term. We get:
\beq
c_E\,=\,-\beta\,E_{nh}\,{r_h^{{\lambda\over 2}}\,(1+\hat b^2\,+\hat d^2)^{{3\over 2}}\over
1+\hat B^2}\,\,,
\eeq
where $\beta$ is given in (\ref{Beta_nonzeroB}). It follows that:
\beq
E^{(0)}\,=\,E_{nh}\,\Big[1\,+\,i\,{k^2\over \omega}\,r_h^{{p-7\over 2}}\,
{(1+\hat B^2+\hat d^2\big)^{{1\over 2}}\over 1+\hat B^2}\,
\,{\cal I}_{\lambda,p}\,\Big]\,\,.
\eeq
Requiring $E^{(0)}=0$ we find a dispersion relation of diffusive type, 
$\hat\omega\,=\,-i \hat D_{\lambda, p}\,\hat k^2$, where $\hat \omega$ and $\hat k$ are rescaled
as in (\ref{hat_omega_k}) and the rescaled diffusion constant $\hat D_{\lambda, p}$ is given by:
\beq
\hat D_{\lambda, p}\,=\,
{(1+\hat B^2+\hat d^2\big)^{{1\over 2}}\over 1+\hat B^2}\,
\int_{1}^{\infty}dx\,\,
{x^{\lambda}+x^{p+\lambda-7}\,\hat B^2\
\over \big(x^{\lambda}+x^{p+\lambda-7}\,\hat B^2+\hat d^2\big)^{{3\over 2}}}\,\,.
\label{Dhat_nonzeroB}
\eeq
We have not been able to compute the integral (\ref{Dhat_nonzeroB})  analytically for arbitrary values of $\lambda$ and $p$. However, when $\lambda+p=7$ this integral can be obtained in terms of a hypergeometric function:
\bear
&&\hat D_{\lambda=7-p, p}\,=\,{2\over \lambda}\,
{(1+\hat B^2+\hat d^2\big)^{{1\over 2}}\over 1+\hat B^2}\,
\Bigg[{\hat d^2\over (\hat d^2+\hat B^2) (1+\hat d^2+\hat B^2)^{{1\over 2}}}\rc
&&\qquad\qquad\qquad
+\Bigg({2\over \lambda-2}\,+\,{\hat B^2\over \hat d^2+\hat B^2}\Bigg)
F\Big({1\over 2}, {1\over 2}-{1\over \lambda};{3\over 2}-{1\over \lambda};
-\hat d^2-\hat B^2\Big)\Bigg]\,\,.\qquad
\label{Dhat_nonzeroB_hyper}
\eear

\begin{figure}[!ht]
\center
 \includegraphics[width=0.65\textwidth]{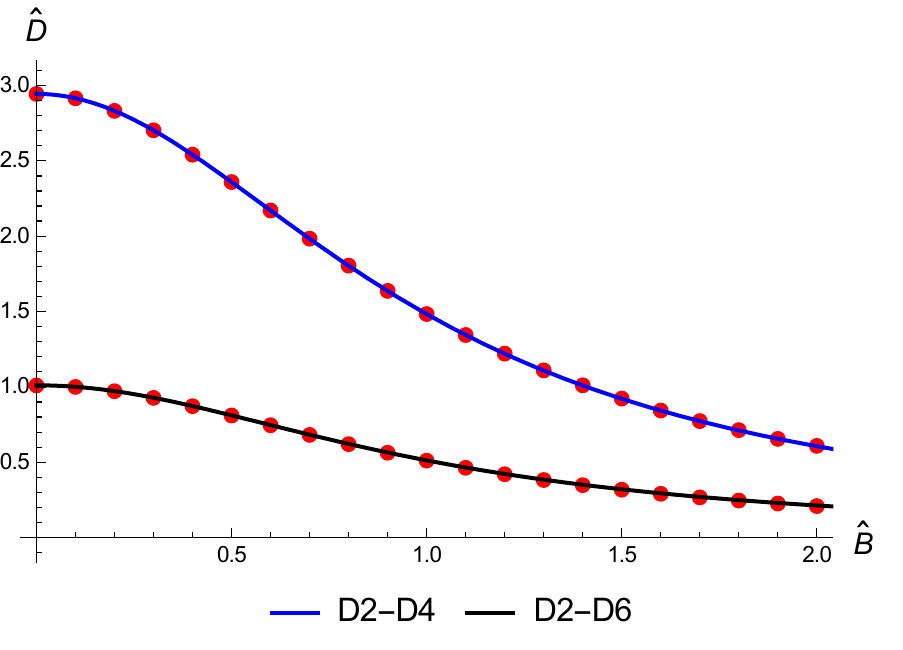}
 \caption{Plot of the diffusion constant for D2-D4 and D2-D6 as a function of $\hat B$ at fixed  $\hat d=10$. The continuous curves  are the analytic results, whereas the points are from the numerics. }
 \label{Diffusion_nonzeroB}
\end{figure}

Notice that the systems for which (\ref{Dhat_nonzeroB_hyper}) holds include the D3-D3 intersection ($(2|3\perp 5)$) and D2-D8' ($(2|2\perp 8)$). 
In Fig.~\ref{Diffusion_nonzeroB} we compare the analytic results to the numerical values of the diffusion constant as a function of $\hat B$. It is also easy to find the limiting value of 
$\hat D_{\lambda, p}$  as $\hat d=0$. Indeed, for any value of $\lambda$ and $p$ we get:
 \beq
\hat D_{\lambda, p}(\hat d=0)\,=\,{2\over \lambda-2}\,
{1\over \sqrt{1+\hat B^2}}\,
F\Big({1\over 2}, {\lambda-2\over 2(7-p)};1+{\lambda-2\over 2(7-p)};
-\hat B^2\Big)\,\,.
\label{hatD_zerohatd_nonzeroB}
\eeq
Notice that when $\hat B=0$  eq. (\ref{hatD_zerohatd_nonzeroB}) reduces to (\ref{hatD_Tinfnity}).

We wish to end this subsection with the following remark. In analogy to the lower bound on the shear viscocity to the entropy ratio, Hartnoll proposed a universal lower bound on the charge diffusivity for incoherent metals at high temperatures \cite{Hartnoll:2014lpa}. Unfortunately, the proposal was not precisely formulated and thus the bound does not have a specific value, and we therefore cannot make a direct comparison. Nevertheless, according to \cite{Hartnoll:2014lpa}, the bound in our language reads
\be
 \hat D \gtrsim \frac{\hbar v_F^2}{k_B}=v_F^2 \ ,
\ee
where $v_F$ is some number of order 1 (for Fermi liquid with quasiparticle description $v_F$ would correspond to the Fermi velocity). The minimum value for $\hat D$ occurs at $\hat d=0$, {\emph{i.e.}}, at high temperature as in the regime of interest. This was computed in (\ref{hatD_Tinfnity}): $\lim_{T\to\infty}\hat D=2/(\lambda-2)$.  At finite magnetic field this is given above in (\ref{hatD_zerohatd_nonzeroB}). In general, turning on the magnetic field the diffusion constant decreases without bound, see Fig.~\ref{Diffusion_nonzeroB} for example. Therefore, holographic metals as modelled in the present work, seem to evade any bound on charge diffusion at least at finite magnetic field.

\subsection{Zero sound with $B$ field}\label{zero_sound_B}

\begin{figure}[ht]
\center
 \includegraphics[width=0.45\textwidth]{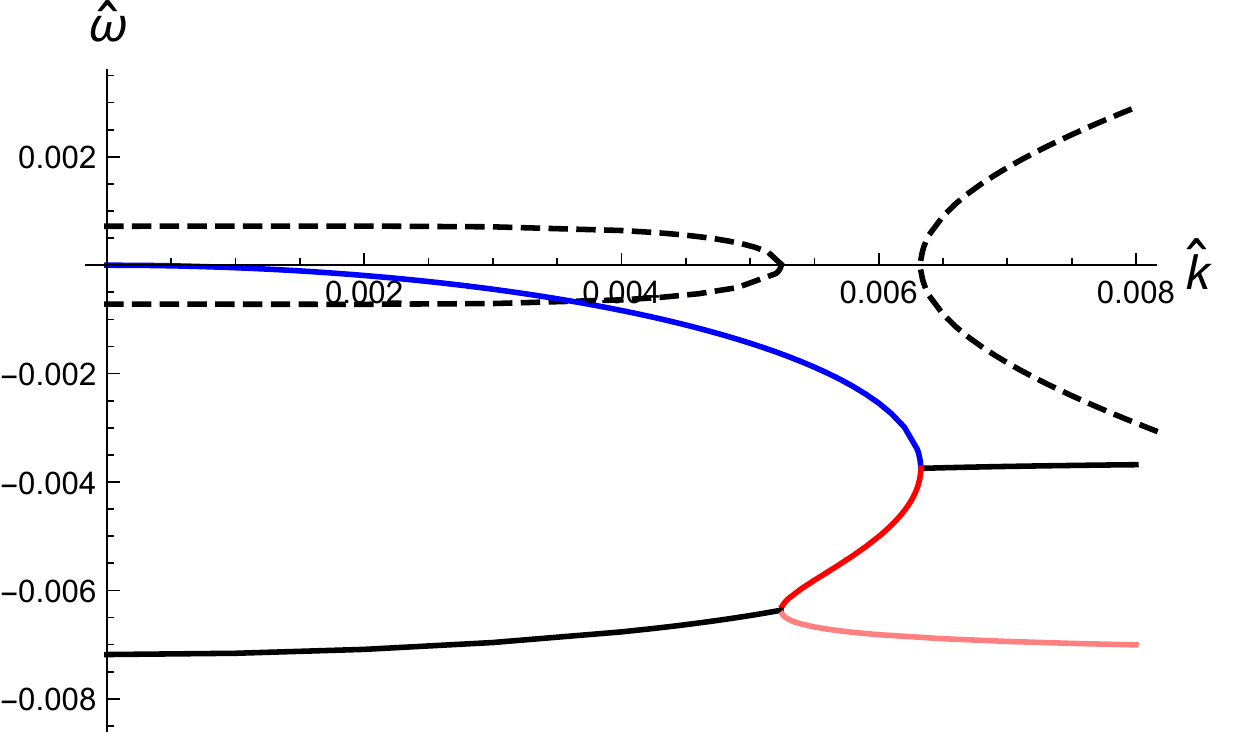}
 \includegraphics[width=0.45\textwidth]{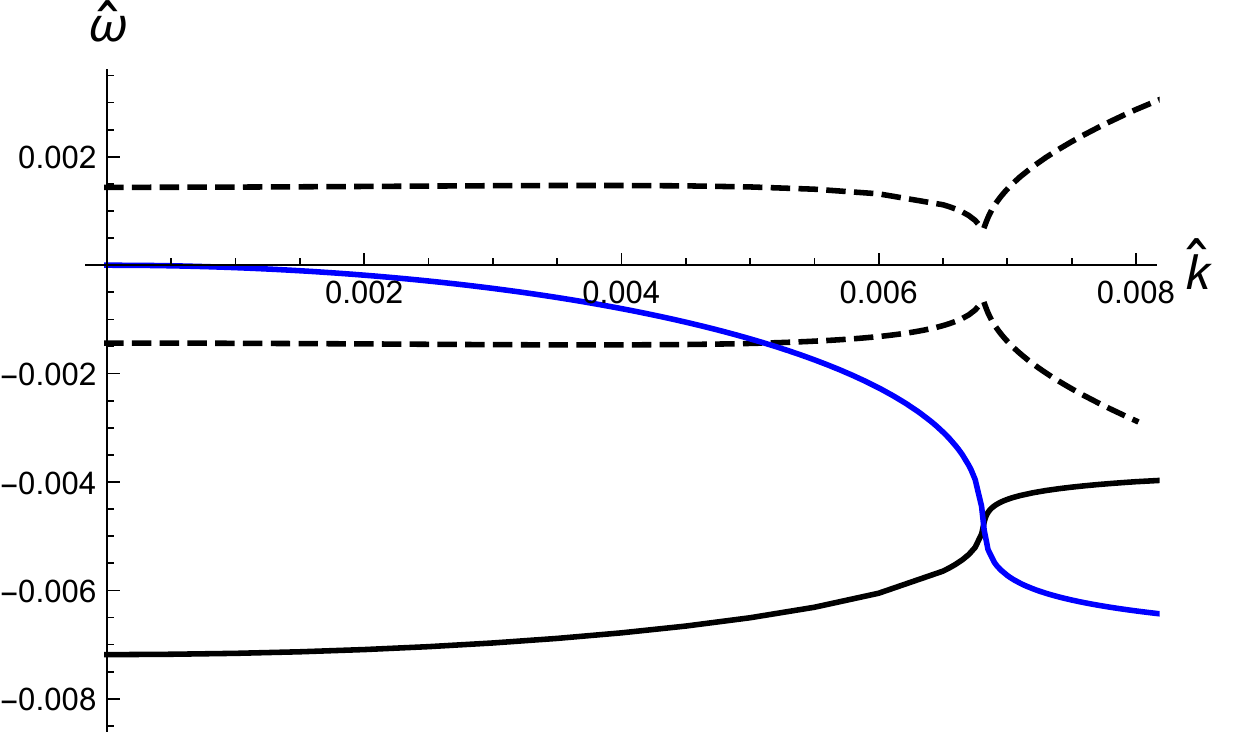}
 \caption{The zero sound becomes massive above a critical magnetic field strength. We depict the dispersions for the D2-D6 model at $\hat B=0.1$ (left panel) and $\hat B=0.2$ (right panel), both at $\hat d=10^6$. In the latter case the zero sound mode has just become massive and starts propagating.}
 \label{fig:zerosound_nonzeroB}
\end{figure}

Let us study the zero sound for intersections with $B$ field. Typical dispersions are depicted in Fig.~\ref{fig:zerosound_nonzeroB} for the D2-D6 model. First of all we analyze the equations of motion of appendix \ref{appendixB} for $E$ and $a_y$ at zero temperature near the horizon $\rho\approx 0$. We will assume that $B$ is small ($\rho^{7-p}\sim B^2$). With these assumptions the equations of motion for $E$ and $a_y$ greatly simplify and become the coupled system:
\bear
E''\,+\,{(7-p)B^2\over \rho(\rho^{7-p}+B^2)}\,E'\,+\,{\omega^2\over \rho^{7-p}}\,E & = &
-i(7-p)\,{B\omega^2\over \rho (\rho^{7-p}+B^2)}\,a_y\rc
a_y''\,+\,{(7-p)B^2\over \rho(\rho^{7-p}+B^2)}\,a_y'\,+\,{\omega^2\over \rho^{7-p}}\,a_y & = &
i(7-p)\,{B\over \rho (\rho^{7-p}+B^2)}\,E\,\,.
\label{E_ay_coupled}
\eear
Let us now define the operator $\hat{\cal O}$, which acts on any function $F(\rho)$  as follows:
\beq
\hat{\cal O}\,F(\rho)\equiv
F''\,+\,{(7-p)B^2\over \rho(\rho^{7-p}+B^2)}\,F'\,+\,{\omega^2\over \rho^{7-p}}\,F\,\,.
\eeq
Then, the  equations  in (\ref{E_ay_coupled}) can be written as:
\beq
\hat{\cal O}\,E\,=\,-i(7-p)\,{B\omega^2\over \rho (\rho^{7-p}+B^2)}\,a_y\,\,,
\qquad\qquad
\hat{\cal O}\,a_y\,=i(7-p)\,{B\over \rho (\rho^{7-p}+B^2)}\,E\,\,.
\eeq
These coupled equations are easy to decouple. 
By defining $y_\pm$ as
\be
 y_\pm = \frac{E}{i \omega}\pm a_y \ ,
\ee
the equations satisfied by $y_{\pm}$ are decoupled, and given by:
\beq
\Big(\hat{\cal O}\pm (7-p)\,{B\omega \over \rho (\rho^{7-p}+B^2)}\Big)\,y_{\pm}\,=\,0\,\,.
\eeq
More explicitly, we have
\beq
y_{\pm}''\,+\,{(7-p)B^2\over \rho(\rho^{7-p}+B^2)}\,y_{\pm}'\,+\,
\Bigg({\omega^2\over \rho^{7-p}}\,\pm {(7-p)\,B\omega\over \rho(\rho^{7-p}+B^2)}\,\Bigg)
y_{\pm}\,=\,0\,\,.
\eeq
We now study the equations for $y_{\pm}$  when the $B$ field is small. Let us neglect the terms that are quadratic in $B$ and approximate these equations as:
\beq
y_{\pm}''\,+\,
\Bigg({\omega^2\over \rho^{7-p}}\,\pm (7-p)\,B {\omega\over \rho^{8-p}}\,\Bigg)
y_{\pm}\approx 0\,\,.
\eeq
Let us expand the solutions of these equations in powers of $B$. Notice that the equation for $y_-$ is obtained from the one for $y_+$ by changing $B$ by -$B$. Accordingly, we write:
\beq
y_{\pm}(\rho)\,=\,y_0(\rho)\pm B\, y_1(\rho)\,\,.
\eeq
The equations for $y_0$ and $y_1$ are independent of $B$ and given by:
\beq
 y_{0}''\,+\,{\omega^2\over \rho^{7-p}}\,y_{0}\,=\,0\,\,,
\qquad\qquad \qquad
 y_{1}''\,+\,{\omega^2\over \rho^{7-p}}\,y_{1}\,=\,-(7-p)\,{\omega\over \rho^{8-p}}\,y_{0}
 \,\,.
\eeq
The solution for the equation of $y_{0}$  with incoming boundary condition is just the Hankel function written in (\ref{Hankel_nh}):
\beq
y_{0}\,=\,
c_+\,\rho^{{1\over 2}}\,H_{{1\over 5-p}}^{(1)}\,
\Big({2\omega\over 5-p}\,\rho^{{p-5\over 2}}\Big)\,\,,
\eeq
where $c_+$ is a constant. Notice that $y_0$ is a source in the equation of $y_1$. Actually, defining the function $y$ as $y_1=c_+\,y$, we find the following inhomogeneous equation for 
$y(\rho)$:
\beq
 y''\,+\,{\omega^2\over \rho^{7-p}}\,y\,=\,-(7-p)\,{\omega\over \rho^{8-p}}\,
\rho^{{1\over 2}}\,H_{{1\over 5-p}}^{(1)}\,
\Big({2\omega\over 5-p}\,\rho^{{p-5\over 2}}\Big)\,\,.
\label{inhomogeneous_y_eq}
\eeq
In the appendix \ref{appendixD} we  find the solution of this equation by using the Wronskian method. This solution is simply:
\beq
y\,=\,-\rho^{{p-6\over 2}}\,H_{{6-p\over 5-p}}^{(1)}\,
\Big({2\omega\over 5-p}\,\rho^{{p-5\over 2}}\Big)\,\,,
\label{y_sol_Bfield}
\eeq
which also satisfies the incoming boundary condition at the horizon.  Let us  now write the solution for $y_{\pm}$. First we define the functions:
\beq
z_1(\rho)\,\equiv \,\rho^{{1\over 2}}\,H_{{1\over 5-p}}^{(1)}\,
\Big({2\omega\over 5-p}\,\rho^{{p-5\over 2}}\Big)\,\,,
\qquad\qquad
z_2(\rho)\,\equiv \,\rho^{{p-6\over 2}}\,H_{{6-p\over 5-p}}^{(1)}\,
\Big({2\omega\over 5-p}\,\rho^{{p-5\over 2}}\Big)\,\,.
\eeq
Then, $y_{\pm(\rho)}$ are given by:
\beq
y_{+}(\rho)\,=\,c_+\,z_1(\rho)\,-\,c_+\,B\,z_2(\rho)\,\,,
\qquad\qquad
y_{-}(\rho)\,=\,c_-\,z_1(\rho)\,+\,c_-\,B\,z_2(\rho)\,\,,
\eeq
where, to obtain $y_{-}$ we changed $B\to -B$ and $c_+\to c_-$. Let us now redefine these constants as follows
\beq
c_1\,=\,{i\omega\over 2}\,(c_++c_-)\,\,,
\qquad\qquad\qquad\qquad
c_2\,=\,-{i\omega\over 2}\,(c_+-c_-)\,\,.
\eeq
Then, $E$ and $a_y$ can be written in matrix form as:
\beq
\begin{pmatrix}
E\\ \\ a_y
\end{pmatrix}
\,=\,
\begin{pmatrix}
  z_1(\rho) &&& B\,z_2(\rho)\\
  {} & {} \\
  {iB\over \omega}\,z_2(\rho) &&& {i\over \omega}\,z_1(\rho)
 \end{pmatrix}\,
 \begin{pmatrix}
c_1\\ \\c_2
\end{pmatrix}\,\,.
\eeq
We now  study the near-horizon solution found in the small $\omega$ limit. At ${\mathcal O}(\omega^{{2\over 5-p}})$ we have (for $p<4$):
\beq
 z_1(\rho)\approx -{2^{\nu}\,\Gamma(\nu)\over \pi\,\alpha^{\nu}}\,i\,\big[\rho+c_p\,
 \omega^{{2\over 5-p}}\big]\,\,,
 \qquad\qquad
  z_2(\rho)\approx-{2^{\nu}\,\Gamma(\nu)\over \pi\, \alpha^{\nu}}\,i\,{1\over \omega}\,\,,
 \eeq
where $\nu$ and $\alpha$ are defined in (\ref{alpha_x_nu}) and $c_p$ is given in (\ref{cp}). Then, after redefining the constants $c_1$ and $c_2$ by absorbing the common factor in $z_1$ and $z_2$, we have:
\beq
\begin{pmatrix}
E\\ \\ a_y
\end{pmatrix}
\,\approx\,
\begin{pmatrix}
  \rho+c_p\,
 \omega^{{2\over 5-p}} &&& {B\over \omega}\\
  {} & {} \\
  {iB\over \omega^2} &&& {i\over \omega}\,\big(\rho+c_p\,
 \omega^{{2\over 5-p}} \big)
 \end{pmatrix}\,
 \begin{pmatrix}
c_1\\ \\c_2
\end{pmatrix}\,\,.
\label{E_ay_nh_low}
\eeq

We will now perform the two limiting operations in the opposite order. As in \cite{Brattan:2012nb}, in  the low-frequency limit we drop all terms not containing derivatives of $E$ and $a_y$, as well as all factors of $B$. Thus, the equation of motion of $E$ reduces to (\ref{eom_E_low_freq}), whose solution is (\ref{E_low_freq_sol}). The corresponding equation for $a_y$ in this limit becomes:
\beq
a_y''\,+\,\partial_{\rho}\,\log\Big(\rho^{\lambda}+d^2\Big)^{{1\over 2}}\,a_y'\approx 0\,\,.
\eeq
This equation can be integrated twice to give:
\beq
a_y\,=\,a_y^{(0)}\,-\,c_y\,{\cal J}_3(\rho)\,\,,
\label{ay_low_frequency}
\eeq
where $c_y$ is a constant and ${\cal J}_3(\rho)$ is the following integral:
\beq
{\cal J}_3(\rho)\,=\,\int_{\rho}^{\infty}\,
{d\bar\rho\over
(\bar\rho^{\lambda}+d^2)^{{1\over 2}}}\,=\,{2\over \lambda-2}\,\rho^{1-{\lambda\over 2}}\,
F\Big(\,{1\over 2}, {1\over 2}-{1\over\lambda}; {3\over 2}\,-\,{1\over\lambda};
- {d^2\over \rho^{\lambda}}\,\Big)\,\,.
\eeq
It is clear from these equations that $a_y^{(0)}$ is the value of $a_y$ at $\rho\to\infty$. Moreover:
\beq
a_y(\rho)\,=\,a_y^{(0)}\,-\,
{2c_y\over \lambda-2}\,\rho^{1-{\lambda\over 2}}\,
F\Big(\,{1\over 2}, {1\over 2}-{1\over\lambda}; {3\over 2}\,-\,{1\over\lambda};
- {d^2\over \rho^{\lambda}}\,\Big)\,\,.
\eeq
Let us now expand $a_y(\rho)$ in powers of $\rho$. First, one can check that, for small $\rho$, 
the integral ${\cal J}_3(\rho)$ can be approximated as:
\beq
{\cal J}_3(\rho)\,\approx\,{\mu\over d}\,-\,{\rho\over d}\,\,,
\eeq
where $\mu$ is the chemical potential (\ref{mu_gamma}). Therefore, for small $\rho$:
\beq
a_y(\rho)\,\approx\,a_y^{(0)}\,-\,c_y\,{\mu\over d}\,+\,c_y\,{\rho\over d}\,\,.
\label{ay_low_nh}
\eeq
The corresponding expansion of $E$ has been written in (\ref{E_low_nh}). We now match (\ref{E_low_nh}) and (\ref{ay_low_nh}) with (\ref{E_ay_nh_low}). By comparing the terms linear in $\rho$ we find the constants $c_1$ and $c_2$ in terms of $c_E$ and $c_y$, namely:
\beq
c_1\,=\,{\omega^2\over d}\,c_E\,\,,
\qquad\qquad
c_2\,=\,-i\,{\omega\over d}\,c_y\,\,,
\eeq
and the comparison of the terms independent of the radial variable leads to the following matrix equation:
\beq
\begin{pmatrix}
E^{(0)}\\ \\ a_y^{(0)}
\end{pmatrix}
\,=\,
\begin{pmatrix}
 {c_p\over d}\,
 \omega^{{2(6-p)\over 5-p}} \,-\,{2\mu\over \lambda d}\,
 \big(k^2-{\lambda\over 2}\,\omega^2\big)
 &&&-i {B\over d}\\
  {} & {} \\
  {iB\over d} &&& {c_p\over d}
 \omega^{{2\over 5-p}}\,+\,{\mu\over d}
 \end{pmatrix}\,
 \begin{pmatrix}
c_E\\ \\c_y
\end{pmatrix}\,\,.
\label{E0_ay0_B}
\eeq
We now require the vanishing of the sources $E^{(0)}$ and $a_y^{(0)}$, which only happens non-trivially if the matrix written above has a vanishing determinant or, equivalently,  when:
\beq
c_p^2\, \omega^{{2(7-p)\over 5-p}}\,+\,2\mu\,c_p\, \omega^{{2(6-p)\over 5-p}}\,-\,
{2\mu c_p\over \lambda}\, \omega^{{2\over 5-p}}\,k^2\,+\,\mu^2\,\omega^2\,-\,
{2\over \lambda}\,\mu^2\,k^2\,-\,B^2\,=\,0\,\,.
\label{poly_omega_k}
\eeq
Eq. (\ref{poly_omega_k})  is a polynomial relation in $\omega$ and $k$ whose solutions determine the dispersion relation of the zero-sound. Let us solve this equation for $\omega$.  At leading order we can neglect all the terms except the last three, which leads to the following gapped dispersion relation:
\beq
\omega=\pm \sqrt{{2\over \lambda}\,k^2\,+\,{B^2\over \mu^2}}\,\,.
\label{gapped_dispersion}
\eeq
In order to solve (\ref{poly_omega_k}) at the next-to-leading order, let us write:
\beq
\omega\,=\,\sqrt{{2\over \lambda}\,k^2\,+\,{B^2\over \mu^2}}\,+\,\delta\omega\,\,.
\label{delta_omega}
\eeq
By plugging (\ref{delta_omega}) into (\ref{poly_omega_k}),
we get at next-to-leading order
\beq
\delta\omega\,=\,-{c_p\over \mu}\,
\Bigg({2\over \lambda}\,k^2+{B^2\over \mu^2}\Bigg)^{{p-3\over 2(5-p)}}\,
\Big[{k^2\over \lambda}+{B^2\over \mu^2}\Big]\,\,.
\eeq
Let us now separate the real and imaginary parts in $\delta\omega$. From the expression of the constant  $c_p$ in (\ref{cp}) we obtain that the imaginary part at this order is given by:
\beq
{\rm Im}\,\omega\,=\,-{\pi\over \mu}\,
{(5-p)^{{3-p\over 5-p}}\over \Big[\Gamma\Big({1\over 5-p}\Big)\Big]^2}\,
\Bigg({2\over \lambda}\,k^2+{B^2\over \mu^2}\Bigg)^{{p-3\over 2(5-p)}}\,
\Big[{k^2\over \lambda}+{B^2\over \mu^2}\Big]\,\,.
\label{Im_omega_B}
\eeq
Moreover, we get the following correction to ${\rm Re}\, \omega$:
\beq
{\rm Re}\,\delta \omega\,=\,{\pi\over \mu}\,
{(5-p)^{{3-p\over 5-p}}\over \Big[\Gamma\Big({1\over 5-p}\Big)\Big]^2}\,
\cot\Big({\pi\over 5-p}\Big)\,
\Bigg({2\over \lambda}\,k^2+{B^2\over \mu^2}\Bigg)^{{p-3\over 2(5-p)}}\,
\Big[{k^2\over \lambda}+{B^2\over \mu^2}\Big]\,\,.
\label{Re_delta_omega_B}
\eeq
Notice that  (\ref{Im_omega_B}) and (\ref{Re_delta_omega_B}) coincide with (\ref{Im_omega}) and (\ref{Re_delta_omega}) when the $B$ vanishes, as it should.

\begin{figure}[ht]
\center
 \includegraphics[width=0.65\textwidth]{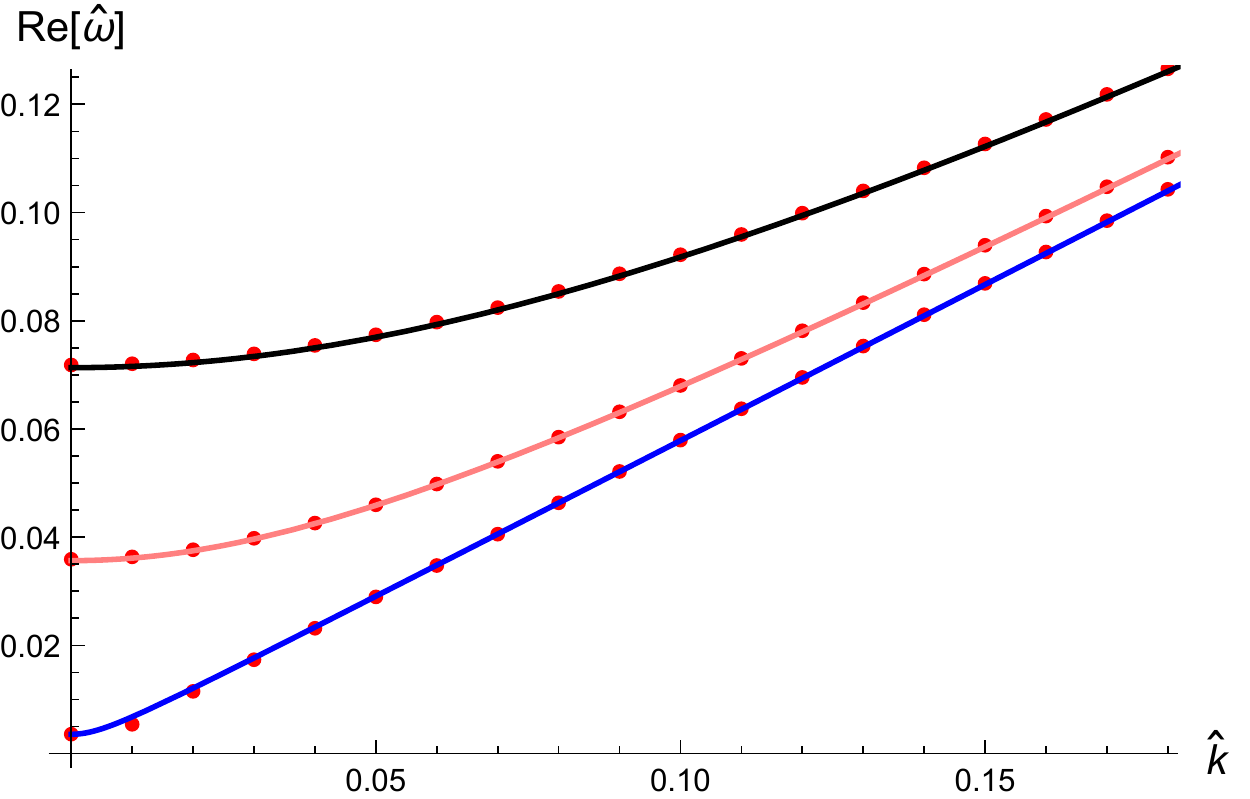}
 \caption{We present the real part of the dispersions in the D2-D6 above the critical value above which the zero sound mode has become massive ($\hat B\sim 0.2$) at various $\hat B=0.5,5,10$ (bottom-up) for $\hat d=10^6$. The points are the numerical values and the continuous curves are the analytic results. Notice that the lowest curve has a small dip at small $\hat k$ diverging from the analytic result. This is easily understood by recalling that the zero sound picked up the mass just slightly earlier.}
 \label{fig:zerosoundgap}
\end{figure}

\subsection{The crossover with $B$ field}\label{crossover}
Now let us finish this section with the following observation. Recall that in section \ref{crossover_zeroB} we were investigating the location where the transition from the collisionless regime to the hydrodynamic regime takes place. By studying the dispersion relations in the complex $\hat\omega$ plane by increasing the real momentum $\hat k$ we found that the diffusion mode met with another purely imaginary mode at some non-zero $(\hat\om_{cr},\hat k_{cr})$, where they merged together to become the pair of zero sound modes propagating in opposite directions. 

\begin{figure}[ht]
\center
 \includegraphics[width=0.45\textwidth]{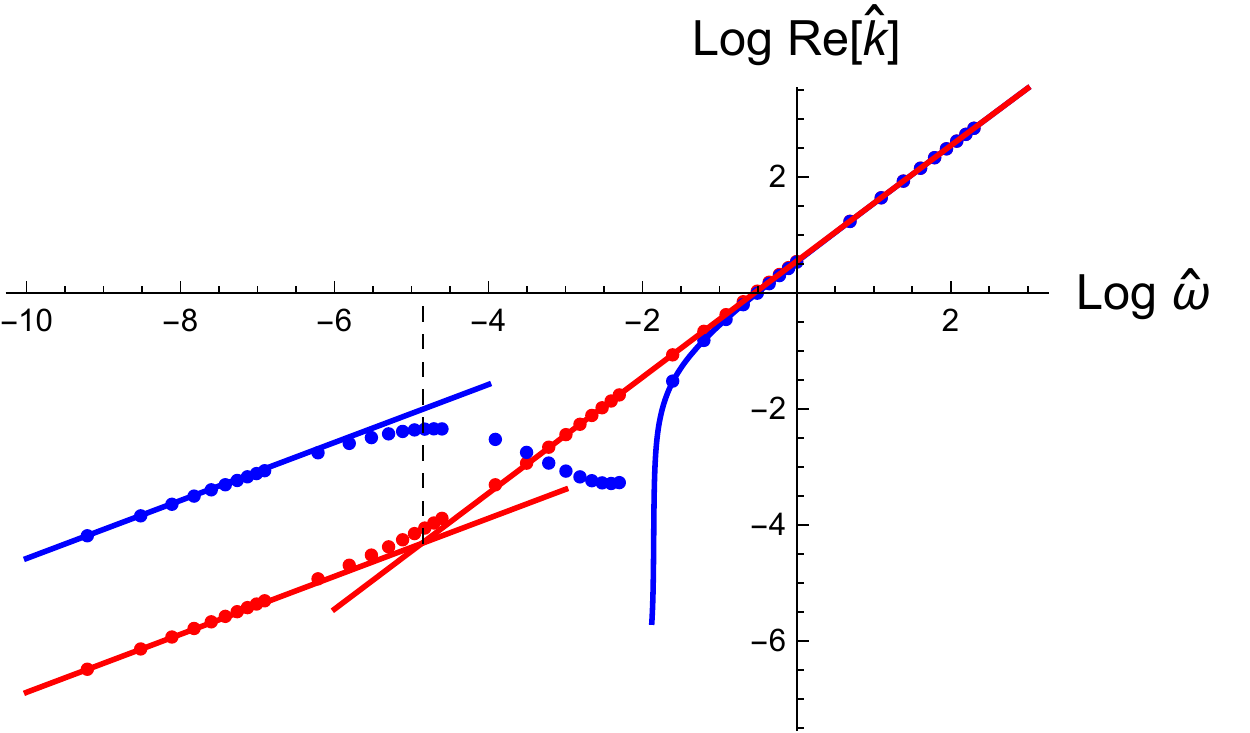}
\qquad
\includegraphics[width=0.45\textwidth]{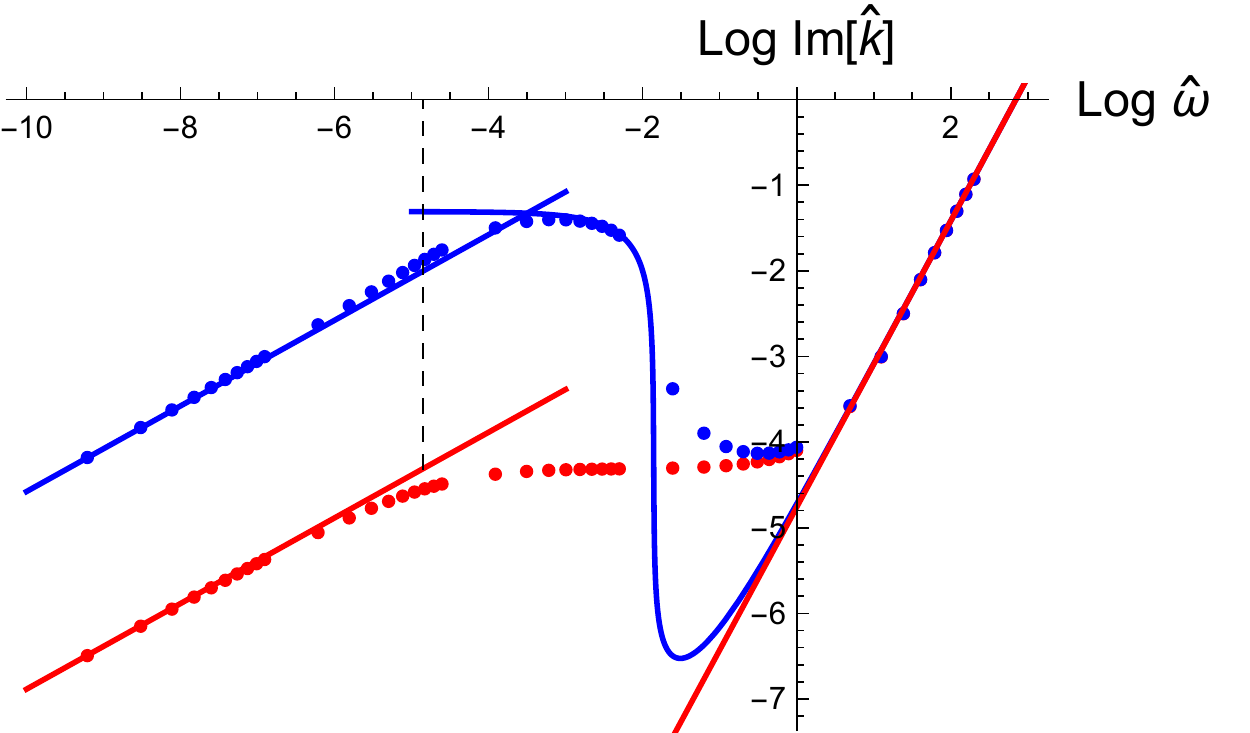}
 \caption{We are depicting the dispersions as complex momentum as a function of frequency for the D2-D6 model. On the left panel we plot the real part of $\hat k$ and on the right ${\rm Im}\,\hat k$. The red, lower, points are numerical data for $\hat B=0$ and the blue, upper, points are for $\hat B=10$. Both cases are for $\hat d=10^5$. The continuous lines starting from the left are analytic results for the diffusion mode (eq. (\ref{Dhat_nonzeroB})) and the continuous curves starting from the right are the analytic results (at $T=0$) for the zero sound (eq. (\ref{delta_omega})). The lower point of the vertical line segment is extracted from the collision of poles in the complex $\om$ plane, indicating the collisionless/hydrodynamic crossover location. We infer that the transition point is essentially insensitive to the inclusion of the magnetic field.}
 \label{fig:crossover_nonzeroB}
\end{figure}

While we now know from several studies that the zero sound generically picks up a mass with a non-zero magnetic field, it is interesting to ask if the location of the transition point is affected by the magnetic field, too. It turns out that the answer to this question, properly defined, is negative. By increasing the magnetic above a critical one, where the zero sound mode has become massive, the diffusive mode never meets with another purely imaginary mode (see Figs.~\ref{fig:dispersion_zeroB} \& \ref{fig:zerosound_nonzeroB} ) and it is unclear how to define the location for the crossover to happen. However, as suggested in \cite{Brattan:2012nb}, one can define it by making the momentum $\hat k$ complex and keeping $\hat\omega$ real. Indeed, we repeated a similar analysis to theirs (D3-D5 and D3-D7 cases) in the D2-D6 model and found out that the transition point is unaffected by the magnetic field strength. This suggests that this phenomenon is rather generic and may happen for all other intersections as well. We have depicted the dispersions in the D2-D6 case in Fig.~\ref{fig:crossover_nonzeroB}.

\section{Alternative quantization}
\label{anyons}

In this section we will not allow the intersection of the probe with the background branes be as general as before, but we will specialize to a $(2+1)$-dimensional intersection. Contrary to the scalar fields, for the gauge fields one can impose mixed Dirichlet-Neumann boundary conditions, {\emph{i.e.}}, adopt an alternative quantization, only when the bulk spacetime is four-dimensional \cite{Witten:2003ya,Yee:2004ju}. From the boundary perspective, a mixed boundary condition for the gauge field can be traced back to a $SL(2,{\mathbb{Z}})$ mapping from the usual Dirichlet condition, where this operation generically corresponds to adding a Chern-Simons term to the action and making the external vector field dynamical.\footnote{Incorporating $SL(2,{\mathbb{Z}})$ duality has sparked many recent studies in holography \cite{Goldstein:2010aw,Bayntun:2010nx,Fujita:2012fp,Jokela:2013hta,Brattan:2013wya,Jokela:2014wsa,Lippert:2014jma,Brattan:2014moa}.} We refrain from presenting a lengthy introduction to the details \cite{Jokela:2013hta} (see also \cite{Brattan:2013wya,Jokela:2014wsa,Brattan:2014moa}). Instead, we will walk the reader through the necessary notations along way as our goal is a simple generalization of the methods developed in \cite{Jokela:2013hta} to tackle all conformally $AdS_4$ backgrounds.

We wish to emphasize the key observation made in \cite{Jokela:2013hta} (allowing the use of alternative quantization) that the bulk action itself does not have to be invariant under $SL(2,{\mathbb{Z}})$ transformation: one only needs the bulk equations of motion for the gauge fields to reduce to free equations near the boundary so that one can choose mixed boundary conditions. 
In other words, since different quantizations only differ by boundary terms, the equations of motion (\ref{eom_E_Bnonzero}) and (\ref{eom_ay_general}) which we plan to solve are still the same. The difference comes in interpreting the different quantities: in general the quantum liquid transforms into an anyonic one at non-zero density and magnetic field. For precise mapping between the parameters, we refer the reader to \cite{Jokela:2013hta}.

Let us thus consider fluctuation modes satisfying the following mixed Dirichlet-Neumann boundary conditions:
\beq
\lim_{\rho\to\infty}\,\Big[\,\textfrak{n}\,\rho^{{\lambda\over 2}}\,f_{\rho\,\mu}\,-\,{1\over 2}\,
\epsilon_{\mu\alpha\beta}\,f^{\alpha\beta}\,\big]\,=\,0\,\,,
\label{bc_with_n}
\eeq
where $\textfrak{n}$ is some constant. 
The Dirichlet boundary conditions considered so far in the normal quantization correspond to $\textfrak{n}=0$. Equivalently, as in \cite{Brattan:2013wya},  we can write the previous condition as: 
\beq
\lim_{\rho\to\infty}\,\Big[\,\rho^{{\lambda\over 2}}\,f_{\rho\,\mu}\,-\,{\textfrak{m}\over 2}\,
\epsilon_{\mu\alpha\beta}\,f^{\alpha\beta}\,\big]\,=\,0\,\,.
\label{bc_with_m}
\eeq
Clearly $\textfrak{m}=1/\textfrak{n}$ and the normal quantization corresponds to $\textfrak{m}=\infty$.  For $\mu=t,x$ the condition (\ref{bc_with_n}) leads to:
\beq
\lim_{\rho\to\infty}\,\big[\,\textfrak{n}\,\rho^{{\lambda\over 2}}\,a_t'\,-\,i\,k\,
a_y\,\big]\,=\,0\,\,,
\qquad\qquad
\lim_{\rho\to\infty}\,\big[\,\textfrak{n}\,\rho^{{\lambda\over 2}}\,a_x'\,+\,i\,\omega\,a_y\,\big]\,=\,0\,\,,
\label{bc_t_x}
\eeq
while for $\mu=y$ it becomes:
\beq
\lim_{\rho\to\infty}\,\big[\,\textfrak{n}\,\rho^{{\lambda\over 2}}\,a_y'\,-\,i\,E\big]\,=\,0\,\,.
\label{bc_y}
\eeq
Let us rewrite (\ref{bc_y}) as:
\beq
\lim_{\rho\to\infty}\,E\,=\,-i\,\textfrak{n}\,\lim_{\rho\to\infty}\,\big[\,\rho^{{\lambda\over 2}}\,a_y'\,\big]\,\,.
\label{lim_E}
\eeq
Moreover, by using the transversality condition  (\ref{at_ax_E}) at $\rho\to\infty$, where $u\to 1$,  the two equations in (\ref{bc_t_x}) reduce to:
\beq
\lim_{\rho\to\infty}\,a_y\,=\,
i\,{\textfrak{n}\over \omega^2-k^2}\,\lim_{\rho\to\infty}\,\big[\,\rho^{{\lambda\over 2}}\,E'\,\big]\,\,.
\label{lim_ay}
\eeq
As stated above, it is clear from (\ref{lim_E})  and (\ref{lim_ay}) that $\textfrak{n}=0$ corresponds to the Dirichlet boundary condition in which $E$ and $a_y$ are fixed at the boundary.

Recall that the horizon radius $r_h$ can be eliminated from the fluctuation equations 
(\ref{eom_E_Bzero}) and (\ref{ay_eq_Bzero}) by rescaling the radial variable $\rho$ as $\hat\rho=\rho/r_h$ and 
rescaling $d$, $\omega$ and $k$ as in (\ref{hat_d_def}) and (\ref{hat_omega_k}). 
We will rescale the gauge potentials as $\hat a_\mu=a_\mu/r_h$, which for the electric field $E$ means the following:
\be
 \hat E = \frac{E}{r_h^{\frac{7-p}{2}}} \ .
\ee
It can be easily checked that we can eliminate $r_h$ from the boundary conditions (\ref{lim_E}) and (\ref{lim_ay}) by rescaling $\textfrak{n}$ as:
\beq
\hat {\textfrak{n}}\,=\,{\textfrak{n}\over r_h^{{7-p-\lambda\over 2}}}\,\,.
\label{hat_n}
\eeq
Equivalently, $\textfrak{m}$ must be rescaled as:
\beq
\hat {\textfrak{m}}\,=\,{\textfrak{m}\over r_h^{{p+\lambda-7\over 2}}}\,\,.
\label{hat_m}
\eeq
Let us now analyze the new boundary conditions at low frequency and momentum. The expressions of $E$ and $a_y$ in this regime have been written in (\ref{E_low_frequency}) and (\ref{ay_low_frequency}). The  UV values at the boundary of these two fields are:
\beq
\lim_{\rho\to\infty}\,E(\rho)\,=\,E^{(0)}\,\,,
\qquad\qquad
\lim_{\rho\to\infty}\,a_y(\rho)\,=\,a_y^{(0)}\,\,.
\eeq
The radial derivatives of $E$ and $a_y$ at the boundary  can be easily obtained from (\ref{Eprime_boundary}) and (\ref{ay_low_frequency}):
\beq
{\partial E\over \partial\rho}\,\Big|_{\rho\to\infty}\approx (\omega^2-k^2)\,c_E\,\rho^{-{\lambda\over 2}}\,\,,
\qquad\qquad
{\partial a_y\over \partial\rho}\,\Big|_{\rho\to\infty}\approx c_y\,
\rho^{-{\lambda\over 2}}\,\,.
\eeq
From these expressions we can recast the boundary conditions for the alternative quantization as a relation between the constants $E^{(0)}$, $a_y^{(0)}$, $c_E$, and 
$c_y$. Indeed, let us define  $E_{\textfrak{n}}^{(0)}$ and $a_{y,\textfrak{n}}^{(0)}$ as:
\beq
E_{\textfrak{n}}^{(0)}\,\equiv E^{(0)}\,+\,i\,\textfrak{n}\, c_y\,\,,
\qquad\qquad
a_{y,\textfrak{n}}^{(0)}\,=\,a_y^{(0)}\,-\,i\,\textfrak{n}\,c_E\,\,.
\eeq
Then, (\ref{lim_E}) and (\ref{lim_ay}) are equivalent to the conditions:
\beq
E_{\textfrak{n}}^{(0)}\,=\,a_{y,\textfrak{n}}^{(0)}\,=\,0\,\,.
\label{En0_ay_n0_def}
\eeq

\subsection{Zero sound}
From the results of section \ref{zero_sound_B} it is straightforward to relate $E_{\textfrak{n}}^{(0)}$ and $a_{y,\textfrak{n}}^{(0)}$ to the constants $c_E$ and $c_y$ for incoming boundary conditions at the horizon at low $\omega$ and $k$. Indeed, from (\ref{E0_ay0_B}) we get:
\beq
\begin{pmatrix}
E_{\textfrak{n}}^{(0)}\\ \\ a_{y,\textfrak{n}}^{(0)}
\end{pmatrix}
\,=\,
\begin{pmatrix}
 {c_p\over d}\,
 \omega^{{2(6-p)\over 5-p}} \,-\,{2\mu\over \lambda d}\,
 \big(k^2-{\lambda\over 2}\,\omega^2\big)
 &&&-i {B\over d}+i\textfrak{n}\\
  {} & {} \\
  {iB\over d}-i\textfrak{n} &&& {c_p\over d}
 \omega^{{2\over 5-p}}\,+\,{\mu\over d}
 \end{pmatrix}\,
 \begin{pmatrix}
c_E\\ \\c_y
\end{pmatrix}\,\,.
\label{En0_ayn0_B}
\eeq
The alternative quantization conditions  (\ref{En0_ay_n0_def}) can be implemented by imposing the vanishing of the determinant of the matrix on the right-hand side of (\ref{En0_ayn0_B}). The corresponding equation is just (\ref{poly_omega_k}) with the substitution $B\to B-\textfrak{n}d$. Solving this equation we get the dispersion relation. At leading order the real part of $\omega$ is just:
\beq
\omega=\pm \sqrt{{2\over \lambda}\,k^2\,+\,{(B-\textfrak{n}d)^2\over \mu^2}}\,\,.
\label{disp_rel_zero_sound}
\eeq
\begin{figure}[ht]
\center
 \includegraphics[width=0.75\textwidth]{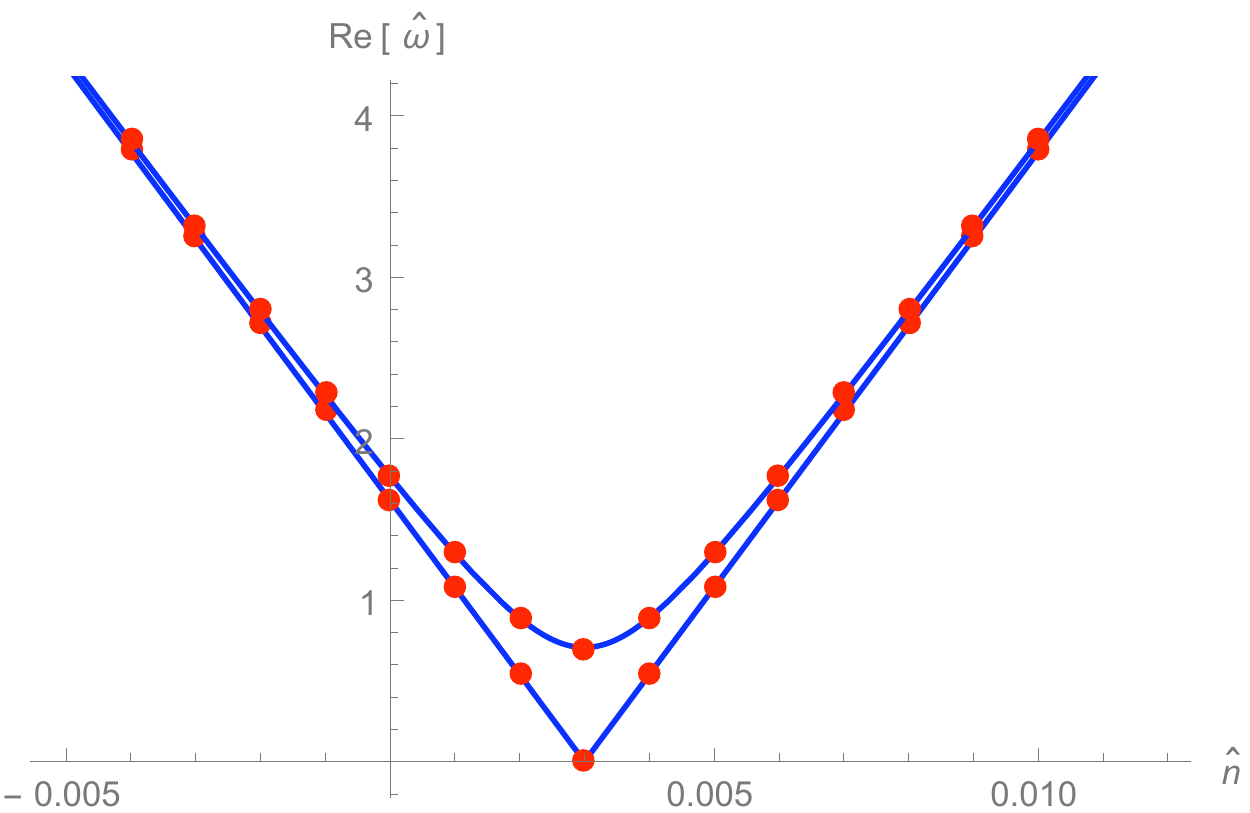}
 \caption{We depict the (positive) real part of the lowest excitation mode in the D$3$-D$5$ model with $\hat d=10^6$ and $\hat B = 3\cdot 10^3$ as a function of $\hat {\textfrak{n}}$. The dots are numerically computed data points whereas the continuous curves stand for the analytic result of (\ref{disp_rel_zero_sound}) for $\hat k=0$ (lower curve) and $\hat k=1$ (upper curve). Notice that the mass gap for the zero sound vanishes precisely when $\hat {\textfrak{n}} = \frac{\hat B}{\hat d}$ for zero momentum.}
 \label{fig:anyon_excitation_spectrum}
\end{figure}
Similarly, we could get the  higher order terms from (\ref{Im_omega_B}) and (\ref{Re_delta_omega_B}).
Notice that the spectrum can be made gapless by adjusting the alternative quantization parameter to cancel the gap induced by the magnetic field. See Fig.~\ref{fig:anyon_excitation_spectrum} for the comparison between (\ref{disp_rel_zero_sound}) with the numerics. This phenomenon was overlooked in the recent paper \cite{Brattan:2014moa}, since the focus was more on the physics about the S-dual $(\mathfrak{m}=0)$ point. 

The closing of the gap was expected since implementing the alternative quantization we allow the gauge field to be dynamical. In particular, this means that we let the external magnetic field to adjust its vacuum expectation value. The particular case where the gap closes corresponds to an anyonic fluid for which the effective magnetic field is vanishing and respectively the zero sound mode becomes gapless. Similar occurrence happened in the D3-D7' model in the incompressible phase, where this mechanism was used to obtaining an anyonic superfluid, with the soft mode in the neutral sector becoming massless precisely when the effective magnetic field vanished \cite{Jokela:2013hta,Jokela:2014wsa}.

\subsection{Conductivities}
\begin{figure}[!ht]
\center
 \includegraphics[width=0.45\textwidth]{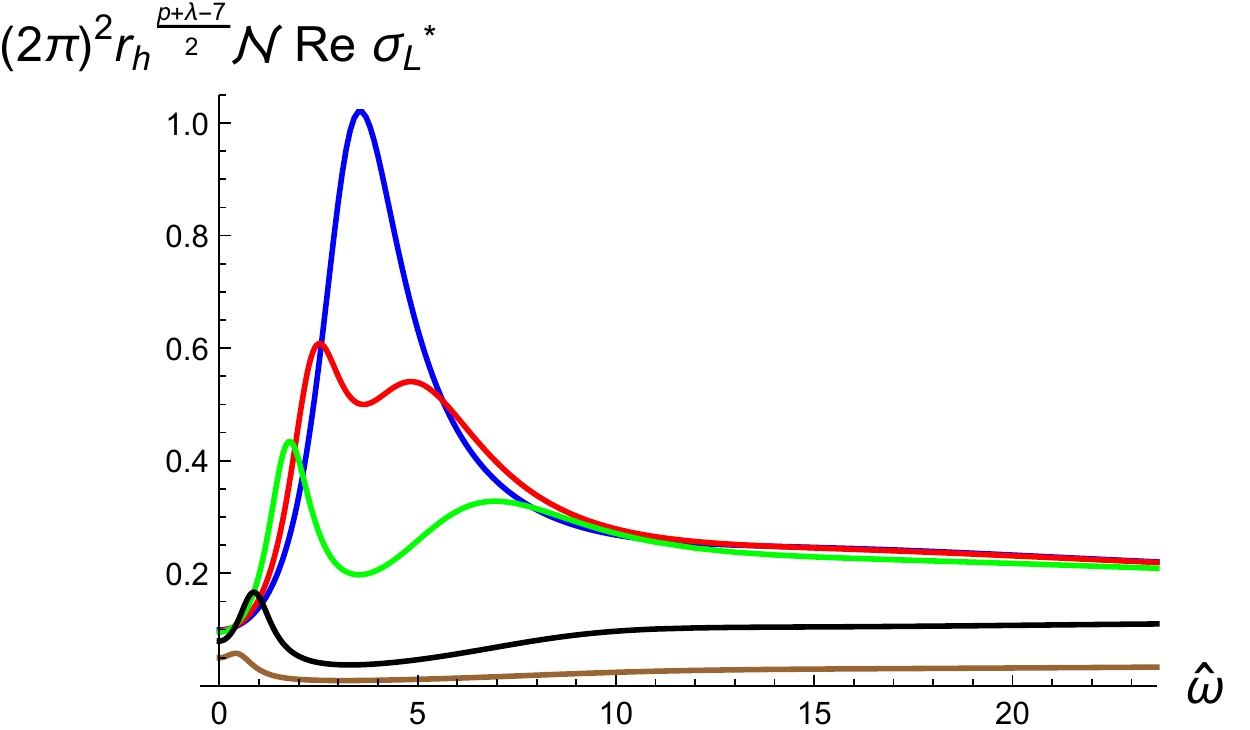}
 \includegraphics[width=0.45\textwidth]{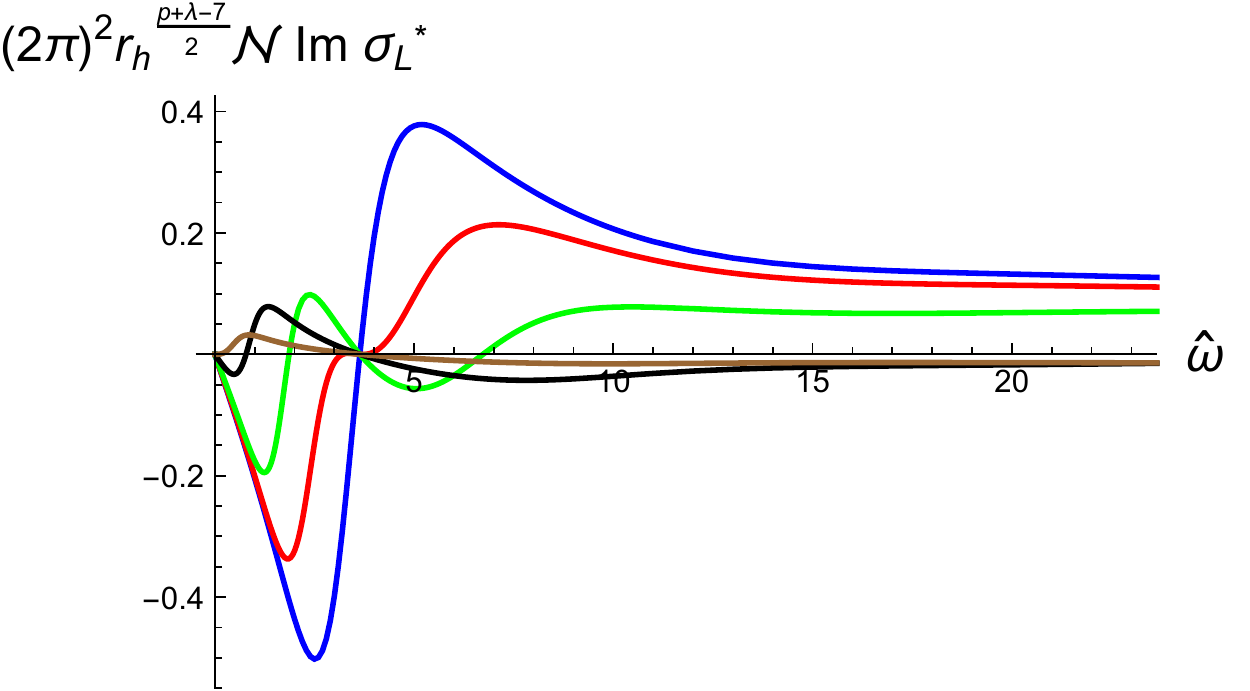}
 \caption{We are plotting the optical conductivities versus $\hat\omega$ for the D2-D6 model at fixed $\hat d=10$ and $\hat B=0$. In the left panel we are plotting the real part of $(2\pi)^2r_h^{{p+\lambda-7\over 2}}{\cal N}\sigma_L^*$ and in the right panel the imaginary part. The different curves correspond to different quantization schemes $\mathfrak m = 0,1,2,5,10$. The bigger amplitudes correspond to smaller $\mathfrak m$'s.}
 \label{fig:AC_with_m}
\end{figure}

Let us obtain the conductivity of the anyonic fluid following the approach of \cite{Brattan:2013wya}. As shown in that paper, the current-current correlator can be parametrized in terms of three functions $C_L$, $C_T$,  and $W$, which transform under S-duality as:
\bear
&&C_L^{*}\,=\,{C_L\over (2\pi)^2\,\big[C_L\,C_T\,+W^2\big]}\,\,,
\qquad\qquad
C_T^{*}\,=\,{C_T\over (2\pi)^2\,\big[C_L\,C_T\,+W^2\big]}\,\,,\rc\rc
&&\qquad\qquad\qquad\qquad
W^{*}\,=\,-{W\over (2\pi)^2\,\big[C_L\,C_T\,+W^2\big]} \ .
\label{CL_CT_W_Sduality}
\eear
Under the $T$ transformation only the parity violation function $W$ transforms as:
\beq
W^{*}\,=\,W+{1\over 2\pi}\,\,.
\label{W_Tduality}
\eeq
The current-current correlator at zero momentum is parametrized in terms $C_L(\omega,0)=C_T(\omega,0)$ and $W(\omega,0)$ as:
\beq
\langle J_i(\omega, 0)\, J_j(\omega, 0)\rangle\,=\,i\omega\,
\big[C_L(\omega,0)\,\delta_{ij}\,+\,W(\omega,0)\,\epsilon_{ij}\big]\,\,,
\eeq
where $i,j$ are spatial indices. The AC conductivities are defined as:
\beq
\sigma_{ij}(\omega)\,=\,
{1\over i\omega}\,\langle J_i(\omega, 0)\, J_j(\omega, 0)\rangle\,\,,
\eeq
or, equivalently
\beq
\sigma_L(\omega)\,=\,C_L(\omega,0)\,\,,
\qquad\qquad
\sigma_H(\omega)\,=\,W(\omega,0)\,\,.
\eeq
In our case we can obtain the value of $\sigma_L$  at $\omega=0$ (the DC conductivity) by looking at the transverse correlator (\ref{transv_corr_Gammas}) at zero momentum. We get that $\sigma_H=0$ and that
\beq
\sigma_L\,=\,{\cal N}\,\Gamma_{\omega} \ \ \ \ ,\ (\om\to 0) \ ,
\eeq
which, after taking (\ref{Gamma_omega_k}) into account, becomes:
\beq
\sigma_L\,=\,r_h^{{p+\lambda-7}\over 2}\,\,{\cal N}\,
\sqrt{1+\hat d^2} \ \ \ \ ,\ (\om\to 0) \ .
\label{sigmaL_nornal_quant}
\eeq
Starting with a theory with $W=0$ and performing the transformation $T^K$ we end up with a theory with a Hall conductivity $\sigma_H=K/(2\pi)$. With a subsequent $S$ transformation the corresponding conductivities can be found from (\ref{CL_CT_W_Sduality}) with 
$C_T(\omega,0)=C_L(\omega,0)=\sigma_L(\omega)$:
\beq
\sigma_L^{*}(\omega)\,=\,{1\over( 2\pi)^2}\,\,{\sigma_L(\omega)\over \sigma_L^2(\omega)\,+\,\Big({K\over 2\pi}\Big)^2}\,\,,
\qquad\qquad
\sigma_H^{*}(\omega)\,=\,-{1\over( 2\pi)^2}\,\,{\sigma_H(\omega)\over \sigma_L^2(\omega)\,+\,\Big({K\over 2\pi}\Big)^2}\,\,.\label{eq:ACcond}
\eeq
For the AC conductivities there are no analytic results available in any model. Had there been for example results for the normal quantization, one could just transfer them to any quantization using the formulas (\ref{eq:ACcond}). Given this shortcoming, we do not wish to make a thorough scan over all the available parameters in the model, but are elated to highlight just one interesting effect upon changing $\mathfrak m$. By numerically extracting the AC conductivities, as in \cite{Brattan:2013wya}, we find that increasing $\mathfrak m$, the Drude-like peak will begin to transform to make a secondary peak at higher $\hat\omega$, similar to what is found in heavy fermion systems; see Fig.~\ref{fig:AC_with_m} for how the longitudinal conductivities vary.

The DC conductivities can be obtained by taking $\omega=0$ in the previous expression, as well as the value (\ref{sigmaL_nornal_quant}) for the DC conductivity $\sigma_L$ of the normally quantized system:
\beq
( 2\pi)^2\,r_h^{{p+\lambda-7\over 2}}\,{\cal N}\,\sigma_L^{*}\,=\,
{\sqrt{1+\hat d^2}\over 1+\hat d^2+\hat {\textfrak{m}}^2}\,\,,
\qquad\qquad
( 2\pi)^2\,r_h^{{p+\lambda-7\over 2}}\,{\cal N}\,\sigma_H^{*}\,=\,-
{\hat {\textfrak{m}}\over 1+\hat d^2+\hat {\textfrak{m}}^2}\,\,,
\eeq
where, as in \cite{Brattan:2013wya}, we identify $\textfrak{m}$ as:
\beq
\textfrak{m}={K\over 2\pi {\cal N}}\,\,,
\label{m_K}
\eeq
and $\hat {\textfrak{m}}$ is related to $\textfrak{m}$ as in (\ref{hat_m}).

\subsection{Diffusion constant}  
Let us consider the current-current correlator with non-zero momentum $k$ following the approach of \cite{Brattan:2013wya}. In this case we have to distinguish between longitudinal and transverse functions $C_L$ and $C_T$. 
The longitudinal function $C_L$ and its S-dual $C_L^{*}$ can be parametrized in terms of the DC conductivities $\sigma_L$ and $\sigma_L^{*}$ and diffusion constants $\hat D$ and $\hat D^{*}$ as:
\beq
C_L\,=\,{\hat k\,\sigma_L\
\over i\hat\omega\,-\,\hat D\,\hat k^2}\,\,,
\qquad\qquad
C_L^{*}\,=\,{\hat k\,\sigma_L^{*}\
\over i\hat\omega\,-\,\hat D^{*}\,\hat k^2} \ .
\label{CL_sigmaL_D}
\eeq
Notice that $C_L$ and $C_L^*$ are related as in (\ref{CL_CT_W_Sduality}), which we now write as:
\beq
{1\over (2\pi)^2\, C_L^*}\,=\, C_T\,+\,{W^2\over C_L} \ .
\label{C_Ls_CT}
\eeq
Moreover, the DC conductivities $\sigma_L$ and $\sigma_L^{*}$  are related as:
\beq
{1\over (2\pi)^2\,\sigma_L^*}\,=\,\sigma_L\,+\,{W^2\over \sigma_L} \ .
\label{S_dual_sigmaL}
\eeq
Let us now obtain the transverse correlation function  $C_T$ in terms of $C_L$ and $C_L^{*}$ from (\ref{C_Ls_CT}). By using  the parametrization (\ref{CL_sigmaL_D}) and the S-dual relation (\ref{S_dual_sigmaL}) we get that:
\beq
C_T\,=\,{\sigma_L\over \hat k}\,\big(i\omega\,-\,H\,\hat k^2\big)\,\,,
\label{CT_H}
\eeq
where $H$ is given by:
\beq
H\,=\,\hat D^{*}\,+\,(\hat D^{*}\,-\,\hat D)\,{W^2\over \sigma_L^2}\,\,.
\label{H_Ds}
\eeq
Inverting this last relation we get $\hat D^{*}$ in terms of $H$ and $\hat D$:
\beq
\hat D^{*}\,=\,{H\,+\,{W^2\over \sigma_L^2}\,\hat D\over 
1+{W^2\over \sigma_L^2}}\,\,.
\label{hatD_H_W}
\eeq
If  parity is conserved in the original theory and, thus,  $W=0$, then $H=\hat D^*$ and we can write $C_T$ as:
\beq
C_T\,=\,{\sigma_L\over \hat k}\,\big(i\omega\,-\,\hat D^{*}\,\hat k^2\big)\,\,.
\label{CT_sigmaL_hatD*}
\eeq
We will use this relation to obtain $\hat D^{*}$ from the transverse correlator calculated in section \ref{Trasnverse_correlator}. Indeed, when $W=0$ the non-vanishing different components  of $\langle J_{\mu}\,J_{\nu}\rangle$ when $k^{\mu}=(\omega, k,0)$ are related to $C_L$ and $C_T$ as:
\bear
&&\langle J_{t}(k)\,J_{t}(-k)\rangle=\,-{k^2\over \sqrt{-\omega^2+k^2}}\,C_L(k)\,\,,
\quad
\langle J_{x}(k)\,J_{x}(-k)\rangle=\,-{\omega^2\over \sqrt{-\omega^2+k^2}}\,C_L(k)
\,\,,\rc\rc
&&\langle J_{x}(k)\,J_{y}(-k)\rangle=-{\omega\, k\over \sqrt{-\omega^2+k^2}}\,C_L(k)\,\,,
\quad
\langle J_{x}(k)\,J_{y}(-k)\rangle=\sqrt{-\omega^2+k^2}\,C_T(k)\,\,\,.\rc\rc
\eear
In the hydrodynamic regime in which $k\sim {\mathcal O}(\epsilon)$ and $\omega\sim {\mathcal O}(\epsilon^2)$, at leading order, we get:
\beq
\langle J_{x}(k)\,J_{y}(-k)\rangle\approx k\,C_T \ .
\eeq
Therefore, the expression of $C_T$ can be found from the correlator  (\ref{transv_corr_Gammas}).  By comparing this equation with (\ref{CT_sigmaL_hatD*}) and using (\ref{sigmaL_nornal_quant}), we find
\beq
\hat D^{*}\,=-{\hat \Gamma_k\over \sqrt{1+\hat d^2}}\,=\,
{2\,\over 2(6-p)-\lambda}\,\,{1\over\sqrt{1+\hat d^2}} \,
F\Big({1\over 2}, {6-p\over \lambda}-{1\over 2}; {6-p\over \lambda}+{1\over 2};
-\hat d^2\Big)\,\,.
\label{hatD*_expression}
\eeq
\begin{figure}[ht]
\center
 \includegraphics[width=0.75\textwidth]{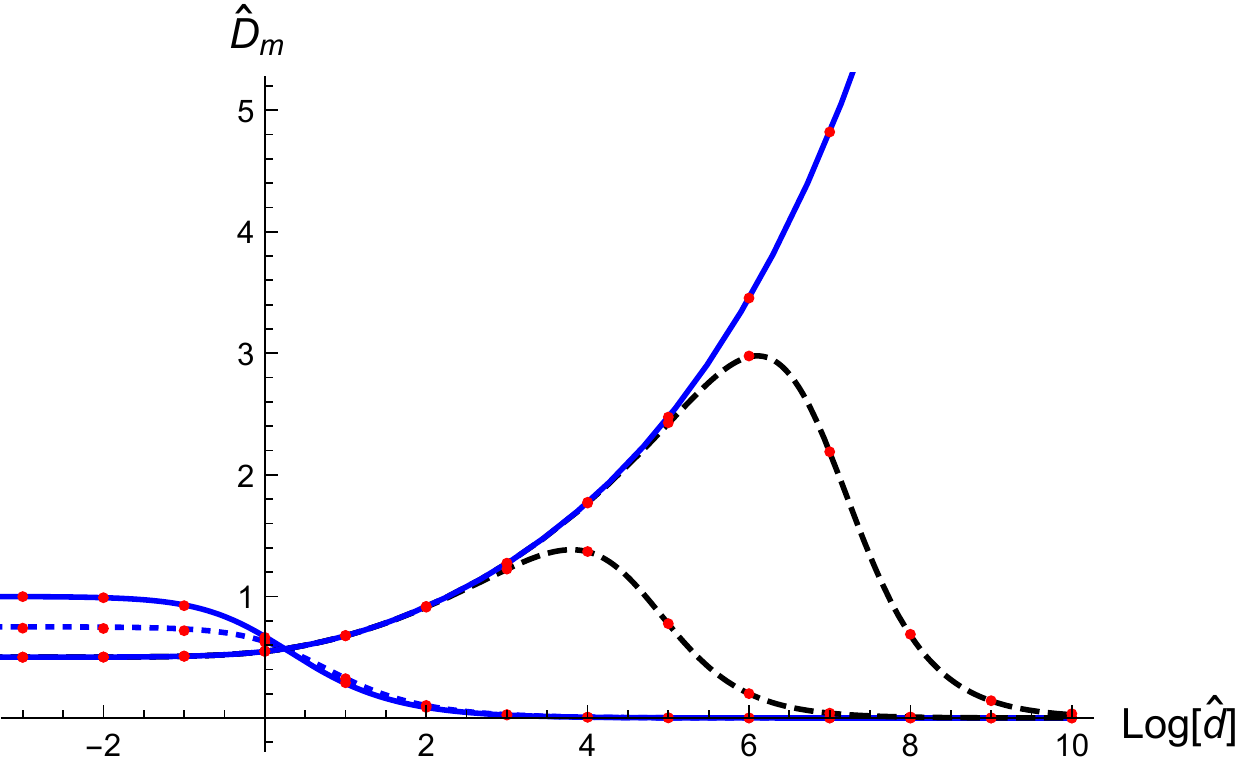}
 \caption{The diffusion constants in the D$2$-D$6$ model with different quantization conditions. The red points represent numerical data. The different curves are the analytic results. The monotonically increasing solid blue curve corresponds to normal quantization, {\emph{i.e.}}, $\hat {\textfrak{n}}=0$ (\ref{hat_D_value}), the others to alternative quantization (\ref{hat_Dm}); the middle two dashed black curves correspond to $\hat {\textfrak{m}}=1000,100$, the dotted blue curve (visible as the middle on the left) to $\hat {\textfrak{m}}=1$, and the monotonically decreasing solid blue curve to $\hat {\textfrak{m}}=0$. Notice that all $\hat D_{\textfrak{m}}$ (iff $\hat {\textfrak{m}}<\infty$) asymptote to S-dual result (\ref{hatD*_expression}) as $\hat d\to\infty$ and at large temperatures they asymptote to constants as given in (\ref{hatDm_hatd_zero}).}
 \label{Diffusion_with_m}
\end{figure}

We can now use this value of $\hat D^{*}$ to  apply the argument of \cite{Brattan:2013wya} and find the diffusion constant after a $S\,T^{K}$ transformation, which is given by:
\beq
\hat D_{{\textfrak{m}}}\,=\,{\hat D^{*}\,+\,\Big({K\over 2\pi\sigma_L}\Big)^2\,\hat D\over
1+\Big({K\over 2\pi\sigma_L}\Big)^2}\,\,,
\label{Dm_D_D*}
\eeq
which, with the identification (\ref{m_K}) and the expression of $\sigma_L$ written in (\ref{sigmaL_nornal_quant}), becomes:
\beq
\hat D_{\textfrak{m}}\,=\,{(1+\hat d^2)\,\hat D^{*}\,+\,\hat {\textfrak{m}}^2\,\hat D\over
1+\hat d^2+\hat {\textfrak{m}}^2}\,\,.
\eeq
By using the values of $\hat D$ (eq. (\ref{hat_D_value})) and $\hat D^{*}$ (eq. (\ref{hatD*_expression})) we get:
\bear
&&\hat D_{\textfrak{m}}\,=\,{2\,\sqrt{1+\hat d^2}\over 1+\hat d^2+\hat {\textfrak{m}}^2}\,\,
\Bigg[{1\,\over 2(6-p)-\lambda}\,\,{1\over\sqrt{1+\hat d^2}} \,
F\Big({1\over 2}, {6-p\over \lambda}-{1\over 2}; {6-p\over \lambda}+{1\over 2};
-\hat d^2\Big)\,+\,\rc\rc
&&\qquad\qquad\qquad\qquad\qquad\qquad
+\,{\hat {\textfrak{m}}^2\over \lambda-2}\,
F\Big(\,{3\over 2}, {1\over 2}-{1\over\lambda}; {3\over 2}\,-\,{1\over\lambda};
-\hat d^2\,\Big)\,\Bigg]\,\,.
\label{hat_Dm}
\eear
From this formula we can get the limiting value of $\hat D_{\textfrak{m}}$ when $\hat d\to 0$:
\beq
\lim_{\hat d\to 0}\,\hat D_{\textfrak{m}}\,=\,{2\over \lambda-2}\,+\,{\lambda-7+p\over (\lambda-2)(12-2p-\lambda)}\,\,.
\label{hatDm_hatd_zero}
\eeq
The numerical values of $\hat D_{\textfrak{m}}$ match extremely well with the analytic results (\ref{hat_Dm}) as represented in Fig.~\ref{Diffusion_with_m} for the D2-D6 model.


\section{Discussion: scaling behavior}\label{sec:discussions}
The energy scale of the bulk theory is given, in terms of the radial coordinate $r$, as
\beq
{\cal E}\sim r^{{5-p\over 2}} \ .
\label{UV/IR/bulk}
\eeq
The coefficient  in (\ref{UV/IR/bulk}) contains $\sqrt{g_{YM}}$, where $g_{YM}$ is the  Yang-Mills coupling of the bulk theory, which is dimensionful when $p\not=3$.  This distance/energy relation was found long time ago in \cite{Peet:1998wn}. It can also be derived by noticing that the D$p$-brane background is conformally AdS. This means that one can perform a Weyl transformation to a dual frame:
\beq
ds^2_{dual}\,=\,e^{2\phi\over p-7}\,ds^2\,\,,
\eeq
followed by a change in the radial coordinate of the form $u\sim r^{{5-p\over 2}}$, after which the metric is that of $AdS_{p+2}\times {\mathbb S}^{8-p}$.  The relation (\ref{UV/IR/bulk}) follows by identifying the new radial coordinate $u$ with the energy scale. We can also extract  (\ref{UV/IR/bulk})  by looking at the relation between the horizon radius $r_h$ and $T$ (\ref{T_rh}).

Let us  now consider rescalings of ${\cal E}$ of the form:
\beq
{\cal E}\to \Lambda\,{\cal E}\,\,.
\eeq
In terms of $\rho$ these rescalings are equivalent to:
\beq
\rho\to \Lambda^{\Delta_{\rho}}\,\rho\,\,,
\qquad\qquad
\Delta_{\rho}={2\over 5-p}\,\,.
\eeq
We will say that $\rho$ transforms with scaling dimension $\Delta_{\rho}$.

Let us obtain the behavior of the density $d$ and magnetic field $B$ by imposing that the three terms in the combination $d^2+\rho^{\lambda}+\rho^{\lambda+p-7}\,B^2$, appearing in the equation of motion for $A_t$ (and also for the fluctuation fields), scale in the same way.  We write
\beq
d\to\Lambda^{\Delta_d}\,d\,\,,
\qquad\qquad
B\to \Lambda^{\Delta_B}\,B\,\,.
\eeq
We find the following values for the scaling dimensions $\Delta_d$ and $\Delta_B$:
\beq
\Delta_d\,=\,{\lambda\over 5-p}\,\,,
\qquad\qquad
\Delta_B\,=\,{7-p\over 5-p}\,\,.
\label{Delta_d_B}
\eeq
Therefore, $\lambda$ determines the scaling dimension $\Delta_d$  of the charge density,\footnote{This was expected as we related $\lambda$ to the polytropic index of the equation of state, see (\ref{eq:polytrope}).} while $\Delta_B$ depends only on $p$. In the conformal case $p=3$ we have $\lambda=2n$ and, thus, $\Delta_d=n$ and $\Delta_B=2$. These values are just the canonical dimensions of these fields in a $(n+1)$-dimensional QFT.  In a non-conformal background $\Delta_d$ and $\Delta_B$ differ from the canonical dimensions. For example,  for the supersymmetric intersections of Table~\ref{table:nd4}, we get:
\beq
\Delta_d^{SUSY}\,=\,{q-p+2\over 5-p}\,=\,{2\over 5-p}\,\big(n+3-p\big)\,\,.
\eeq
Curiously, when $p\not =3$, $\Delta_d^{SUSY}=n$ only for $(2+1)$-dimensional intersections (\ie, with $n=2$ for any $p\not=3$, corresponding to D2-D6 and D4-D4 in Table~\ref{table:nd4}). 

The scaling dimensions of the different quantities determine the way in which they must be rescaled with the temperature to make them temperature independent. Indeed, given a quantity $O$, its rescaled quantity $\hat O$ is defined in such a way that $\hat O$ is invariant under rescalings. As $\Delta_T=1$, it is clear that the relation between $O$ and $\hat O$ must be of the type: 
\beq
\hat O\sim {O\over T^{\Delta_O}}\,\,,
\label{hat_O}
\eeq
with $\Delta_O$ being the scaling dimension of $O$. Notice that this agrees with our definition of 
$\hat \omega$ and $\hat k$ in (\ref{hat_omega_k}) (since $\Delta_{\omega}=\Delta_{k}=1$) and also with the definition of $\hat d$ in (\ref{hat_d_def}). Recall the expression of $\hat B$ written in (\ref{hat_B_def}):
\beq
\hat B\,=\,{B\over r_{h}^{{7-p\over 2}}}\,=\,
\Big({7-p\over 4\pi}\Big)^{{7-p\over 5-p}}\,\,
{B\over T^{{7-p\over 5-p}}} \ .
\eeq
A posteriori, we notice that the rescaling in (\ref{hat_B_def}) was chosen to agree with the general definition (\ref{hat_O}) when $\Delta_B$ is given by (\ref{Delta_d_B}). 

The behavior under rescalings can also be used to determine the general formula of the invariant quantities normalized by the density. These quantities will be denoted by a tilde and are generally given by:
\beq
\tilde O\,=\,{O\over d^{{\Delta_O\over  \Delta_d}}}\,\,.
\eeq
Let us first apply this definition to the temperature $T$, frequency $\omega$, and momentum $k$. We have:
\beq
\tilde T\,=\,{T \over d^{{1\over \Delta_d}}}\,=\,
{T \over d^{{5-p\over \lambda}}}\,\,\,,
\qquad\qquad
\tilde \omega\,=\,{\omega \over d^{{1\over \Delta_d}}}\,=\,
{\omega \over d^{{5-p\over \lambda}}}\,\,\,,
\qquad\qquad
\tilde k \,=\,{k \over d^{{1\over \Delta_d}}}
\,=\,{k \over d^{{5-p\over \lambda}}}\,\,.
\eeq
For $p=3$ the exponent of the density in these definition is $1/ n$, in agreement with the 
definitions used in \cite{Brattan:2012nb}. In the case of the magnetic field we have:
\beq
\tilde B\,=\,{B\over d^{{7-p\over \lambda}}}\,\,.
\eeq
In the conformal case $p=3$ the power of $d$ is $2/n$, again in agreement with \cite{Brattan:2012nb}.


\section{Conclusions and outlook}
\label{conclusions}

In this paper we studied the collective excitations of holographic matter engineered as intersection of two stacks of D-branes.  One type of branes (the color branes) were substituted by the geometry they generated, while the flavor branes were considered as probes and their dynamics were governed by the DBI action.  We analyzed these systems at high baryonic density, both at zero and non-zero temperature, and we also studied the influence of the magnetic field. 

We determined (both analytically and  numerically) the dispersion relation of the holographic zero sound at low temperature, as well as the diffusion constant at higher temperature.  We also studied numerically the crossover between the collisionless regime at low temperature and the hydrodynamic regime at higher temperature. When the intersection is $(2+1)$-dimensional, one can further study the anyonic degrees of freedom in the system by performing an alternative quantization. We implemented this  procedure for a general $(2+1)$-dimensional brane intersection and computed the corresponding anyonic correlators.

Our results apply to a large number of brane intersections, characterized by  the three numbers $p$, $q$, and $n$. However, we found that they only depend on $p$ (the dimensionality of the background branes) and the index $\lambda$ defined in (\ref{lambda_n}). Indeed, this $(p,\lambda)$ universality showed up in the different rescalings we performed and in the relations that the rescaled quantities satisfied. Thus, our analysis unified and extended previous results in the literature.  

We are continuing our efforts in generalizing the results presented here to the case, where we allow a non-zero mass for the fundamentals. This is a much more involved study, since then also the fluctuations of the scalar field has to be taken care of. However, our initial exploration suggest that most of the results in the present paper can be made available also in the presence of the mass.  Actually, for the supersymmetric intersections at  zero temperature and nonvanishing chemical potential, one can choose a system of coordinates such that the embedding function of the probe is also a cyclic variable \cite{Karch:2007br}. This has allowed to compute analytically the dispersion relation of the zero sound for massive quarks in the case of the D3-D7 \cite{Kulaxizi:2008kv,Davison:2011ek}, D3-D5, and D3-D3 intersections\cite{Ammon:2012je}. Interestingly, the speed of zero sound for these systems vanishes when the quark mass equals the chemical potential, signaling a phase transition with violation of hyperscaling, whose critical exponents have been evaluated in \cite{Ammon:2012je}. We intend to analyze this phenomenon for arbitrary intersections in the near future.

Another interesting avenue to pursue would be to allow an internal flux in the worldvolume of the probe D-brane. In the D$p$-D$(p+2)$ systems, turning on the appropriate internal flux induces a non-trivial profile of some scalar.  This corresponds, in the field theory dual, to moving to the Higgs branch of the theory (see \cite{Arean}).  The analysis of the collective excitations of these intersections is an interesting open problem which we will try to address in the future.


\vspace{0.5cm}

{\bf \large Acknowledgments}
We thank Danny Brattan, Georgios Itsios, Gilad Lifschytz, and Matthew Lippert for useful comments and for careful readings of the manuscript. 
N.J. is supported by the Academy of Finland grant no. 1268023. 
A.~V.~R. is funded by the Spanish grant FPA2011-22594, by the Consolider-Ingenio 2010 Programme CPAN (CSD2007-00042), by Xunta de
Galicia (Conselleria de Educaci\'on, grant INCITE09-206-121-PR and grant PGIDIT10PXIB206075PR), and by FEDER. 

\appendix

\vskip 1cm
\renewcommand{\theequation}{\rm{A}.\arabic{equation}}
\setcounter{equation}{0}

\section{Fluctuation equations of motion}
\label{appendixA}
Let us write the equations of motion (\ref{eom_general})  with $B$ field. The equation of motion for $a_t$ becomes:
\bear
&&\partial_{\rho}\,\Bigg[{|g_{tt}|\,\sqrt{H}\over u^3\,\sqrt{g_{rr}}}\,
\Bigg({g_{xx}\,f_p\over g_{xx}^{\,2}+B^2}\Bigg)^{3\over 2}\,a_t'\,\Bigg]\,+\,
{1\over u}\,\sqrt{{g_{rr}\,H\over f_p}}\,
\Bigg({g_{xx}\over g_{xx}^2+B^2}\Bigg)^{3\over 2}
\partial_x
\big(\partial_x\,a_t\,-\,\partial_t\,a_x\big)\rc
&&\qquad\qquad\qquad\qquad\qquad\qquad\qquad\qquad
+\partial_{\rho}\Bigg({d\,B\over g_{xx}^2+B^2}\Bigg)\,\partial_{x}\,a_y\,=\,0
\eear
Let us write this equation in momentum space. By using  (\ref{at_ax_E}) we can write it in terms of the electric field $E$. The result is:
\bear
&&E''\,+\,\partial_{\rho}\log\Bigg[
{\sqrt{|g_{tt}|}\over \sqrt g_{rr}}\,{g_{xx}\,f_p\over  g_{xx}^2+B^2}\,
{\sqrt{H+d^2}\over \omega^2-k^2\,u^2}\,\Bigg]\,E'\,+\,
{g_{rr}\over |g_{tt}|\,f_p^2}\,(\omega^2-k^2\,u^2)\,E\rc
&&\qquad\qquad\qquad\qquad
=i\,B\,d\,{\sqrt{g_{rr}}\over \sqrt{|g_{tt}|}}\,{g_{xx}^{\,2}+B^2\over g_{xx}\,f_p}\,
{\omega^2-k^2\,u^2\over \sqrt{H+d^2}}\,
\partial_{\rho}\Bigg({1\over g_{xx}^2+B^2}\Bigg)\,a_y \ .
\eear
For a D$p$-brane background the metric elements and the blackening factor  are given in (\ref{Dp-metric_dilaton}) and (\ref{blackeningDp}) while the function $H$ has been written in (\ref{H_Dp}). Moreover, when $B\not=0$ we have
\beq
\omega^2\,-\,k^2\,u^2\,=\,
{(\omega^2-f_p\,k^2)\,\rho^{\lambda}+\omega^2\,B^2\,\rho^{\lambda+p-7}+
\omega^2\,d^2
\over 
\rho^{\lambda}+\rho^{\lambda+p-7}\,B^2+d^2}\,\,.
\eeq
Plugging these values in the previous equation, we get:
\bear
&&E''+\partial_{\rho}\log\,\Bigg[
{\rho^{7-p}\over \rho^{7-p}+B^2}\,
{(\rho^{\lambda}+\rho^{\lambda+p-7}\,B^2\,+d^2)^{{3\over 2}}\,f_p\over
(\omega^2-f_p\,k^2)\,\rho^{\lambda}+\omega^2\,B^2\,\rho^{\lambda+p-7}+
\omega^2\,d^2}\Bigg]\,E'\rc
&&\qquad\qquad
+{1\over \rho^{7-p}\,f_p^2}\,
{(\omega^2-f_p\,k^2)\,\rho^{\lambda}+\omega^2\,B^2\,\rho^{\lambda+p-7}+
\omega^2\,d^2
\over 
\rho^{\lambda}+\rho^{\lambda+p-7}\,B^2+d^2}\,E \nonumber\\
&&\qquad\qquad
=-{(7-p)\,i\,B\,d\over \rho f_p\,(\rho^{7-p}\,+\,B^2)}\,
{(\omega^2-f_p\,k^2)\,\rho^{\lambda}+\omega^2\,B^2\,\rho^{\lambda+p-7}+
\omega^2\,d^2\over (\rho^{\lambda}+\rho^{\lambda+p-7}\,B^2+d^2)^{{3\over 2}}}\,a_y\,\,.
\qquad\qquad
\label{eom_E_Bnonzero}
\eear
The equation for $a_y$ can be written as:
\bear
&&a_y''\,+\,\partial_{\rho}\,\log\,\Bigg[{g_{xx}\,\sqrt{|g_{tt}|}\over \sqrt{g_{rr}}}\,f_p\,
{\sqrt{H+d^2}\over g_{xx}^2+B^2}
\Bigg]\,a_y'\,+\,
{g_{rr}\over f_p^2\,|g_{tt}|}\,
(\omega^2-k^2\,u^2)\,a_y \nonumber\\
&&\qquad\qquad\qquad\qquad
\,=\,-id\,B\,{\sqrt{g_{rr}}\over g_{xx}\,\sqrt{|g_{tt}|}}\, {1\over f_p}\,
{g_{xx}^2+B^2\over \sqrt{H+d^2}}\,
\partial_{\rho}\Bigg({1\over g_{xx}^2+B^2}\Bigg)\,E
\,\,.
\eear
For the D$p$-brane background this equation becomes:
\bear
&&a_y''\,+\,\partial_{\rho}\,\log\Big[{\rho^{7-p}\over \rho^{7-p}+B^2}\,
\sqrt{\rho^{\lambda}+\rho^{\lambda+p-7}\,B^2+d^2}\,f_p\Big]\,a_y'\rc
&&\qquad\qquad\qquad\qquad
+{1\over \rho^{7-p}\,f_p^2}\,
{(\omega^2-f_p\,k^2)\,\rho^{\lambda}+\omega^2\,B^2\,\rho^{\lambda+p-7}+
\omega^2\,d^2
\over 
\rho^{\lambda}+\rho^{\lambda+p-7}\,B^2+d^2}\,a_y \nonumber\\
&&\qquad\qquad\qquad\qquad
=\,{(7-p)\,i\,B\,d\over \rho f_p}\,
{E\over (\rho^{7-p}+B^2)\,\sqrt{\rho^{\lambda}+\rho^{\lambda+p-7}\,B^2+d^2}}\,\,.
\label{eom_ay_general}
\eear

\vskip 1cm
\renewcommand{\theequation}{\rm{B}.\arabic{equation}}
\setcounter{equation}{0}

\section{Indicial equation}
\label{appendixB}
Let us consider a differential equation for the function $E=E(\rho)$ of the type:
\beq
E''\,+\,\Big({1\over \rho-r_h}\,+\,c_1\Big)\,E'\,+\,
\Big({A\over (\rho-r_h)^2}\,+\,{c_2\over \rho-r_h}\Big)\,E\,=\,0\,\,,
\label{E_eq_near-horizon}
\eeq
where $r_h$, $A$, $c_1$, and $c_2$ are constants. We want to find a solution in Frobenius series for $\rho$ close to $r_h$, of the type:
\beq
E(\rho)\,=\,E_{nh}\,(\rho-r_h)^{\alpha}\,\Big[\,1\,+\beta\,(\rho-r_h)\,+\,\ldots\,\Big]\,\,.
\label{Frobenius_E_nh}
\eeq
By substituting this expansion  in (\ref{E_eq_near-horizon}) and comparing the different powers of $\rho-r_h$, we get that $\alpha$ and $\beta$ must satisfy the equations:
\beq
\alpha^2\,+A\,=\,0\,\,,
\qquad\qquad
\beta (\alpha^2+2\alpha+A+1 )\,+\,\alpha c_1\,+\,c_2\,=\,0\,\,.
\eeq
By choosing infalling boundary conditions, we get that $\alpha$ (when $A>0$) must be:
\beq
\alpha\,=\,-i\,\sqrt{A}\,\,.
\label{sol_indicial_eq}
\eeq
Moreover, $\beta$ is given by:
\beq
\beta\,=\,-{\alpha\,c_1+c_2\over 1+2\alpha}\,\,.
\label{beta_general}
\eeq

\vskip 1cm
\renewcommand{\theequation}{\rm{C}.\arabic{equation}}
\setcounter{equation}{0}

\section{Transverse correlators}
\label{Trasnverse_correlator}

We now study the equation of motion (\ref{ay_eq_Bzero}) for $a_y$ in the low frequency regime in which $k\sim {\mathcal O}(\epsilon)$ and $\omega\sim {\mathcal O}(\epsilon^2)$. First, we expand (\ref{ay_eq_Bzero}) near the horizon $\rho=r_h$. The coefficient of the term without derivatives is just the same as in (\ref{eom_E_Bzero}) and can be expanded as in (\ref{nh_expansion_Eeq}) and (\ref{A_c1_c2}). Moreover, the coefficient multiplying $a_y'$ in (\ref{ay_eq_Bzero}) can be represented near $\rho=r_h$ as:
\beq
\partial_{\rho}\,\log\Big[\sqrt{\rho^{\lambda}+d^2}\,f_p\Big]\,=\,{1\over \rho-r_h}\,+\,d_1\,+\,
\ldots\,\,,
\eeq
where $d_1$ is given by:
\beq
d_1\,=\,{p+\lambda-8\over 2\,r_h}\,-\,
{\lambda\over 2\,r_h }\,{d^2\over d^2+r_h^{\lambda}}\,\,.
\label{d1}
\eeq
Let us now solve for $a_y$ in Frobenius series around $\rho=r_h$:
\beq
a_y(\rho)\,=\,(\rho-r_h)^{\alpha}\,(1+\beta\,(\rho-r_h)+\ldots)\,\,.
\label{ay_nh}
\eeq
From the equations of appendix \ref{appendixA} one can show that the exponent $\alpha$ is just the same as in (\ref{exponent_alpha}). The coefficient $\beta$ is given by (\ref{beta_general}) with $c_1$ changed by $d_1$ (and given by (\ref{d1})). Since 
$\alpha\,,\,c_2\sim {\mathcal O}(\epsilon^2)$ and $d_1\sim {\mathcal O}(1)$, we have that, at order $\epsilon^2$, $\beta$ can be written as:
\beq
\beta\approx -(\alpha\,d_1+\,c_2)\,\,.
\label{beta_alpha_d_c}
\eeq
Plugging the values of $\alpha$, $d_1$, and $c_2$ (written in (\ref{exponent_alpha}), (\ref{d1}), and (\ref{A_c1_c2}) respectively), we find that $\beta$ is given by:
\beq
\beta\,=\,{i\over 2(7-p)\,r_h^{{7-p\over 2}}}\,
{(p-8)\,d^2\,+\,(p+\lambda-8)\,r_h^{\lambda}\over  d^2+r_h^{\lambda}}\,\,\omega\,+\,
{r_h^{p+\lambda-6}\over (7-p)\,(d^2+r_h^{\lambda})}\,k^2\,\,.
\label{beta_ay}
\eeq
We now take the near-horizon and low-frequency limits in the opposite order. First of all, let us write the equation of motion of $a_y$ in the form:
\beq
a_y''\,+\,{G'\over G}\,a_y'\,+\,Q\,a_y\,=\,0 \ ,
\label{a_y_eq_G_Q}
\eeq
where $G=G(\rho)$ is defined as:
\beq
G(\rho)\equiv \sqrt{\rho^{\lambda}+d^2}\,f_p(\rho)\,\,.
\label{G_def}
\eeq
The function $Q$ in (\ref{a_y_eq_G_Q}) can be read from (\ref{ay_eq_Bzero}). Notice that, at order $\epsilon^2$, this function is simply:
\beq
Q(\rho)\approx -{\rho^{\lambda+p-7}\over (\rho^{\lambda}+d^2)\,f_p(\rho)}\,\,k^2\,\,.
\eeq
We want to match the near-horizon expansion (\ref{ay_nh}). Therefore, it is better to redefine $a_y$ in the form:
\beq
a_y(\rho)\,=\,F(\rho)\,\alpha_y(\rho)\,\,,
\label{ay_alphay}
\eeq
where $\alpha_y(\rho)$ should be regular at $\rho=r_h$ and $F(\rho)$ is given by:
\beq
F(\rho)=(\rho-r_h)^{\alpha}\,\,.
\label{F_value}
\eeq
The resulting equation for $\alpha_y$ is:
\beq
\alpha_y''\,+\,\Big({G'\over G}\,+\,2\,{F'\over F}\,\epsilon^2\Big)\,\alpha_y'\,+\,
\epsilon^2\,(P+Q)\,\alpha_y\,=\,0\,\,,
\label{alpha_y_eq}
\eeq
where we have explicitly introduced the powers of $\epsilon$ to keep track of the low frequency expansion and we have defined the new function $P(\rho)$ as:
\beq
P(\rho)\equiv {F''\over F}\,+\,{G'\over G}\,{F'\over F}\,\,.
\label{P_def}
\eeq
We will solve (\ref{alpha_y_eq}) order by order in  a series expansion in $\epsilon$ of the form:
\beq
\alpha_y\,=\,\alpha_0\,+\,\epsilon^2\,\alpha_1\,+\,\ldots\,\,.
\label{epsilon_expansion_alphay}
\eeq
The equation for $\alpha_0$ is:
\beq
\alpha_0''\,+\,{G'\over G}\,\alpha_0'\,=\,0\,\,,
\eeq
whose first integration yields:
\beq
\alpha_0'\,=\,{c_0\over  \sqrt{\rho^{\lambda}+d^2}\,f_p(\rho)}\,\,,
\eeq
where $c_0$ is a constant. When $c_0\not=0$ the function $\alpha_0'$ blows up at $\rho=r_h$. Thus, we choose $c_0=0$ and therefore $\alpha_0={\rm constant}$. Without loss of generality we can take 
\beq
\alpha_0=1\,\,.
\eeq
The equation for $\alpha_1$ is
\beq
\alpha_1''\,+\,{G'\over G}\,\alpha_1'\,=\,-P\,-\,Q \ .
\label{alpha1_eq}
\eeq
Let us solve this equation by variation of constants. We put:
\beq
\alpha_1'\,=\,{\Lambda(\rho)\over  \sqrt{\rho^{\lambda}+d^2}\,f_p(\rho)}\,=\,
{\Lambda(\rho)\over G(\rho)}\,\,,
\eeq
where $\Lambda(\rho)$ is a function to be determined. By direct substitution into (\ref{alpha1_eq}) we get that $\Lambda(\rho)$ must satisfy:
\beq
\Lambda'\,=\,-G\,(P+Q)\,\,.
\eeq
To integrate this equation we notice that the first term on the right-hand-side is, actually, a total derivative. Indeed, at ${\mathcal O}(\epsilon^2)$ we have that
\beq
\big(\log F\big)''\,=\,{F''\over F}\,-\,\Big({F'\over F}\Big)^2\approx {F''\over F}\,\,,
\eeq
since $F'/F\sim {\mathcal O}(\epsilon^2)$ when $F$ is the function (\ref{F_value}). Using this result in the definition of $P$ we conclude that:
\beq
G\,P\,=\,\Big[\,G\,(\log F)'\Big]'\,+\,{\mathcal O}(\epsilon^4)\,\,.
\eeq
Therefore, 
\beq
\Lambda(\rho)\,=\,-G\, (\log F)'\,-\,c\,-\int_{r_h}^{\rho}\,G(\bar \rho)\,Q(\bar \rho)\,d\bar \rho\,\,,
\eeq
where $c$ is a constant to be determined. Let us next define the integral ${\cal I}(\rho)$ as:
\beq
k^2\,{\cal I}(\rho)\,\equiv\,-\int_{r_h}^{\rho}\,G(\bar \rho)\,Q(\bar \rho)\,d\bar\rho\,\,,
\eeq
or, more explicitly:
\beq
{\cal I}(\rho)\,=\,\int_{r_h}^{\rho}\,{\bar\rho^{\lambda+p-7}\over 
\sqrt{\bar\rho^{\lambda}+d^2}}\,d\bar\rho\,\,.
\eeq
Then, $\alpha_1'$ can be written as:
\beq
\alpha_1'\,=\,-\Bigg[\,{c\over \sqrt{\rho^{\lambda}+d^2}\,f_p}\,+\,
{\alpha\over \rho-r_h}\,\Bigg]\,+\,{{\cal I}(\rho)\,k^2\over \sqrt{\rho^{\lambda}+d^2}\,f_p}\,\,.
\label{alpha_1_prime_integral}
\eeq
We will fix the constant $c$ by imposing that $\alpha_1'$ be regular at $\rho=r_h$. This is equivalent to require that the term with $c$ cancels the term with $1/(\rho-r_h)$. Notice that, by construction, ${\cal I}(\rho)$ vanishes linearly at the horizon and the last term in (\ref{alpha_1_prime_integral}) is therefore regular at the horizon. Taking into account the expansion of $1/G$ around $\rho=r_h$:
\beq
{1\over \sqrt{\rho^{\lambda}+d^2}\,f_p}\,\approx\,{r_h\over (7-p)\,\sqrt{r_h^{\lambda}+d^2}} \,{1\over \rho-r_h}\,-\,
{(p-8)\,d^2\,+\,(p+\lambda-8)\,r_h^{\lambda}\over  
2(7-p)(r_h^{\lambda}+d^2)^{{3\over 2}}} \ ,
\label{1_over_G_nh_expansion}
\eeq
we get:
\beq
c\,=\,-{7-p\over r_h}\,\sqrt{r_h^{\lambda}+d^2}\,\,\alpha\,=\,i\,
{\sqrt{r_h^{\lambda}+d^2}\over r_h^{{7-p\over 2}}}\,\omega\,\,.
\eeq
Thus, $\alpha_y'\,=\,\alpha_1'\,+{\mathcal O}(\epsilon^4)$ and is given by:
\beq
\alpha_y'\,=\,-{i\over (7-p)\, r_h^{{5-p\over 2}}}\Bigg[\,{7-p\over r_h}\,
{\sqrt{r_h^{\lambda}+d^2}\over \sqrt{\rho^{\lambda}+d^2}\,\,f_p}\,-\,{1\over \rho-r_h}\,\Bigg]\,
\omega\,+\,{{\cal I}(\rho)\over \sqrt{\rho^{\lambda}+d^2}\,f_p}\,k^2\,\,.
\eeq
In order to match this solution with (\ref{ay_nh}), let us expand ${\cal I}(\rho)$ near $\rho=r_h$. We easily get:
\beq
{\cal I}(\rho)={r_h^{\lambda+p-7}\over \sqrt{r_h^{\lambda}+d^2}}\,(\rho-r_h)\,+\,
{\mathcal O}((\rho-r_h)^2)\,\,.
\eeq
From this result and the expansion of $1/G$ in (\ref{1_over_G_nh_expansion}), it is  easy to check that, indeed,
\beq
\alpha_y'(\rho=r_h)\,=\,\beta\,\,,
\eeq
where $\beta$ is given in (\ref{beta_ay}). Moreover, since $\alpha_y\approx 1+\alpha_1$ and $F\approx 1+\alpha\,\log(\rho-r_h)$, we conclude that, at ${\mathcal O}(\epsilon^2)$:
\beq
a_y\,\approx \,1+\alpha_1\,+\,\alpha \log(\rho-r_h)\,\,.
\eeq
Therefore, it follows that:
\beq
a_y'\,=\, \alpha_1'\,+\,{\alpha \over \rho-r_h}\,+\,{\mathcal O}(\epsilon^4)\,=\,
\alpha_y'\,+\,{\alpha\over  \rho-r_h}\,+\,{\mathcal O}(\epsilon^4)\,\,.
\eeq
Explicitly, 
\beq
a_y'\,=\,-{1\over G(\rho)}\,
\Bigg[ i\,{\sqrt{r_h^{\lambda}+d^2}\over r_h^{{7-p\over 2}}}\,\,\omega\,-\,
{\cal I}(\rho)\,k^2\,\Bigg]\,\,,
\eeq
where $G(\rho)$ has been defined in (\ref{G_def}). 

Let us now extract the $\left\langle J_y\,J_y\right\rangle $ correlator from the previous results. Recall that the term depending on $a_y$ of the Lagrangian density is of the form:
\beq
{\cal L}(a_y)\,=\,{\cal F}\,(f_{y\rho})^2\,=\,{\cal F}\, (a_y')^2\,\,,
\eeq
where ${\cal F}$ is given by:
\beq
{\cal F}\,=\,-{\cal N}\,{\sqrt{g_{rr}\,|g_{tt}|}\over \sqrt{H+d^2}}\,H\,{\cal G}^{yy}\,
{\cal G}^{\rho\rho} = -{\cal N}\,G(\rho) \ ,
\eeq
with ${\cal N}$ being a normalization constant. 
The on-shell action of $a_y$ is:
\beq
S_{{\rm on-shell}}(a_y)\,=\,\int\,d^n\,x\,\Big( {\cal F}\,a_y\,a_y'\Big)_{\rho\to\infty}\,\,.
\eeq
With the normalization condition we are using ($\alpha_0=1$), the two-point function of $J_y$ is given by the standard AdS/CFT prescription:
\beq
\left\langle J_y(p)\,J_y(-p)\right\rangle\,=\,\Big( {\cal F}\,a_y'\Big)_{\rho\to\infty}\,\,.
\label{correlator_formula}
\eeq
From the explicit expressions of ${\cal F}$ and $a_y$, we get:
\beq
{\cal F}\,a_y'\,=\,{\cal N}\,
\Bigg[ i\,{\sqrt{r_h^{\lambda}+d^2}\over r_h^{{7-p\over 2}}}\,\,\omega\,-\,
{\cal I}(\rho)\,k^2\,\Bigg]\,\,.
\eeq
In order to get the value of the integral ${\cal I}$ at the boundary, let us rewrite its expression as:
\beq
{\cal I}(\rho)\,=\,{r_h^{{\lambda\over 2}+p-6}\over \lambda}\,\,
\int_1^{\big({\rho\over r_h}\big)^{\lambda}}\,\,
{\zeta^{{p-6\over \lambda}}\over \sqrt{\zeta+\hat d^2}}\,d\zeta\,\,,
\eeq
with $\hat d$ given by  (\ref{hat_d_def}). 
Then, it is easy to prove that, when $2(6-p)>\lambda$ the $\rho\to\infty$ limit of ${\cal I}$ converges and is given by:
\beq
{\cal I}(\rho\to\infty)\,=\,{2\,\over 2(6-p)-\lambda}\,\,r_h^{{\lambda\over 2}+p-6}\,
F\Big({1\over 2}, {6-p\over \lambda}-{1\over 2}; {6-p\over \lambda}+{1\over 2};
-\hat d^2\Big)\,\,.
\eeq
It follows that the $\left\langle J_y\,J_y\right\rangle $ correlator takes the form:
\beq
\left\langle J_y(p)\,J_y(-p)\right\rangle\,=\,{\cal N}\,\Big[\,\Gamma_{\omega}\,i\omega\,+\,\Gamma_k\,k^2\,\Big]\,\,,
\label{transv_corr_Gammas}
\eeq
where the coefficients $\Gamma_{\omega}$  and $\Gamma_k$ are:
\bear
\Gamma_{\omega} & = & r_h^{{\lambda+p-7\over 2}}\,\sqrt{1+\hat d^2}\rc
\Gamma_k & = & -{2\,\over 2(6-p)-\lambda}\,\,r_h^{{\lambda\over 2}+p-6}\,
F\Big({1\over 2}, {6-p\over \lambda}-{1\over 2}; {6-p\over \lambda}+{1\over 2};
-\hat d^2\Big)\,\,.
\label{Gamma_omega_k}
\eear
When $p=3$ and $\lambda=4$ the result written above coincides with the one in \cite{Brattan:2013wya}. 

Let us rewrite the correlator in terms of the rescaled frequency and momentum $\hat\omega$ and $\hat k$ defined in (\ref{hat_omega_k}):
\beq
\left\langle J_y(p)\,J_y(-p)\right\rangle\,=\,\hat{\cal N}\,\Big[\,\hat\Gamma_{\omega}\,i\hat\omega\,+\,\hat\Gamma_k\,\hat k^2\,\Big]\,\,,
\eeq
where $\hat{\cal N}$ is related to ${\cal N}$ as:
\beq
\hat{\cal N}\,=\,{\cal N}\,r_h^{{\lambda\over 2}-1}\,\,,
\eeq
and the rescaled coefficients $\hat\Gamma_{\omega}$  and $\hat\Gamma_k$ are:
\bear
\hat\Gamma_{\omega} &  =& \sqrt{1+\hat d^2}\rc
\hat\Gamma_k & =& -{2\,\over 2(6-p)-\lambda}\,
F\Big({1\over 2}, {6-p\over \lambda}-{1\over 2}; {6-p\over \lambda}+{1\over 2};
-\hat d^2\Big)\,\,.
\eear

\vskip 1cm
\renewcommand{\theequation}{\rm{D}.\arabic{equation}}
\setcounter{equation}{0}

\section{Wronskian method}
\label{appendixD}

In this appendix we solve the inhomogeneous equation (\ref{inhomogeneous_y_eq}) by the Wronskian method. We start by defining a new function ${\cal Y}$ as:
\beq
{\cal Y}\,\equiv \rho^{-{1\over 2}}\,y(\rho)\,\,,
\eeq
and a new independent variable $x$ as:
\beq
x\,\equiv\,{2\omega\over 5-p}\,\rho^{{p-5\over 2}}\,\,.
\eeq
In what follows we consider ${\cal Y}$ as a function of $x$. After this change of variables, eq. (\ref{inhomogeneous_y_eq}) becomes:
\beq
{d^2\,{\cal Y}\over d\,x^2}\,+\,{1\over x}\,{d\,{\cal Y}\over dx}\,+\,
\Big(1\,-\,{\nu_p^2\over x^2}\Big)\,{\cal Y}\,=\,f(x)\,\,,
\label{eq_cal_Y}
\eeq
where $f(x)$ is given by:
\beq
f(x)=d_p(\omega)\,x^{2\nu_p}\,
H_{\nu_p}^{(1)}(x)\,\,,
\eeq
and $\nu_p$ and $d_p$ are defined as:
\beq
\nu_p\,\equiv\,{1\over 5-p}\,\,,
\qquad\qquad\qquad
d_p(\omega)\,\equiv\,-(7-p)\,
\Big({5-p\over 2}\Big)^{{2\over 5-p}}\,\omega^{-{7-p\over 5-p}}\,\,.
\eeq
Eq. (\ref{eq_cal_Y}) is a linear inhomogeneous equation, whose solutions can be obtained from those of the homogeneous equation. Let ${\cal Y}_1(x)$ and ${\cal Y}_2(x)$ be two independent solutions of (\ref{eq_cal_Y}) with $f(x)=0$. Then, the solution of 
(\ref{eq_cal_Y}) for $f(x)\not=0$ can be written as:
\beq
{\cal Y}(x)\,=\,I_1(x)\,{\cal Y}_1(x)\,+\,I_2(x)\,{\cal Y}_2(x)\,\,,
\eeq
where $I_1(x)$ and $I_2(x)$ are the following indefinite integrals:
\beq
I_1(x)\,=\,-\int {{\cal Y}_2(x)\,f(x)\over W({\cal Y}_1, {\cal Y}_2)}\,dx\,\,,
\qquad\qquad
I_2(x)\,=\,\int {{\cal Y}_1(x)\,f(x)\over W({\cal Y}_1, {\cal Y}_2)}\,dx\,\,,
\eeq
and $ W({\cal Y}_1, {\cal Y}_2)$ is the Wronskian function of ${\cal Y}_1$ and ${\cal Y}_2$:
\beq
W({\cal Y}_1, {\cal Y}_2)\,=\,{\cal Y}_1\, {\cal Y}_2{'}\,-\,{\cal Y}_2\, {\cal Y}_1{'}\,\,.
\eeq
The homogeneous version of (\ref{eq_cal_Y}) is just the Bessel equation. Therefore, we can take the Hankel functions of index $\nu_p$ as the two independent solutions 
${\cal Y}_1$ and  ${\cal Y}_2$:
\beq
{\cal Y}_i(x)\,=\,H_{\nu_p}^{(i)}\,\,,
\qquad\qquad
(i=1,2)\,\,.
\eeq
The Wronskian of two Hankel functions is rather simple, namely:
\beq
W(H_{\nu_p}^{(1)}(x), H_{\nu_p}^{(2)}(x))\,=\,-{4i\over \pi x}\,\,.
\eeq
Therefore, $I_1(x)$ and $I_2(x)$ are given by:
\bear
&&I_1(x)\,=\,-i{\pi d_p\over 4}\,\int x^{2\nu_p+1}\,
H_{\nu_p}^{(2)}(x)\,H_{\nu_p}^{(1)}(x)\,dx\rc
&&I_2(x)\,=\,i{\pi d_p\over 4}\,\int x^{2\nu_p+1}\,
H_{\nu_p}^{(1)}(x)\,H_{\nu_p}^{(1)}(x)\,dx\,\,. 
\eear
Taking into account that (for $\nu+\mu\not=1$):
\beq
 \int x^{\mu+\nu+1}\,
H_{\mu}^{(\alpha)}(x)\,H_{\nu}^{(\beta)}(x)\,dx\,=\,
{x^{\mu+\nu+2}\over 2(\mu+\nu+1)}\,\Big[
H_{\mu}^{(\alpha)}(x)\,H_{\nu}^{(\beta)}(x)\,+\,
H_{\mu+1}^{(\alpha)}(x)\,H_{\nu+1}^{(\beta)}(x)\,\Big]\,\,,
\eeq
we get that $I_1(x)$ and $I_2(x)$ are given by:
\bear
I_1(x) & =&-i{\pi d_p\over 8}\,{5-p\over 7-p}\,x^{{2(6-p)\over 5-p}}\,
\Big[\,H_{\nu_p}^{(2)}(x)\,H_{\nu_p}^{(1)}(x)\,+\,
H_{\nu_p+1}^{(2)}(x)\,H_{\nu_p+1}^{(1)}(x)\,\Big]\rc
I_2(x)&=&i{\pi d_p\over 8}\,{5-p\over 7-p}\,x^{{2(6-p)\over 5-p}}\,
\Big[\,H_{\nu_p}^{(1)}(x)\,H_{\nu_p}^{(1)}(x)\,+\,
H_{\nu_p+1}^{(1)}(x)\,H_{\nu_p+1}^{(1)}(x)\,\Big]\,\,. 
\eear
It then follows that:
\beq
{\cal Y}(x)\,=\,i{\pi d_p\over 8}\,{5-p\over 7-p}\,x^{{2(6-p)\over 5-p}}\,
\Big[\,H_{\nu_p+1}^{(1)}(x)\, H_{\nu_p}^{(2)}(x)\,-\,
H_{\nu_p}^{(1)}(x)\, H_{\nu_p+1}^{(2)}(x)\,\Big]\,H_{\nu_p+1}^{(1)}(x)\,\,.
\eeq
By using the  following property of the Hankel functions:
\beq
H_{\nu_p+1}^{(1)}(x)\, H_{\nu_p}^{(2)}(x)\,-\,
H_{\nu_p}^{(1)}(x)\, H_{\nu_p+1}^{(2)}(x)\,=\,-{4i\over x}\,\,,
\eeq
we arrive at:
\beq
{\cal Y}(x)\,=\,-\Big[{(5-p)x\over 2\omega}
\Big]^{{7-p\over 5-p}}\,H_{{6-p\over 5-p}}^{(1)}(x)\,\,,
\eeq
which matches with the expression for $y(\rho)$ in (\ref{y_sol_Bfield}).



\begin{thebibliography}{99}

\bibitem{Landau} L. D. Landau, ``The theory of a Fermi liquid", Zh. Eksp. Teor. Fiz. {\bf 30}, 1058 (1956) [Soviet Phys. JETP {\bf 3}, 920 (1957)]. 

\bibitem{LFL}
See, for example: 
E. M. Lifshitz, L. P. Pitaevskii, ``Statistical Physics", Part 2, Pergamon Press, Oxford 1980;
D.Pines and P. Nozi\`eres, ``The theory of quantum liquids", Benjamin, New York 1966. 



\bibitem{AdS_CFT_reviews} For reviews see: 
 J.~Casalderrey-Solana, H.~Liu, D.~Mateos, K.~Rajagopal and U.~A.~Wiedemann,
  ``Gauge/String Duality, Hot QCD and Heavy Ion Collisions,''
  arXiv:1101.0618 [hep-th];
 J.~McGreevy,
  ``Holographic duality with a view toward many-body physics,''
  Adv.\ High Energy Phys.\  {\bf 2010}, 723105 (2010)
  [arXiv:0909.0518 [hep-th]];
    A.~V.~Ramallo,
  ``Introduction to the AdS/CFT correspondence,''
  Springer Proc.\ Phys.\  {\bf 161} (2015) 411
  [arXiv:1310.4319 [hep-th]].
 

\bibitem{Karch:2002sh}
  A.~Karch and E.~Katz,
  ``Adding flavor to AdS / CFT,''
  JHEP {\bf 0206} (2002) 043
  [hep-th/0205236].
  



\bibitem{Kobayashi:2006sb}
  S.~Kobayashi, D.~Mateos, S.~Matsuura, R.~C.~Myers and R.~M.~Thomson,
  ``Holographic phase transitions at finite baryon density,''
  JHEP {\bf 0702} (2007) 016
  [hep-th/0611099].
  




\bibitem{Karch:2008fa}
  A.~Karch, D.~T.~Son and A.~O.~Starinets,
  ``Zero Sound from Holography,''
  arXiv:0806.3796 [hep-th].
 
\bibitem{Karch:2009zz}
  A.~Karch, D.~T.~Son and A.~O.~Starinets,
  ``Holographic Quantum Liquid,''
  Phys.\ Rev.\ Lett.\  {\bf 102} (2009) 051602.


\bibitem{Bergman:2011rf}
  O.~Bergman, N.~Jokela, G.~Lifschytz and M.~Lippert,
  ``Striped instability of a holographic Fermi-like liquid,''
  JHEP {\bf 1110} (2011) 034
  [arXiv:1106.3883 [hep-th]].
  
\bibitem{Jokela:2012vn}
  N.~Jokela, G.~Lifschytz and M.~Lippert,
  ``Magnetic effects in a holographic Fermi-like liquid,''
  JHEP {\bf 1205} (2012) 105
  [arXiv:1204.3914 [hep-th]].
  
  
\bibitem{Kulaxizi:2008kv}
  M.~Kulaxizi and A.~Parnachev,
  ``Comments on Fermi Liquid from Holography,''
  Phys.\ Rev.\ D {\bf 78} (2008) 086004
  [arXiv:0808.3953 [hep-th]].
 
\bibitem{Kim:2008bv}
  K.~Y.~Kim and I.~Zahed,
  ``Baryonic Response of Dense Holographic QCD,''
  JHEP {\bf 0812} (2008) 075
  [arXiv:0811.0184 [hep-th]].
 
\bibitem{Kulaxizi:2008jx} 
  M.~Kulaxizi and A.~Parnachev,
  ``Holographic Responses of Fermion Matter,''
  Nucl.\ Phys.\ B {\bf 815}, 125 (2009)
  [arXiv:0811.2262 [hep-th]].

\bibitem{Hung:2009qk}
  L.~Y.~Hung and A.~Sinha,
  ``Holographic quantum liquids in 1+1 dimensions,''
  JHEP {\bf 1001} (2010) 114
  [arXiv:0909.3526 [hep-th]].
 
\bibitem{Edalati:2010pn}
  M.~Edalati, J.~I.~Jottar and R.~G.~Leigh,
  ``Holography and the sound of criticality,''
  JHEP {\bf 1010} (2010) 058
  [arXiv:1005.4075 [hep-th]].
 
\bibitem{Lee:2010uy}
  B.~H.~Lee and D.~W.~Pang,
  ``Notes on Properties of Holographic Strange Metals,''
  Phys.\ Rev.\ D {\bf 82} (2010) 104011
  [arXiv:1006.4915 [hep-th]].
    
\bibitem{HoyosBadajoz:2010kd}
  C.~Hoyos-Badajoz, A.~O'Bannon and J.~M.~S.~Wu,
  ``Zero Sound in Strange Metallic Holography,''
  JHEP {\bf 1009} (2010) 086
  [arXiv:1007.0590 [hep-th]].

\bibitem{Lee:2010ez}
  B.~H.~Lee, D.~W.~Pang and C.~Park,
  ``Zero Sound in Effective Holographic Theories,''
  JHEP {\bf 1011} (2010) 120
  [arXiv:1009.3966 [hep-th]].

\bibitem{Ammon:2011hz}
  M.~Ammon, J.~Erdmenger, S.~Lin, S.~Muller, A.~O'Bannon, J.~P.~Shock, J.~Erdmenger and S.~Lin {\it et al.},
  ``On Stability and Transport of Cold Holographic Matter,''
  JHEP {\bf 1109} (2011) 030
  [arXiv:1108.1798 [hep-th]].
  
\bibitem{Davison:2011ek}
  R.~A.~Davison and A.~O.~Starinets,
  ``Holographic zero sound at finite temperature,''
  Phys.\ Rev.\ D {\bf 85} (2012) 026004
  [arXiv:1109.6343 [hep-th]].
  
\bibitem{Goykhman:2012vy}
  M.~Goykhman, A.~Parnachev and J.~Zaanen,
  ``Fluctuations in finite density holographic quantum liquids,''
  JHEP {\bf 1210} (2012) 045
  [arXiv:1204.6232 [hep-th]].
  
\bibitem{Gorsky:2012gi}
  A.~Gorsky and A.~V.~Zayakin,
  ``Anomalous Zero Sound,''
  JHEP {\bf 1302} (2013) 124
  [arXiv:1206.4725 [hep-th]].

\bibitem{Brattan:2012nb}
  D.~K.~Brattan, R.~A.~Davison, S.~A.~Gentle and A.~O'Bannon,
  ``Collective Excitations of Holographic Quantum Liquids in a Magnetic Field,''
  JHEP {\bf 1211} (2012) 084
  [arXiv:1209.0009 [hep-th]].
  
\bibitem{Jokela:2012se}
  N.~Jokela, M.~J\"arvinen and M.~Lippert,
  ``Fluctuations and instabilities of a holographic metal,''
  JHEP {\bf 1302} (2013) 007
  [arXiv:1211.1381 [hep-th]].

\bibitem{Davison:2013bxa}
  R.~A.~Davison and A.~Parnachev,
  ``Hydrodynamics of cold holographic matter,''
  JHEP {\bf 1306} (2013) 100
  [arXiv:1303.6334 [hep-th]].
    
\bibitem{Pang:2013ypa}
  D.~W.~Pang,
  ``Probing holographic semilocal quantum liquids with D-branes,''
  Phys.\ Rev.\ D {\bf 88} (2013) 4,  046002
  [arXiv:1306.3816 [hep-th]].
 
\bibitem{Dey:2013vja}
  P.~Dey and S.~Roy,
  ``Zero sound in strange metals with hyperscaling violation from holography,''
  Phys.\ Rev.\ D {\bf 88} (2013) 046010
  [arXiv:1307.0195 [hep-th]].

\bibitem{Edalati:2013tma}
  M.~Edalati and J.~F.~Pedraza,
  ``Aspects of Current Correlators in Holographic Theories with Hyperscaling Violation,''
  Phys.\ Rev.\ D {\bf 88} (2013) 086004
  [arXiv:1307.0808 [hep-th]].
  
\bibitem{Davison:2013uha}
  R.~A.~Davison, M.~Goykhman and A.~Parnachev,
  ``AdS/CFT and Landau Fermi liquids,''
  JHEP {\bf 1407} (2014) 109
  [arXiv:1312.0463 [hep-th]].
  
\bibitem{DiNunno:2014bxa}
  B.~S.~DiNunno, M.~Ihl, N.~Jokela and J.~F.~Pedraza,
  ``Holographic zero sound at finite temperature in the Sakai-Sugimoto model,''
  JHEP {\bf 1404} (2014) 149
  [arXiv:1403.1827 [hep-th]].
   
\bibitem{Arias:2014msa}
  R.~E.~Arias and I.~S.~Landea,
  ``Hydrodynamic Modes of a holographic $p-$ wave superfluid,''
  JHEP {\bf 1411} (2014) 047
  [arXiv:1409.6357 [hep-th]].
  
\bibitem{Jensen:2010vd}
  K.~Jensen, A.~Karch and E.~G.~Thompson,
  ``A Holographic Quantum Critical Point at Finite Magnetic Field and Finite Density,''
  JHEP {\bf 1005} (2010) 015
  [arXiv:1002.2447 [hep-th]].
  
  
  
  
  
  
\bibitem{Khon}
 W. Kohn,  ``Cyclotron resonance and the Haas-van Alphen oscillations of an interacting electron gas", Phys. Rev. {\bf 123} (1961) 1242. 
 
 
\bibitem{Jokela:2013hta}
  N.~Jokela, G.~Lifschytz and M.~Lippert,
  ``Holographic anyonic superfluidity,''
  JHEP {\bf 1310} (2013) 014
  [arXiv:1307.6336 [hep-th]].

 
 
\bibitem{Jokela:2014wsa}
  N.~Jokela, G.~Lifschytz and M.~Lippert,
  ``Flowing holographic anyonic superfluid,''
  JHEP {\bf 1410} (2014) 21
  [arXiv:1407.3794 [hep-th]].

\bibitem{Brattan:2013wya}
  D.~K.~Brattan and G.~Lifschytz,
  ``Holographic plasma and anyonic fluids,''
  JHEP {\bf 1402} (2014) 090
  [arXiv:1310.2610 [hep-th]].

 
\bibitem{Brattan:2014moa}
  D.~K.~Brattan,
  ``A strongly coupled anyon material,''
  arXiv:1412.1489 [hep-th].


\bibitem{Karch:2009eb}
  A.~Karch, M.~Kulaxizi and A.~Parnachev,
  ``Notes on Properties of Holographic Matter,''
  JHEP {\bf 0911} (2009) 017
  [arXiv:0908.3493 [hep-th]].
 
 
 
 
\bibitem{Hartnoll:2009ns}
  S.~A.~Hartnoll, J.~Polchinski, E.~Silverstein and D.~Tong,
  ``Towards strange metallic holography,''
  JHEP {\bf 1004} (2010) 120
  [arXiv:0912.1061 [hep-th]].
  
\bibitem{Bigazzi:2013jqa}
  F.~Bigazzi, A.~L.~Cotrone and J.~Tarrio,
  ``Charged D3-D7 plasmas: novel solutions, extremality and stability issues,''
  JHEP {\bf 1307} (2013) 074
  [arXiv:1304.4802 [hep-th]].
 
 
 
 
\bibitem{Arean:2006pk}
  D.~Arean and A.~V.~Ramallo,
  ``Open string modes at brane intersections,''
  JHEP {\bf 0604} (2006) 037
  [hep-th/0602174].
 
 
 
 
\bibitem{Sakai:2004cn}
  T.~Sakai and S.~Sugimoto,
  ``Low energy hadron physics in holographic QCD,''
  Prog.\ Theor.\ Phys.\  {\bf 113} (2005) 843
  [hep-th/0412141].

\bibitem{Bergman:2010gm}
  O.~Bergman, N.~Jokela, G.~Lifschytz and M.~Lippert,
  ``Quantum Hall Effect in a Holographic Model,''
  JHEP {\bf 1010} (2010) 063
  [arXiv:1003.4965 [hep-th]].

 
\bibitem{Myers:2008me}
  R.~C.~Myers and M.~C.~Wapler,
  ``Transport Properties of Holographic Defects,''
  JHEP {\bf 0812} (2008) 115
  [arXiv:0811.0480 [hep-th]].


 
\bibitem{Jokela:2011eb}
  N.~Jokela, M.~J\"arvinen and M.~Lippert,
  ``A holographic quantum Hall model at integer filling,''
  JHEP {\bf 1105} (2011) 101
  [arXiv:1101.3329 [hep-th]].
  
\bibitem{Itsios:2015kja}
  G.~Itsios, N.~Jokela and A.~V.~Ramallo,
  ``Cold holographic matter in the Higgs branch,''
  Phys.\ Lett.\ B {\bf 747} (2015) 229
  [arXiv:1505.02629 [hep-th]].
  
\bibitem{Hohler:2009tv} 
  P.~M.~Hohler and M.~A.~Stephanov,
  ``Holography and the speed of sound at high temperatures,''
  Phys.\ Rev.\ D {\bf 80}, 066002 (2009)
  [arXiv:0905.0900 [hep-th]].

\bibitem{Cherman:2009tw}
  A.~Cherman, T.~D.~Cohen and A.~Nellore,
  ``A Bound on the speed of sound from holography,''
  Phys.\ Rev.\ D {\bf 80} (2009) 066003
  [arXiv:0905.0903 [hep-th]].
  
\bibitem{Cherman:2009kf}
  A.~Cherman and A.~Nellore,
  ``Universal relations of transport coefficients from holography,''
  Phys.\ Rev.\ D {\bf 80} (2009) 066006
  [arXiv:0905.2969 [hep-th]].
 
  
  
  

\bibitem{Amado:2009ts}
  I.~Amado, M.~Kaminski and K.~Landsteiner,
  ``Hydrodynamics of Holographic Superconductors,''
  JHEP {\bf 0905} (2009) 021
  [arXiv:0903.2209 [hep-th]].
  
\bibitem{Kaminski:2009dh}
  M.~Kaminski, K.~Landsteiner, J.~Mas, J.~P.~Shock and J.~Tarrio,
  ``Holographic Operator Mixing and Quasinormal Modes on the Brane,''
  JHEP {\bf 1002} (2010) 021
  [arXiv:0911.3610 [hep-th]].


  
\bibitem{Hartnoll:2014lpa}
  S.~A.~Hartnoll,
  ``Theory of universal incoherent metallic transport,''
  Nature Phys.\  {\bf 11} (2015) 54
  [arXiv:1405.3651 [cond-mat.str-el]].

 
 
 
\bibitem{Witten:2003ya}
  E.~Witten,
  ``SL(2,Z) action on three-dimensional conformal field theories with Abelian symmetry,''
  In *Shifman, M. (ed.) et al.: From fields to strings, vol. 2* 1173-1200
  [hep-th/0307041].

\bibitem{Yee:2004ju}
  H.~U.~Yee,
  ``A Note on AdS / CFT dual of SL(2,Z) action on 3-D conformal field theories with U(1) symmetry,''
  Phys.\ Lett.\ B {\bf 598} (2004) 139
  [hep-th/0402115].


\bibitem{Goldstein:2010aw}
  K.~Goldstein, N.~Iizuka, S.~Kachru, S.~Prakash, S.~P.~Trivedi and A.~Westphal,
  ``Holography of Dyonic Dilaton Black Branes,''
  JHEP {\bf 1010} (2010) 027
  [arXiv:1007.2490 [hep-th]].

\bibitem{Bayntun:2010nx}
  A.~Bayntun, C.~P.~Burgess, B.~P.~Dolan and S.~S.~Lee,
  ``AdS/QHE: Towards a Holographic Description of Quantum Hall Experiments,''
  New J.\ Phys.\  {\bf 13} (2011) 035012
  [arXiv:1008.1917 [hep-th]].

\bibitem{Fujita:2012fp}
  M.~Fujita, M.~Kaminski and A.~Karch,
  ``SL(2,Z) Action on AdS/BCFT and Hall Conductivities,''
  JHEP {\bf 1207} (2012) 150
  [arXiv:1204.0012 [hep-th]].



\bibitem{Lippert:2014jma}
  M.~Lippert, R.~Meyer and A.~Taliotis,
  ``A holographic model for the fractional quantum Hall effect,''
  JHEP {\bf 1501} (2015) 023
  [arXiv:1409.1369 [hep-th]].



 
 
\bibitem{Peet:1998wn}
  A.~W.~Peet and J.~Polchinski,
  ``UV / IR relations in AdS dynamics,''
  Phys.\ Rev.\ D {\bf 59} (1999) 065011
  [hep-th/9809022].

   
  
  
\bibitem{Karch:2007br}
  A.~Karch and A.~O'Bannon,
  ``Holographic thermodynamics at finite baryon density: Some exact results,''
  JHEP {\bf 0711} (2007) 074
  [arXiv:0709.0570 [hep-th]].
 

\bibitem{Ammon:2012je}
  M.~Ammon, M.~Kaminski and A.~Karch,
  ``Hyperscaling-Violation on Probe D-Branes,''
  JHEP {\bf 1211} (2012) 028
  [arXiv:1207.1726 [hep-th]].
  


\bibitem{Arean}
  D.~Arean, A.~V.~Ramallo and D.~Rodriguez-Gomez,
  ``Mesons and Higgs branch in defect theories,''
  Phys.\ Lett.\ B {\bf 641} (2006) 393
  [hep-th/0609010];
   ``Holographic flavor on the Higgs branch,''
  JHEP {\bf 0705} (2007) 044
  [hep-th/0703094.
  



  
  \end{thebibliography}
\end{document}